\documentclass[aps,showpacs,nofootinbib,superscriptaddress,eqsecnum]{revtex4}
\usepackage{natbib}
\usepackage{amsbsy,amsmath,amsfonts}
\usepackage{graphicx}
\usepackage{wick}
\usepackage{float}
\usepackage{rotating}
\usepackage{epsfig}
\usepackage{bm}
\usepackage{color}
\usepackage{multirow}

\newcommand{\MeV}{{\rm \,MeV}}
\newcommand{\tr}{{\rm tr }}

\newcommand{\thrur}[1]{\mathrel{\mathop{#1\!\!\!/}}}
\newcommand{\HQSS}{{\rm HQSS}}
\newcommand{\WT}{{\rm WT}}
\newcommand{\SU}{\mbox{SU}}
\newcommand{\U}{\mbox{U}}
\newcommand{\bc}{{\bar{c}}}
\newcommand{\bq}{{\bar{q}}}
\newcommand{\bU}{\bar{U}}

\newcommand{\bQ}{\bar{Q}}

\newcommand{\pv}{i\overset{\leftrightarrow}{\partial}\!{}_v}

\newcommand{\ben}{\begin{enumerate}}
\newcommand{\een}{\end{enumerate}}

\newcommand{\be}{\begin{equation}}
\newcommand{\ee}{\end{equation}}
\newcommand{\bea}{\begin{eqnarray}}
\newcommand{\eea}{\end{eqnarray}}
\newcommand{\ds}{\begin{displaystyle}}
\renewcommand{\ss}{\begin{scriptstyle}}

\newcommand{\Eq}[1]{Eq.~(\ref{eq:#1})}

\newcommand{\ignore}[1]{}

\newcommand{\ba}{\begin{eqnarray}}
\newcommand{\ea}{\end{eqnarray}}

\newcommand{\Db}{{\bar{D}}}

\newcommand{\Dsb}{{\bar{D}_s}}

\newcommand{\DSb}{{\bar{D}^*}}

\newcommand{\DsSb}{{\bar{D}_s^*}}

\newcommand{\etac}{{\eta_c}}
\newcommand{\Jpsi}{{\psi}}
\newcommand{\Nucleon}{{N}}
\newcommand{\Lambdac}{{\Lambda_c}}
\newcommand{\SigmaS}{{\Sigma^*}}
\newcommand{\Sigmac}{{\Sigma_c}}
\newcommand{\SigmacS}{{\Sigma_c^*}}
\newcommand{\Cascada}{{\Xi}}
\newcommand{\CascadaS}{{\Xi^*}}
\newcommand{\Cascadac}{{\Xi_c}}
\newcommand{\CascadacS}{{\Xi_c^*}}

\newcommand{\Cascadacp}{{\Xi_c^\prime}}
\newcommand{\Omegac}{{\Omega_c}}

\newcommand{\OmegacS}{{\Omega_c^*}}

\newcommand{\GeV}{{\rm \,GeV}}

\begin{document} 

\title{
Hidden charm $N$ and $\Delta$  resonances with heavy-quark symmetry}

\author{C.~Garc\'{\i}a-Recio}
\affiliation{Departamento~de~F\'{\i}sica~At\'omica, Molecular~y~Nuclear, and
  Instituto Carlos I de F\'{\i}sica Te\'orica y Computacional,
  Universidad~de~Granada, E-18071~Granada, Spain}

\author{J.~Nieves}
\affiliation{Instituto~de~F\'{\i}sica~Corpuscular~(centro~mixto~CSIC-UV),
  Institutos~de~Investigaci\'on~de~Paterna, Aptdo.~22085,~46071,~Valencia,
  Spain}

\author{O.~Romanets}
\affiliation{KVI, University~of~Groningen, Zernikelaan~25,~9747AA~Groningen,
  The~Netherlands}

\author{L.~L.~Salcedo}
\affiliation{Departamento~de~F\'{\i}sica~At\'omica, Molecular~y~Nuclear, and
  Instituto Carlos I de F\'{\i}sica Te\'orica y Computacional,
  Universidad~de~Granada, E-18071~Granada, Spain}

\author{L.~Tolos}
\affiliation{Institut~de~Ci\`encies~de~l'Espai~(IEEC/CSIC),
  Campus~Universitat~Aut\`onoma~de~Barcelona, Facultat~de~Ci\`encies,
  Torre~C5,~E-08193~Bellaterra,~Spain}
\affiliation{Frankfurt Institute for Advanced Studies,\\ Johann Wolfgang
  Goethe University, Ruth-Moufang-Str.~1, 60438 Frankfurt am Main}

\date{\today}

\begin{abstract}
We develop a model to describe odd parity baryon resonances generated
dynamically through a unitary baryon-meson coupled-channels approach. The
scheme applies to channels with light and/or heavy quark content. Distinct
features of the model are that, i) the interaction is an $S$-wave contact one,
ii) it reduces to the SU(3) Weinberg-Tomozawa Hamiltonian when light
pseudoscalar mesons are involved, thus, respecting chiral symmetry, iii)
spin-flavor in preserved in the light quark sector, and iv) heavy quark spin
symmetry is fulfilled in the heavy quark sector. In particular, baryon-meson
states with different content in $c$ or in $\bar{c}$ do not mix.  The model is
a minimal one and it contains no free parameters. In this work, we focus on
baryon resonances with hidden-charm (at least one $\bar{c}$ and one $c$
quarks). We analyze several possible sectors and, for the sector with zero net
charm, we write down the most general Lagrangian consistent with SU(3) and
heavy quark spin symmetry. We explicitly study the $N$ and $\Delta$ states,
which are produced from the $S$-wave interaction of pseudoscalar and vector
mesons with $1/2^+$ and $3/2^+$ baryons within the charmless and strangeless
hidden charm sector. We predict seven odd parity $N$-like and five
$\Delta$-like states with masses around $4\GeV$, most of them as bound states.
These states form heavy-quark spin multiplets, which are almost degenerate in
mass. The predicted new resonances definitely cannot be accommodated by quark
models with three constituent quarks and they might be looked for in the
forthcoming $\bar {\mathrm P}$ANDA experiment at the future FAIR facility.
\end{abstract}

%\keywords{}

\pacs{14.20.Gk 14.20.Pt 11.10.St}

\maketitle
\tableofcontents

%\newpage

\section{Introduction}
\label{sec:1}

The possible observation of new states with the charm degree of freedom has
attracted a lot of attention over the past years in connection with many
experiments such as CLEO, Belle, BaBar, and others
\cite{Aubert:2003fg,Briere:2006ff,Krokovny:2003zq,Abe:2003jk,Choi:2003ue,%
  Acosta:2003zx,Abazov:2004kp,Aubert:2004ns,Abe:2007jn,Abe:2007sya,Abe:2004zs,%
  Aubert:2007vj,Uehara:2005qd,Albrecht:1997qa,Frabetti:1995sb,Aubert:2006sp,%
  Abe:2006rz, Artuso:2000xy,Albrecht:1993pt,Frabetti:1993hg,Edwards:1994ar,%
  Ammar:2000uh,Brandenburg:1996jc, Ammosov:1993pi, Aubert:2008if,%
  Mizuk:2004yu,Lesiak:2008wz, Frabetti:1998zt, Gibbons:1996yv, Avery:1995ps,%
  Csorna:2000hw, Alexander:1999ud, Aubert:2007dt, Chistov:2006zj,%
  Jessop:1998wt, Aubert:2006je}. Moreover, the future $\bar {\mathrm P}$ANDA
and CBM experiments at FAIR facility of GSI \cite{cbm,Lutz:2009ff} will aim at
obtaining data in the heavy-flavor sector, thus opening the possibility for
observation of new exotic states with quantum numbers of charm and
strangeness. In this regard, a clear goal would be to understand the nature of
these states, and in particular, whether they can be described with the usual
three-quark baryon or quark-antiquark meson interpretation, or qualify better
as hadron molecules within a baryon-meson coupled-channels description.

Unitarized coupled-channels approaches have shown to be very successful in
describing some of the existing experimental data. These schemes include, for
example, models based on the chiral perturbation amplitudes for $S$-wave
scattering of $0^-$ octet Goldstone bosons off baryons of the nucleon $1/2^+$
multiplet~\cite{Kaiser:1995eg,Kaiser:1995cy, Oset:1997it, Krippa:1998us,
  Krippa:1998ix, Nacher:1999vg, Meissner:1999vr,Oller:2000fj, Jido:2003cb,
  Nieves:2001wt, Inoue:2001ip, Lutz:2001yb, Garcia-Recio:2002td, Oset:2001cn,
  Ramos:2002xh, Tolos:2000fj, Tolos:2002ud, GarciaRecio:2003ks, Oller:2005ig,
  Borasoy:2005ie, Borasoy:2006sr, Hyodo:2008xr}.  Recently the charm degree of
freedom has been incorporated in these models and several experimental states
have been described as dynamically-generated baryon molecules
\cite{Tolos:2004yg, Tolos:2005ft, Lutz:2003jw, Lutz:2005ip, Hofmann:2005sw,
  Hofmann:2006qx, Lutz:2005vx, Mizutani:2006vq, Tolos:2007vh,
  JimenezTejero:2009vq, Haidenbauer:2007jq, Haidenbauer:2008ff,
  Haidenbauer:2010ch, Wu:2010jy, Wu:2010vk, Wu:2012md, Oset:2012ap}.  This is
the case, for example, of the $\Lambda_c(2595)$, which is the charm sector
counterpart of the $\Lambda(1405)$. Some of these approaches are based on a
bare baryon-meson interaction saturated with the $t$-channel exchange of
vector mesons between pseudoscalar mesons and baryons \cite{Tolos:2004yg,
  Tolos:2005ft, Lutz:2003jw, Lutz:2005ip, Hofmann:2005sw, Hofmann:2006qx,
  Lutz:2005vx, Mizutani:2006vq, Tolos:2007vh, JimenezTejero:2009vq}, others
make use of the J\"ulich meson-exchange model \cite{Haidenbauer:2007jq,
  Haidenbauer:2008ff, Haidenbauer:2010ch} or some rely on the hidden gauge
formalism \cite{Wu:2010jy, Wu:2010vk, Wu:2012md, Oset:2012ap}.

Nevertheless, these models do not explicitly incorporate heavy-quark spin
symmetry (HQSS)~\cite{Isgur:1989vq,Neubert:1993mb,Manohar:2000dt}, and
therefore, it is unclear whether this symmetry is respected. HQSS is a QCD
symmetry that appears when the quark masses, such as the charm mass, become
larger than the typical confinement scale. HQSS predicts that all types of
spin interactions involving heavy quarks vanish for infinitely massive
quarks. Thus, HQSS connects vector and pseudoscalar mesons containing charmed
quarks. On the other hand, chiral symmetry fixes the lowest order interaction
between Goldstone bosons and other hadrons in a model independent way; this is
the Weinberg-Tomozawa (WT) interaction. Thus, it is appealing to have a
predictive model for four flavors including all basic hadrons (pseudoscalar
and vector mesons, and $1/2^+$ and $3/2^+$ baryons) which reduces to the WT
interaction in the sector where Goldstone bosons are involved and which
incorporates HQSS in the sector where charm quarks participate. This model was
developed in Ref.~\cite{GarciaRecio:2008dp, Gamermann:2010zz,
  Romanets:2012hm}, following the steps of the SU(6) approach in the light
sector of Refs.~\cite{GarciaRecio:2005hy, GarciaRecio:2006wb, Toki:2007ab,
  Gamermann:2011mq}. In these works, several resonances have been analyzed and
compared to experimental states, such as the $S$-wave states with charm
$C=1,2,3$ \cite{GarciaRecio:2008dp, Romanets:2012hm} together with $C=-1$
states \cite{Gamermann:2010zz}. Also this scheme has been recently extended to
incorporate the bottom degree of freedom \cite{GarciaRecio:2012db} in order to
study the nature of the newly discovered $\Lambda_b(5912)$ and
$\Lambda^*_b(5920)$ resonances \cite{Aaij:2012da} as possible molecular
states.
 
In this paper we aim at continuing those studies on dynamically-generated
baryon resonances using HQSS constraints. We will discuss extensively the
details of the model and how heavy-quark spin symmetry is implemented. The
model respects spin-flavor symmetry in the light sector and HQSS in the heavy
sector, and it reduces to SU(3) WT in the light sector respecting chiral
symmetry. Moreover, we will focus on the dynamical generation of hidden
charmed states. The coupled-channels in the hidden charm sectors are
characterized by containing equal number of $c$- and $\bar c$- quarks.  As we
shall discuss, HQSS does not mix sectors with different number of $c$- or
$\bar c$- quarks. Thus, this model has a symmetry $\SU(6)\times\HQSS$, with
$\HQSS=\SU_c(2)\times\SU_\bc(2)\times \U_c(1)\times \U_\bc(1)$. We will pay a
special attention to the charmless ($C=0$) and strangeless ($S=0$)
sector. Recent works \cite{Wu:2010jy, Wu:2010vk, Wu:2012md, Oset:2012ap}
predict the existence of a few nucleon-like states with masses around $4\GeV$
which result from the baryon-meson scattering in this hidden charm sector. In
this paper we will analyze these results within our model, and predict the
existence of several odd parity $\Delta$- and $N$- like bound states with
various spins. These resonances can be organized in heavy-quark spin
multiplets, which are almost degenerate in mass and that can be subject to
experimental detection.

The predicted new resonances might be subject to experimental detection in the
forthcoming $\bar {\mathrm P}$ANDA/FAIR experiment. If confirmed, they
definitely cannot be accommodated by quark models with three constituent
quarks.

The paper is organized as follows. In Sec. \ref{sec:2} we present the WT
interaction implementing heavy-quark spin symmetry, and analyze the different
hidden charm sectors, classified according to their charmed content. In
Sec.~\ref{sec:uni} we introduce the unitarized coupled-channels approach used
throughout this work.  Sec.~\ref{sec:results} is devoted to present our
results and in Sec.~\ref{sec:conc} we summarize the conclusions. In Appendix
\ref{app:spin-flavor} we give details for the construction of the meson and
baryon tensors and the computation of the different matrix elements of the
interaction.  The tables of the interaction matrices for the different
baryon-meson channels are collected in Appendix \ref{app:tables}.

\section{The model}
\label{sec:2}

\subsection{Weinberg-Tomozawa interaction}

The theoretical model we use has been developed in previous works for
baryon-meson sectors involving light and/or heavy quarks, but not for those
with hidden charm, so we devote this section to fully specify the model.

The guiding principle is to blend several well established hadronic symmetries
in a model as simple as possible. Specifically, to comply with chiral
symmetry, $\SU(N_F)_L\times\SU(N_F)_R$ for $N_F$ flavors, we require the
interaction to reproduce the Weinberg-Tomozawa (WT) Hamiltonian
\cite{Weinberg:1966kf,Tomozawa:1966jm,Birse:1996hd}, a contact $S$-wave
interaction, when light pseudoscalar mesons are involved. The low energy
interaction of soft pseudo-Nambu-Goldstone bosons of the spontaneously broken
chiral symmetry off any (flavored) target takes the WT universal
form\footnote{An extra factor $1/2$ is to be added if the projectile and
  target are identical particles.}
\begin{equation}
V_\WT = \frac{K(s)}{4f^2} 2 J_P^i J_T^i,
\qquad
i=1,\ldots,N_F^2-1 ,
\label{eq:2.1}
\end{equation}
where $f$ is the decay constant of the pseudoscalar meson ($\sim 93\MeV$), and
$J_P^i$, $J_T^i$ are the $\SU(N_F)$ group generators (with the standard
normalization $f_{ikl}f_{jkl}=N_F\delta_{ij}$, where $f_{ijk}$ are the
structure constants) for pseudoscalar meson and
target, respectively. Further, $K(s)=k^0+k^\prime{}^0$ represents the sum of
the incoming and outgoing energies of the meson. In the center-of mass frame
(CM)
\begin{equation}
K(s)= \frac{s-M^2+m^2}{\sqrt{s}},
\end{equation}
where $\sqrt{s}$ is the total CM energy, $M$ the mass of the target, and $m$
the mass of the pseudo-Nambu-Goldstone meson.

$V_\WT$ is the tree level on-shell interaction. The normalization we use is
such that the corresponding $T$-matrix for elastic scattering is related to
the scattering amplitude by
\begin{equation}
f(s)=-\frac{1}{8\pi}\frac{2M}{\sqrt{s}}T(s),
\quad
f(s)=\frac{e^{2i\delta}-1}{2ik}
,
\end{equation}
where $k$ is the CM momentum.\footnote{$T=i\mathcal{M}=-\mathcal{T}$,
  $\mathcal{M}$ and $\mathcal{T}$ being the amplitudes defined in
  \cite{Mandl:1985bg} and \cite{Itzykson:1980rh}, respectively. $2MT$ equals 
  $\mathcal{M}$ in \cite{Beringer:1900zz}.}

For three flavors and baryons in the $\frac{1}{2}^+$ (nucleon) octet, the
Hamiltonian density of the WT interaction takes the form (we assume exact
$\SU(3)$ symmetry for simplicity) \cite{Nieves:2001wt}
\begin{equation}
{\mathcal H}_\WT(x) = -\frac{i}{4f^2}:\tr\left(\bar{B}\gamma^\mu
[[\phi,\partial_\mu\phi],B]\right): \qquad (N_F=3) ,
\end{equation}
where $\phi$ and $B$ are the meson and baryon matrices in the adjoint
representation of SU(3). On account of the trace cyclic property, and
neglecting the meson momentum, the WT interaction can be recast as
\begin{equation}
\frac{i}{4f^2}:\tr\left([\phi,\partial_0\phi]\{B^\dagger,B\}\right):
\label{eq:2.3}
\end{equation}
From the symmetry group point of view, the interaction in \Eq{2.3} is,
schematically,
\begin{equation}
{\mathcal H}_\WT = \frac{1}{f^2}\left((M^\dagger\otimes M)_{\rm
  adjoint,antisymmetric} \otimes (B^\dagger\otimes B)_{\rm adjoint}\right)_{\rm
  singlet}
,
\label{eq:2.6}
\end{equation}
where adjoint and singlet refers to the adjoint and singlet representations of
the flavor group. This corresponds with the structure in \Eq{2.1}.  $M$,
$M^\dagger$, $B$ and $B^\dagger$ refer to the matrices of annihilation and
creation operators of meson and baryons (see below).

It is worth noticing that the eigenvalues of the relevant operator in
\Eq{2.1}, $2 J_P^i J_T^i$, can be written using the quadratic Casimir
operator of $\SU(N_F)$ (see e.g.  \cite{Hyodo:2006kg})
\begin{equation}
(2 J_P^i J_T^i)_\mu = C_2(\mu;N_F)-C_2(\mu_P;N_F)-C_2(\mu_T;N_F) ,
\end{equation}
where $\mu$, $\mu_P$ and $\mu_T$ are, respectively, the $\SU(N_F)$ irreducible
representations (irreps) of the system, the Nambu-Goldstone boson (i.e., the
adjoint representation) and the target. We use the normalization $C_2({\rm
  adjoint};N)=N$ for $\SU(N)$.  In our convention a positive eigenvalue
indicates repulsion and a negative one attraction.

\subsection{Spin-flavor extended Weinberg-Tomozawa interaction}
\label{sec:2.B}

Next we turn to spin-flavor (SF) symmetry, $\SU(2N_F)$,
\cite{Gursey:1992dc,Pais:1964,Sakita:1964qq}. This symmetry has been
phenomenologically successful in the classification of lowest-lying hadrons as
well as in uncovering regularities present in the masses and other hadron
properties \cite{Hey:1982aj,Lebed:1994ga}. This is particularly true for
baryons, a fact that can be understood from the large $N_C$ (number of colors)
limit of QCD. In that limit SF becomes exact for the baryon sector
\cite{Dashen:1993jt}. As for mesons, the lowest-lying states can also be
classified quite naturally according to SF multiplets, but the symmetry works
worse for the meson spectrum. A prime example of this is provided by the pion
and rho mesons. They belong to the same multiplet of SU(6) and this would
require these two mesons to be approximately degenerated in mass. Also, the
pion is a collective state identified as the pseudo-Nambu-Goldstone boson of
the spontaneously broken approximate chiral symmetry, whereas the rho meson
mass fits well with a pair of constituent quark-antiquark. Vector dominance
also suggests that the rho meson should belong to the same chiral
representation of the vector current $(8,1)+(1,8)$, which is different from
the chiral representation of the pion, $(3,3^*)+(3^*,3)$. The apparent
conflict was solved by Caldi and Pagels in two insightful papers
\cite{Caldi:1975tx,Caldi:1976gz}, where a number of related puzzles are
clarified. These authors noted that chiral and SF symmetries are compatible,
as they can be regarded as subgroups of a larger symmetry group,
$\SU(2N_F)_L\times\SU(2N_F)_R$, actually a realization of the
Feynman--Gell-Mann--Zweig algebra \cite{Feynman:1964fk}. In their solution,
the SF extended chiral symmetry is spontaneously broken, the rho meson being a
dormant Goldstone boson of this breaking. The collective nature of the rho
meson has been confirmed in lattice QCD \cite{Smit:1980nf}. As it is well
known, exact SF invariance is not compatible with exact relativistic
invariance \cite{Coleman:1967ad}. In the Caldi-Pagels scenario, vector mesons
acquire mass through SF--breaking relativistic corrections which restore
Poincar{\'e} invariance.

While not an immediate consequence of the QCD Lagrangian, SF symmetry emerges
in some limits, such as large $N_C$ for baryons, as already noted, and
partially also in heavy quark limits. We will turn to this point below. The
lack of exact relativistic invariance is not unusual in other treatments
involving hadron or quark interactions to form new hadrons, either bound
states or resonances. A good example would be the successful ``relativized
quark model'' of Isgur and coworkers \cite{Godfrey:1985xj,Capstick:1986bm}.
Similarly, in our approach the breaking of relativistic invariance is very
mild: the spin is treated as just an internal label (as another kind of
flavor) and our fields are effectively spinless as regards to the
kinematics. So we have fully relativistic kinematics with SF as a purely
internal symmetry. This entails the following. Because angular momentum can be
transferred between orbital and spin components, and spin is not conserved
under Lorentz transformations, a strict relativistic treatment requires fields
with different spin to behave differently. In turn this yields differences in
the off-shell propagators for each spin and this breaks strict SF symmetry. To
the extent that we consider on-shell particles the various fields behave in
the same way, and only the off-shell baryon-meson propagator (or loop
function) would depend on the spin of the particles involved. We disregard
this effect.\footnote{However, any such spin dependence would not be easy to
  extract from phenomenology as it will be masked by the intrinsic ambiguity
  of the loop function, which has to be renormalized using some
  phenomenological prescription.}  On the other hand, because we consider pure
$S$-wave interactions, the spin is separately conserved, and also, near
threshold, the case of interest to us, the specifically relativistic
properties of the spin become irrelevant. It should be noted that even
approaches with formally relativistic Lagrangians are in practice subject to
simplifying approximations in vertices and propagators which break exact
relativistic invariance without relevant phenomenological implications. Also
we remark that SF invariance is just our starting point for modeling the
interaction. Modifications will be introduced below to account for other
established properties of QCD, and more importantly, we use physical values
for hadron masses and meson decay constants in our kinematics.

The compatibility between SF and chiral symmetries implies that the WT
interaction can be extended to enjoy SF invariance, and this can be done in a
unique way \cite{GarciaRecio:2005hy}. For the on-shell vertex the extension is
simply
\begin{equation}
V_\WT^{\rm sf} = \frac{K(s)}{4 f^2} 4 J_M^i J_B^i,
\qquad
i=1,\ldots,(2N_F)^2-1,
\label{eq:2.8}
\end{equation}
where $J_M^i$ and $J_B^i$ are the $\SU(2N_F)$ generators on mesons and
baryons.  Mesons consists of $0^-$ $(P)$ and $1^-$ $(V)$ lowest-lying states,
while baryons contain $\frac{1}{2}^+$ $(B)$ and $\frac{3}{2}^+$ $(B^*)$
lowest-lying states. When this interaction is restricted to the sector $PB\to
PB$ it reproduces the standard WT off $B$ targets. Its SF extension
automatically yields the standard WT for $PB^*\to P B^*$ (hence the
compatibility between the two symmetries). Additionally, the extended
interaction provides contact $S$-wave vertices for $VB\to VB$, $VB^*\to VB^*$,
$PB\leftrightarrow VB$, $PB\leftrightarrow VB^*$, $PB^*\leftrightarrow VB$,
$PB^*\leftrightarrow VB^*$, and $VB\leftrightarrow VB^*$. As we have tried to
argue above, these new vertices are well defined predictions of an approximate
emergent symmetry of hadrons. So we adopt them as our starting point to
describe interactions involving vector mesons.

The Hamiltonian corresponding to the vertex in \Eq{2.8} can be written for any
number of flavors and colors $N_C$ \cite{GarciaRecio:2006wb}. For the physical
case $N_C=3$
\begin{equation}
{\mathcal H}_\WT^{\rm sf}(x)
= -\frac{{\rm i}}{4f^2} :[\Phi, \partial_0 \Phi]^A{}_B
{\cal B}^\dagger_{ACD} {\cal B}^{BCD}:
,
\quad
A,B,\ldots = 1,\ldots,2N_F
.
\label{eq:2.9}
\end{equation}

The indices $A,B,\ldots$, denote spin and flavor, and so they take $2N_F$
values. $\Phi^A{}_B(x)$ is the meson field, a $2N_F\times 2N_F$ Hermitian
matrix which contains the fields of $0^-$ (pseudoscalar) and $1^-$ (vector)
mesons. This matrix is not traceless; for later convenience it includes the
$\SU(2N_F)$ singlet meson (the mathematical $\eta_1$). The contribution of
$\eta_1$ to $\Phi$ is proportional to the identity matrix and so it does not
couple in ${\mathcal H}_\WT^{\rm sf}$. The normalization of $\Phi(x)$ is such
that a mass term (with a common mass $m$ for all mesons) would read
$\frac{1}{2}m^2\tr(\Phi^2)$.

${\cal B}(x)$ is the baryon field. ${\cal B}^{ABC}$ is a completely symmetric
tensor, that is, in the irrep $[3]$ of $\SU(2N_F)$. It has 56 components for
$N_F=3$, and 120 components for $N_F=4$, and contains the lowest-lying baryons
with $J^P=\frac{1}{2}^+$ and $\frac{3}{2}^+$. The normalization of the field
${\cal B}$ is such that a mass term (with a common mass $M$ for all baryons)
would take the form $M\frac{1}{3!}{\cal B}^\dagger_{ABC} {\cal B}^{ABC}$. E.g.
the fields ${\cal B}^{123}(x)$, ${\cal B}^{112}(x)/\sqrt{2}$, and ${\cal
  B}^{111}(x)/\sqrt{6}$ have the standard normalization of a fermionic field.
We refer to the Appendix for the detailed construction of $\Phi^A{}_B(x)$ and
${\cal B}^{ABC}(x)$ in terms of the individual meson and baryon fields for
$N_F=4$.

The Hamiltonian ${\mathcal H}_\WT^{\rm sf}$ has precisely the same structure
displayed in \Eq{2.6}, this time for the SF group $\SU(2N_F)$.\footnote{Note
  that the singlet part in ${\cal B}^\dagger_{ACD} {\cal B}^{BCD}$ does not
  couple since the matrix $[\Phi, \partial_0 \Phi]$ is traceless}

The predictions of the SF extended WT model for $N_F=3$ have been worked out
in \cite{Gamermann:2011mq} for baryonic resonances, and in
\cite{GarciaRecio:2010ki} for the mesonic version. Applications involving
charm have been given in
\cite{GarciaRecio:2008dp,Gamermann:2010zz,Romanets:2012hm}.

Before closing this subsection, we note that the eigenvalues of the relevant
operator $4 J_M^i J_B^i$ in \Eq{2.8} can also be written using the quadratic
Casimir operator of $\SU(2N_F)$
\begin{equation}
(4 J_M^i J_B^i)_\mu = 2\left(C_2(\mu;2N_F)-C_2(\mu_M;2N_F)-C_2(\mu_B;2N_F)\right)
.
\end{equation}
For baryons in the irrep (with Young tableau) $[3]$ of $\SU(2N_F)$ ($\bm{56}$
or $\bm{120}$ for $N_F=3$ or $4$, respectively) and mesons in $[2,1^{2N_F-2}]$
(the adjoint representation of $\SU(2N_F)$, $\bm{35}$ or $\bm{63}$ for $N_F=3$
or $4$, respectively), the baryon-meson states lie in the irreps $[3]$,
$[2,1]$, $[5,1^{2N_F-2}]$ and $[4,2,1^{2N_F-3}]$, which correspond to
$\bm{56}$, ${\bf 70}$, $\bm{700}$ and $\bm{1134}$ for $N_F=3$, and to
$\bm{120}$, ${\bf 168}$, $\bm{2520}$ and $\bm{4752}$ for $N_F=4$. The
corresponding eigenvalues are \cite{GarciaRecio:2006wb}
\begin{equation}
  \lambda_{[3]} = -4N_F,\quad
  \lambda_{[2,1]} = -4N_F-6,\quad
  \lambda_{[5,1^{2N_F-2}]} = 6,\quad
  \lambda_{[4,2,1^{2N_F-3}]} = -2
.
\label{eq:2.11a}
\end{equation}
The SF extended WT interaction is attractive in three multiplets and repulsive
in the remaining one. The sum of all eigenvalues with their multiplicity,
i.e. the trace of $4 J_M^i J_B^i$, is zero, as follows e.g. from
$\tr(J_M^i)=0$. The operator can be written as
\begin{equation}
4 J_M^i J_B^i = \sum_\mu \lambda_\mu P_\mu,
\end{equation}
where $\mu$ are the four baryon-meson irreps of $\SU(2N_F)$, $\lambda_\mu$ the
eigenvalues and $P_\mu$ the orthogonal projectors. This allows to compute the
matrix elements using the $\SU(2N_F)$ Clebsch-Gordan coefficients
\cite{GarciaRecio:2010vf}.

\subsection{Heavy-quark spin symmetry implementation}

Whereas the model just described can be used directly for three flavors, the
extension to include charm requires more care. One reason is that chiral
symmetry and SU(4) invariance are less reliable for fixing the interaction in
the sectors involving charm. At the same time, new specific symmetries of non
relativistic type arise when heavy quarks are involved
\cite{Isgur:1989vq,Neubert:1993mb,Manohar:2000dt}. In the heavy quark limit
the number of charm quarks and the number of charm antiquarks are separately
conserved. This implies a symmetry $\U_c(1)\times \U_\bc(1)$. Likewise, the
terms in the QCD Hamiltonian which depend on the heavy quark or antiquark spin
are suppressed, being of order $1/m_h$, where $m_h$ is the mass of the heavy
quark. Therefore, in the heavy quark limit, arbitrary rotations of the spin
carried by the $c$ quarks and, independently, of the spin carried by the $\bc$
antiquarks, would leave unchanged the energy of the hadronic
state.\footnote{However all $c$ quarks present in the state, being identical
  particles, are rotated by a common rotation, and similarly for the $\bc$.}
This implies a symmetry $\SU_c(2)\times \SU_\bc(2)$ in the heavy quark
limit. These invariances are aspects of heavy-quark spin symmetry (HQSS).  In
what follows we refer to $\SU_c(2) \times \SU_\bc(2) \times \U_c(1)\times
\U_\bc(1)$ as the HQSS group.

The approximate HQSS reflects on the hadronic spectrum and for charm it has a
level of accuracy similar to that of flavor $\SU(3)$. Taking HQSS into account
implies that the model described in the previous subsection, the SF extended
WT or just SU(8)-WT model (for four flavors), has to be slightly modified.  In
order to keep the model simple, we will impose exact HQSS on it. The
alternative would be to introduce instead $1/m_h$ suppressions in some
amplitudes, but such improvement is beyond the scope of the present work.

First, it should be noted that SF by itself already guarantees HQSS in many
sectors. Consider for instance, the couplings involving the channels $ND$ and
$ND^*$. These channels are related through HQSS since there should be
invariance under rotations of the $c$ quark spin (leaving the light quarks
unrotated), and this mixes $D$ and $D^*$. But the same invariance is already
implied by SF, which requires symmetry under independent rotations of spin for
each flavor separately. The only cases where SF does not by itself guarantee
HQSS is when there are simultaneously $c$ quarks and $\bc$ antiquarks: SF
implies invariance under equal rotations for $c$ and $\bc$, but HQSS requires
also invariance when the two spin rotations are different.

To be more specific, let us consider baryon-meson channels, and let $N_c$ be
the number of $c$ quarks and $N_\bc$ the number of $\bc$ antiquarks.  $N_c$
ranges from $0$ to $4$, and $N_\bc$ from $0$ to $1$.  SF guarantees HQSS in
the sectors $(N_c, N_\bc) = (0,0), (0,1), (1,0), (2,0), (3,0), (4,0)$, but not
in the sectors $(1,1), (2,1), (3,1), (4,1)$.  As compared to the former
sectors, the latter ones contain extra $c\bc$ pairs. For the present
discussion, we refer collectively to these sectors as sectors with ``hidden
charm'', regardless of whether they have net charm or not. The hidden charm
sectors are the main subject of the present work. We note that, for $S$-wave
interactions (the ones of interest here), even SU(6) SF, rather than SU(8), is
sufficient to guarantee HQSS in the sectors without hidden charm: a rotation
of the single heavy quark (or antiquark) can be produced by a light sector
rotation followed by a global rotation, without changing the energy. In order
words, in those sectors and for $S$-wave, any SF invariant interaction enjoys
HQSS automatically.

It is perfectly possible to write down non trivial models enjoying
simultaneously SU(8) and HQSS invariances (namely, by requiring
$\SU_q(8)\times \SU_\bq(8)$) but they would not reduce to WT in the light
sector. Concretely SU(8)-WT conserves $C=N_c-N_\bc$ but not $N_c$ and $N_\bc$
separately. Of course, one could impose this by hand, but it is automatically
taken care of by our modified interaction below (\Eq{2.21}). Also, the
restrictions of SU(8)-WT to the sectors $(N_c, N_\bc)= (1,1), (2,1), (3,1),
(4,1)$ turn out to violate HQSS.

In order to implement HQSS in the model let us analyze its content. We extract
the trivial kinematic part and work directly in the space with only spin and
flavor as degrees of freedom. Let
\begin{equation}
H_\WT = 4 J_M^i J_B^i
.
\end{equation}
This operator can be written in terms of meson and baryon operators
\cite{GarciaRecio:2006wb,GarciaRecio:2008dp}, and it contains two distinct
mechanisms which stem from expanding the meson commutator in \Eq{2.9},
\begin{eqnarray}
H_\WT &=& H_{\rm ex} + H_{\rm ac}
,
\nonumber \\
H_{\rm ex} &=&
:M^A{}_C M^{\dagger C}{}_B B^{BDE} B^\dagger_{ADE}:
,
\nonumber \\
H_{\rm ac} &=& 
  -: M^\dagger{}^A{}_C M^C{}_B B^{BDE} B^\dagger_{ADE}:
,
\quad
A,\ldots,E = 1,\ldots,2N_F
.
\end{eqnarray}
Here $M^A{}_B$ and $B^{ABC}$ are the annihilation operators of mesons and
baryons, respectively, with $M^\dagger{}^A{}_B = (M^B{}_A)^\dagger$, and
$B^\dagger_{ABC}=(B^{ABC})^\dagger$. $B^{ABC}$ is a completely symmetric
tensor. They are normalized as
\begin{eqnarray}
[ M^A{}_B, M^{\dagger C}{}_D] &=& \delta^A_D\delta^C_B
,\nonumber \\
\{ B^{ABC}, B^\dagger_{{A^\prime}{B^\prime}{C^\prime}} \}
&=&
\delta^A_{A^\prime} \delta^B_{B^\prime} \delta^C_{C^\prime}
+ \cdots \mbox{~~(6 permutations)}
.
\label{eq:2.11}
\end{eqnarray}

Note that in the SU(8)-WT model, the $\eta_1$ (SU(8) singlet meson) does not
couple and could be ignored, however, this meson has to be present in the
corrected interaction since it mixes with the other mesons under HQSS.

%-------------------------------------------------------------
\begin{figure}[h]
\begin{center}
\includegraphics[width=0.5\textwidth]{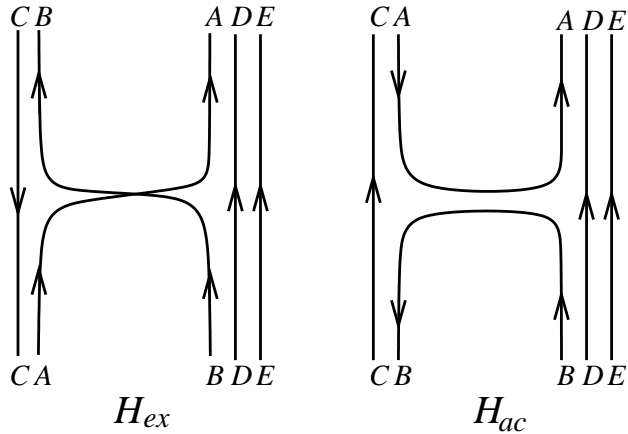}
\caption{\small The two mechanisms acting in the spin-flavor extended WT
  interaction. $H_{\rm ex}$ (exchange of quarks) and $H_{\rm ac}$
  (annihilation and creation of quark-antiquark pairs).  In the HQSS corrected
  version of the interaction, \Eq{2.21a}, the labels $A$ and $B$ in $H_{\rm
    ac}$ only take light-flavor values.}
\label{fig:1}
\end{center}
\end{figure}
%-------------------------------------------------------------

Schematically, representing the quark and antiquark operators by $Q^A$ and
$\bQ_A$,
\begin{equation}
M^A{}_B\sim Q^A\bQ_B, \quad M^\dagger{}^A{}_B\sim \bQ^\dagger{}^A
Q^\dagger_B,\quad B^{ABC}\sim Q^A Q^B Q^C, \quad B^\dagger_{ABC}\sim Q^\dagger_A
Q^\dagger_B Q^\dagger_C \ .
\end{equation}
So, an upper index in $M$ or $B$ represents the SF of a quark to be
annihilated, whereas in $M^\dagger$ it represents that of an antiquark to be
created.  Likewise, a lower index in $M^\dagger$ or $B^\dagger$ represents the
SF of a quark to be created while in $M$ it represents that of an antiquark to
be annihilated. From this identification it is immediate to interpret the two
mechanisms $H_{\rm ex}$ and $H_{\rm ac}$ in terms of quark and antiquark
propagation.

\begin{figure}[h]
\begin{center}
\includegraphics[width=0.60\textwidth]{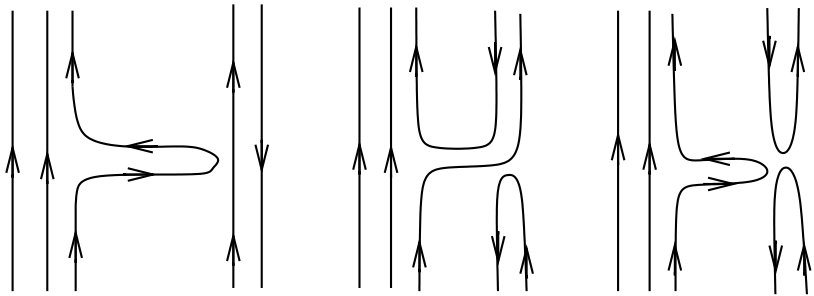}
\caption{\small OZI rule violating mechanisms. Gluons are implicit.}
\label{fig:2}
\end{center}
\end{figure}

The two mechanisms involved, $H_{\rm ex}$ and $H_{\rm ac}$ are displayed in
Fig. \ref{fig:1}. In $H_{\rm ex}$ (exchange) the quark with spin-flavor $A$ is
transferred from the meson to the baryon, as is the quark with label $B$ from
the baryon to the meson. On the other hand, in $H_{\rm ac}$
(annihilation-creation) an antiquark with spin-flavor $B$ in the meson
annihilates with a similar quark in the baryon, with subsequent creation of a
quark and an antiquark with spin-flavor $A$.  In both mechanisms the quarks or
antiquarks $C$, $D$ and $E$ are spectators from the point of their spin-flavor
(the ubiquitous gluons are not explicited). Also in both mechanisms
effectively a meson is exchanged. In passing we note that the OZI rule is
automatically fulfilled as regards to the exchanged meson. OZI rule violating
mechanisms would be of the type depicted in Fig. \ref{fig:2} and are not
present in WT.
%-------------------------------------------------------------
%-------------------------------------------------------------

It appears that $H_{\rm ac}$ can violate HQSS when the annihilation or
creation of $q\bq$ pairs involves heavy quarks, whereas $H_{\rm ex}$ would not
be in conflict with HQSS. This is indeed correct. To expose this fact more
clearly, let us consider the symmetries of these two interaction
mechanisms. Let $N_F=4$ and let $U$ be a matrix of SF $\SU(8)$. Upper indices
transform with $U^\dagger$ and lowers indices with $U$
\begin{eqnarray}
&&
Q^A\to {U^\dagger}{}^A{}_B Q^B, \quad
\bQ^\dagger{}{}^A\to {U^\dagger}{}^A{}_B\bQ^\dagger{}^B,
\nonumber \\ 
&& \bQ_A\to U{}^B{}_A\bQ_B, \quad
Q^\dagger_A\to U{}^B{}_A
Q^\dagger_B
.
\label{eq:2.17}
\end{eqnarray}
Therefore (with obvious matrix/tensor notation)
\begin{eqnarray}
&&
M \to U^\dagger MU
,
\quad
M^\dagger \to U^\dagger M^\dagger U,
\nonumber \\ 
&&
B\to (U^\dagger\otimes U^\dagger\otimes U^\dagger)B
,\quad
B^\dagger \to B^\dagger(U\otimes U\otimes U)
.
\end{eqnarray}

The indices are so contracted that $H_{\rm ex}$ and $H_{\rm ac}$ are both
invariant under these SU(8) transformations.  However, HQSS requires also
invariance when the charm quark and the charm antiquark receive different
rotations. To examine this, let us consider the transformation under
$U\in\SU_q(8)$ and $\bU\in\SU_\bq(8)$, i.e., different transformations for
quarks and antiquarks (previously $U=\bU$). In this case
\begin{eqnarray}
&&
Q\to U^\dagger Q, \quad
\bQ^\dagger\to \bU^\dagger\bQ^\dagger,
%\nonumber \\ && 
\quad
\bQ\to \bQ \bU, \quad
Q^\dagger \to  Q^\dagger U
,
\end{eqnarray}
and therefore
\begin{eqnarray}
&&
M \to U^\dagger M \bU
,
\quad
M^\dagger \to \bU^\dagger M^\dagger U,
\nonumber \\ 
&&
B \to (U^\dagger \otimes U^\dagger \otimes U^\dagger)B
,\quad
B^\dagger \to B^\dagger(U\otimes U\otimes U)
.
\end{eqnarray}

Clearly, the mechanism $H_{\rm ex}$, which depends on the combination
$MM^\dagger$, is still invariant under this larger group $\SU_q(8)\times
\SU_\bq(8)$.\footnote{Note that the commutation relations \Eq{2.11} are also
  preserved by this symmetry.} It certainly preserves SF and HQSS. On the
other hand, $H_{\rm ac}$ depends on the combination $M^\dagger M$ which
transforms with $\bU$, while $BB^\dagger$ transforms with $U$. $H_{\rm ac}$ is
SF invariant ($U=\bU$) but not HQSS invariant. A simple solution to enforce
HQSS with minimal modifications is to remove just the offending contributions
in $H_{\rm ac}$, which come from creation or annihilation of charm
quark-antiquark pairs. This implies to remove the interaction when the labels
$A$ or $B$ are of heavy type in $H_{\rm ac}$.\footnote{Keeping the
  contributions with $A=B$ of heavy type would preserve $\U_c(1)\times
  \U_\bc(1)$, i.e., conservation of $N_c$ and $N_\bc$, but not $\SU_c(2)\times
  \SU_\bc(2)$.}  That is, we adopt the following modified version of the
$H_{\rm ac}$ mechanism
\begin{equation}
H^\prime_{\rm ac} =
  -: M^\dagger{}^{\hat{A}}{}_C M^C{}_{\hat{B}} B^{{\hat{B}}DE} B^\dagger_{{\hat{A}}DE}:
,
\quad
C,D,E = 1,\ldots,8
,\quad
\hat{A},\hat{B} = 1,\ldots,6
.
\label{eq:2.21a}
\end{equation}
The indices with hat are restricted to $\SU(6)$. By construction $N_c$ and
$N_\bc$ are exactly conserved in $H^\prime_{\rm ac}$. Also $\SU_c(2)\times
\SU_\bc(2)$ is conserved: when $U$ and $\bU$ act only on the heavy sector
$M^\dagger{}^{\hat{A}}{}_C M^C{}_{\hat{B}}$ and $B^{{\hat{B}}DE}
B^\dagger_{{\hat{A}}DE}$ are unchanged. So HQSS is preserved. On the other
hand, when $U=\bU$ and this matrix acts on the light sector, $H^\prime_{\rm
  ac}$ is unchanged, so it enjoys SU(6) symmetry. Exact SF SU(8) and flavor
SU(4) is no longer present. Presumably this breaking of SU(4) is comparable to
the breaking through the kinematics due to the substantially heavier mass of
the charmed quark.

To summarize, our model (in all sectors) is given by
\begin{equation}
V = \frac{K(s)}{4f^2} H^\prime_\WT
,\qquad
H^\prime_\WT = H_{\rm ex}+H^\prime_{\rm ac}
.
\label{eq:2.21}
\end{equation}
This model fulfills some desirable requirements: i) It has symmetry
$\SU(6)\times \HQSS$, i.e., SF symmetry in the light sector and HQSS in the
heavy sector, the two invariances being compatible. ii) It reduces to SU(6)-WT
in the light sector, so it is consistent with chiral symmetry in that
sector. And, iii) is a minimal modification that preserves simplicity and does
not introduce new adjustable parameters.

\subsection{The model in the various sectors}

We can analyze the model in the different $(N_c,N_\bc)$ sectors, which as
already noted, do not mix due to HQSS.

In all the sectors without hidden charm, namely, $(N_c,N_\bc)=(0,1), (0,0),
(1,0), (2,0), (3,0), (4,0)$, $H^\prime_{\rm ac}$ produces the same amplitudes
as $H_{\rm ac}$ when the latter is restricted to the corresponding sector.
Indeed these interactions vanish unless the state contains a quark-antiquark
pair with quark and antiquark of the same type. In the absence of hidden
charm, the pair must be light and in this case the two operators produce the
same result. This is consistent with our previous observation that, when there
are only heavy quarks or heavy antiquarks but not both, SF already implies
HQSS.  So in all these sectors, our model produces the same amplitudes as
SU(8)-WT after removing channels involving hidden charm. This observation has
been applied in \cite{GarciaRecio:2008dp,Gamermann:2010zz,Romanets:2012hm,GarciaRecio:2012db}.

It is noteworthy that, in the sectors $(N_c,N_\bc)=(0,1)$ and $(4,0)$,
corresponding to $C=-1$ and $C=4$, $H^\prime_{\rm ac}=H_{\rm ac}=0$ as they do
not contain light quark-antiquark pairs. Also these two sectors cannot couple
to any other $(N_c,N_\bc)$ sector in the baryon-meson case. Therefore for
them, our model coincides directly with SU(8)-WT.

Let us turn now to the sectors with hidden charm. These are
$(N_c,N_\bc)=(1,1), (2,1), (3,1), (4,1)$. For all these sectors $H^\prime_{\rm
  ac}$ vanishes. The reason is that in these sectors the relevant
quark-antiquark pair (quark and antiquark with equal labels) is necessarily of
heavy type, and such amplitude has been removed from $H^\prime_{\rm
  ac}$. (Note that $H_{\rm ac}$ does not vanish in these sectors.) So for the
hidden charm sectors $H^\prime_\WT$ reduces to the exchange mechanism $H_{\rm
  ex}$.

The interaction is effectively $H_{\rm ex}$ for the four hidden charm sectors
and also for $C=-1$ and $C=4$. This has the immediate consequence that the
interaction $H^\prime_\WT$ has only two eigenvalues, namely $-2$ (attractive)
and $6$ (repulsive), in those sectors.\footnote{Remarkably, these two
  eigenvalues are the only $N_F$-independent ones in \Eq{2.11a}.} This follows from the fact that $H_{\rm
  ex}$ has a large (accidental) symmetry group, $\SU_q(8)\times \SU_\bq(8)$,
which produces only two large multiplets of degenerated states in the
baryon-meson coupled-channels space. Under $\SU_q(8)\times \SU_\bq(8)$ the
baryons fall in the irrep $(\bm{120},\bm{1})$ ($\bm{120}$ being the symmetric
representation of three quarks with $8$ possible spin-flavor
states).\footnote{Fermi-Dirac statistic is taken care of by the antisymmetric
  color wavefunction.} Likewise, the mesons belong to $(\bm{8},\bm{8}^*)$
(being $q\bq$ states). Therefore, the baryon-meson states form two
$\SU_q(8)\times \SU_\bq(8)$ multiplets:
\begin{equation}
(\bm{120},\bm{1})\otimes (\bm{8},\bm{8}^*)
= 
(\bm{330},\bm{8}^*)\oplus (\bm{630},\bm{8}^*)
,
\end{equation}
($\bm{330}$ is the symmetric representation $[4]$ of SU(8) while $\bm{630}$
corresponds to the mixed symmetry $[3,1]$). The two corresponding eigenvalues
are\footnote{They can be obtained by applying $H_{\rm ex}$ to two suitable
  states, e.g., $(M^\dagger{}^1{}_1B^\dagger_{211} \pm
  M^\dagger{}^1{}_2B^\dagger_{111})|0\rangle$.  }
\begin{equation}
\lambda_{(\bm{330},\bm{8}^*)} = 6
,
\quad
\lambda_{(\bm{630},\bm{8}^*)}= -2
.
\end{equation}
These two eigenvalues are also present in the original SU(8)-WT interaction
(namely, $\lambda_{\bm{2520}}=6$ and $\lambda_{\bm{4752}}=-2$ from \Eq{2.11a})
since the two interactions coincide for $C=-1$ or $C=4$. It can also be noted
that under SF SU(8), these multiplets reduce as follows
\begin{equation}
(\bm{330},\bm{8}^*) = \bm{120} \oplus \bm{2520}
,\qquad
(\bm{630},\bm{8}^*) = \bm{120} \oplus \bm{168} \oplus \bm{4752}
.
\end{equation}
Of course, these are the same SU(8) irreps obtained from $\bm{120}\otimes
\bm{63}$ (baryon-meson except $\eta_1$) and $\bm{120}\otimes {\bf 1}$
(baryon-$\eta_1$). Therefore, the interaction can be written using SU(8)
Clebsch-Gordan coefficients by means of\footnote{One exception comes from the
  two $\bm{120}$ irreps, which differ by the type of symmetry of the four
  quarks. This information is not contained in the SU(8) Clebsch-Gordan
  coefficients.}
\begin{equation}
H_{\rm ex} = 6 (P^\prime_{\bm{120}} +P_{\bm{2520}}) -2 (P_{\bm{120}} +
  P_{\bm{168}} + P_{\bm{4752}}).
\end{equation}

We emphasize that the large multiplets $(\bm{330},\bm{8}^*)$ and
$(\bm{630},\bm{8}^*)$ are not realized in our model. First of all, they
contain sectors without hidden charm, for which the interaction does not
reduce to the exchange mechanism $H_{\rm ex}$. And second, the eigenvalues $6$
and $-2$ refer only to the driving operator $4J_M^iJ_B^i$. The vertex $V$ in
\Eq{2.21} depends also on hadron masses and meson decay constants, for which
we use physical values in $V$ and in the propagators.  The values that we use
for these magnitudes are collected in Table II of \cite{Romanets:2012hm}.

\subsection{Analysis of the hidden charm sectors}
\label{subsec:SymStr}

Here we consider hidden charm sectors with $C=0,1,2,3$, i.e.  $N_{\bc}=1$ and
$N_c=1,2,3,4$, respectively. We want to classify the possible states under the
symmetry group $\SU(6)\times \HQSS$, with
$\HQSS=\SU_c(2)\times\SU_\bc(2)\times \U_c(1)\times \U_\bc(1)$.  Since in the
hidden charm sectors there is exactly one heavy antiquark, it is not necessary
to specify the irrep of the factor $\SU_\bc(2)\times \U_\bc(1)$ and we can use
the notation ${\bm R}_{2J_c+1,C}$ for the irreps of $\SU(6)\times \HQSS$,
${\bm R}$ being the $\SU(6)$ irrep of the light sector, $C$ the charm quantum
number and $J_c$ the total spin carried by the one or more $c$ quarks (not
including the spin of the $\bc$ antiquarks). The corresponding dimension is
$\mbox{dim}\,{\bm R}\times (2J_c+1)\times 2$ (the last factor coming from the
two possible spin states of the $\bc$).

Subsequently, we study the breaking of light SF down to $\SU(3)\times \SU(2)$
keeping HQSS, and enumerate the number of attractive channels in each
$(C,\bm{r},J)$ sector, where $\bm{r}$ is the SU(3) irrep and $J$ the total
spin.

In practice we will assume exact isospin and spin $\SU(2)_I\times \SU(2)_J$,
as well as conservation of each flavor, but not exact SU(3) and HQSS, for the baryons
and mesons composing the coupled-channels space. Therefore the sectors are
labeled by $(C,S,I,J)$, $S$ being the strangeness quantum number and $I$ the
isospin. This implies a further breaking of each $(C,\bm{r},J)$ sector into
$(C,S,I,J)$ subsectors.

\subsubsection{$C=0$}

For $C=0$, the quark content is $\ell\ell\ell c\bc$, with two possibilities of
grouping into baryon-meson: $(\ell\ell\ell) (c\bc)$ and $(\ell\ell c)(\ell
\bc)$. (Here $\ell$ denotes any light flavor quark, $u,d,s$.)  The total
dimension of the space is $56\times 2 \times 2 + 21 \times 2 \times 6 \times 2
= 728$, and contains the following $\SU(6)\times \HQSS$
multiplets\footnote{$(\ell\ell\ell) (c\bc)$ is purely $\bm{56}_{2,0}$ from the
  symmetry of the three light quarks. The two light quarks in $(\ell\ell c)$
  are symmetric giving a $\bm{21}$ of $\SU(6)$, coupled to the further light
  quark in $(\ell \bc)$ gives $\bm{21}\otimes \bm{6} = \bm{56}\oplus
  \bm{70}$. These two $\bm{56}_{2,0}$ are not directly those in \Eq{2.26}.}
\begin{equation}
\mathcal{H}_{C=0} = \bm{56}_{2,0} \oplus \bm{56}_{2,0} \oplus \bm{70}_{2,0}
\qquad
(\SU(6)\times\HQSS)
.
\label{eq:2.26bis}
\end{equation}
The eigenvalues turn out to be
\begin{equation}
\lambda_{\bm{56}_{2,0}} = 
\lambda_{\bm{70}_{2,0}} = -2
,\qquad
\lambda^\prime_{\bm{56}_{2,0}} = 6
.
\label{eq:2.26}
\end{equation}
The accidental degeneracy between $\bm{70}_{2,0}$ and one $\bm{56}_{2,0}$
takes place in our model and it is not a necessary consequence of
$\SU(6)\times\HQSS$.  This symmetry does not fix the three possible
eigenvalues and the precise splitting between the two copies of
$\bm{56}_{2,0}$. In general, the accidental degeneracy will be lifted in $V$
and the $T$-matrix even when an exact $\SU(6)\times\HQSS$ invariance is
assumed in the hadron masses and meson decay constants.

Next, we consider the breaking of light SF SU(6) down to
$\SU(3)\times\SU_{J_\ell}(2)$. E.g. $\bm{56} = \bm{8}_2 \oplus \bm{10}_4$.
HQSS is unbroken. After recoupling the spin carried by light and heavy quarks
and antiquarks to yield the total spin $J$, we obtain representations of
$\SU(3)\times\SU(2)_J$ labeled as $\bm{r}_{2J+1}$, where $\bm{r}$ is the SU(3)
irrep. This yields the following reductions (the two $\bm{56}_{2,0}$ have the
same reduction)
\begin{eqnarray}
\bm{56}_{2,0}
&=&
(\bm{8}_2 \oplus \bm{10}_4)_{2,0}
=
(\bm{8}_2 \oplus\bm{8}_2 \oplus \bm{8}_4)
\oplus
(\bm{10}_2 \oplus \bm{10}_4 \oplus\bm{10}_4 \oplus \bm{10}_6)
,
\label{eq:2.29}
 \\
\bm{70}_{2,0}
&=&
(\bm{1}_2 \oplus \bm{8}_2 \oplus\bm{8}_4 \oplus\bm{10}_2)_{2,0}
=
(\bm{1}_2 \oplus \bm{1}_2 \oplus \bm{1}_4)
\oplus
(\bm{8}_2 \oplus\bm{8}_2 \oplus \bm{8}_4)
\oplus
(\bm{8}_2 \oplus \bm{8}_4 \oplus\bm{8}_4 \oplus \bm{8}_6)
\oplus
(\bm{10}_2 \oplus \bm{10}_2 \oplus\bm{10}_4 )
.
\nonumber
\end{eqnarray}

In the reduction $(\bm{8}_2)_{2,0}= (\bm{8}_2 \oplus\bm{8}_2 \oplus \bm{8}_4)$
in $\bm{56}_{2,0}$, the three octets only differ in how the light sector spin
is coupled to the heavy sector spin, therefore these three irreps are
degenerated if exact HQSS is assumed. $(\bm{8}_2 \oplus\bm{8}_2 \oplus
\bm{8}_4)$ is a multiplet of $\SU(3)\times \HQSS$. Similar statements hold in
the other cases: each $\bm{56}_{2,0}$ produces two such multiplets and
$\bm{70}_{2,0}$ produces four. Consequently, in the hidden charm sector with
$C=0$ we expect to find 8 different eigenvalues after $\SU(6)\times \HQSS$ is
broken down to $\SU(3)\times \HQSS$. Let $\lambda_1$, $\lambda_2$ be the
eigenvalues of the two multiplets in the repulsive $\bm{56}_{2,0}$,
$\lambda_3$, $\lambda_4$ in the attractive $\bm{56}_{2,0}$, and $\lambda_5$,
$\lambda_6$, $\lambda_7$, $\lambda_8$ those in $\bm{70}_{2,0}$.  In this case,
the spectra in each $(C,{\bm r},J)$ sector is as follows:
\begin{eqnarray}
\bm{1}_2 &:&
(\lambda_5,\lambda_5)
,
\nonumber\\
\bm{1}_4 &:&
(\lambda_5)
,
\nonumber\\
\bm{8}_2 &:&
(\lambda_1,\lambda_1,\lambda_3,\lambda_3,\lambda_6,\lambda_6,\lambda_7)
,
\nonumber\\
\bm{8}_4 &:&
(\lambda_1,\lambda_3,\lambda_6,\lambda_7,\lambda_7)
,
\nonumber\\
\bm{8}_6 &:&
(\lambda_7)
,
\nonumber\\\bm{10}_2 &:&
(\lambda_2,\lambda_4,\lambda_8,\lambda_8)
,
\nonumber\\
\bm{10}_4 &:&
(\lambda_2,\lambda_2,\lambda_4,\lambda_4,\lambda_8)
,
\nonumber\\
\bm{10}_6 &:&
(\lambda_2,\lambda_4) \label{eq:c0lambda}
.
\end{eqnarray}
In the SU(6) limit, $\lambda_1=\lambda_2$, $\lambda_3=\lambda_4$,
$\lambda_5=\lambda_6 = \lambda_7=\lambda_8$. Breaking down the symmetry to
SU(3), one expects
\begin{equation}
\lambda_{3,4,5,6,7,8} < 0 < \lambda_{1,2}
.
\end{equation}

Each negative eigenvalue can give rise to a resonance or bound state. Each
such state is a full multiplet of $\SU(3)\times \SU(2)_J$.  This implies the
following expected number of states in each $(C,\bm{r},J)$ sector: up to two
states in $\bm{1}_2$, one in $\bm{1}_4$, five in $\bm{8}_2$, four in
$\bm{8}_4$, one in $\bm{8}_6$, three in ${\bf 10}_2$, three in $\bm{10}_4$,
and one in $\bm{10}_6$, all of them with $C=0$.

\subsubsection{$C=1$}

For $C=1$ there are two baryon-meson structures, namely, $(\ell\ell c) (c\bc)$
and $(\ell c c) (\ell\bc)$. The total dimension of the space is $384$, with
the following reduction under $\SU(6)\times\HQSS$
\begin{equation}
\mathcal{H}_{C=1} = 
\bm{21}_{3,1} \oplus \bm{21}_{3,1} \oplus  \bm{21}_{1,1} \oplus \bm{15}_{3,1}
\qquad
(\SU(6)\times\HQSS)
,
\end{equation}
and eigenvalues
\begin{equation}
\lambda_{\bm{21}_{3,1}}=\lambda_{\bm{21}_{1,1}}=\lambda_{\bm{15}_{3,1}}=-2
,
\qquad
\lambda^\prime_{\bm{21}_{3,1}}=6
.
\end{equation}
Once again the accidental degeneracy beyond $\SU(6)\times\HQSS$ is lifted in
the $T$-matrix.

After the breaking  $\SU(6)\supset \SU(3)\times\SU(2)_{J_\ell}$, and
recoupling to $J$, one finds
\begin{eqnarray}
\bm{21}_{3,1}
&=&
(\bm{6}_3 \oplus \bm{3}^*_1)_{3,1}
=
(\bm{6}_2 \oplus \bm{6}_2 \oplus \bm{6}_4 \oplus \bm{6}_4 \oplus \bm{6}_6)
\oplus
(\bm{3}^*_2 \oplus \bm{3}^*_4)
,
\nonumber \\
\bm{21}_{1,1}
&=&
(\bm{6}_3 \oplus \bm{3}^*_1)_{1,1}
=
(\bm{6}_2 \oplus \bm{6}_4)
\oplus
(\bm{3}^*_2 )
,
\nonumber \\
\bm{15}_{3,1}
&=&
(\bm{6}_1 \oplus \bm{3}^*_3)_{3,1}
=
(\bm{6}_2 \oplus \bm{6}_4)
\oplus
(\bm{3}^*_2 \oplus \bm{3}^*_2 \oplus \bm{3}^*_4 \oplus \bm{3}^*_4 \oplus \bm{3}^*_6)
.
\end{eqnarray}
Thus for $C=1$, each of the four SU(6) irreps split into two
$\SU(3)\times\HQSS$ multiplets.\footnote{Note that the two $(\bm{6}_2 \oplus
  \bm{6}_4)$ multiplets above do not mix if HQSS holds, as they carry
  different heavy quark spin labels. Such label has been left implicit to
  avoid clumsiness.}

In principle there are eight different eigenvalues. Denoting by
$\lambda_1,\lambda_2$ the eigenvalues of the two multiplets in the repulsive
$\bm{21}_{3,1}$, $\lambda_3,\lambda_4$ for the attractive $\bm{21}_{3,1}$,
$\lambda_5,\lambda_6$ for $\bm{21}_{1,1}$, and $\lambda_7,\lambda_8$ for
$\bm{15}_{3,1}$, the spectra in each $(C,{\bm r},J)$ sector is as follows:
\begin{eqnarray}
\bm{3}^*_2 &:&
(\lambda_2,\lambda_4,\lambda_6,\lambda_8,\lambda_8)
,
\nonumber\\
\bm{3}^*_4 &:&
(\lambda_2,\lambda_4,\lambda_8,\lambda_8)
,
\nonumber\\
\bm{3}^*_6 &:&
(\lambda_8)
,
\nonumber\\
\bm{6}_2 &:&
(\lambda_1,\lambda_1,\lambda_3,\lambda_3,\lambda_5,\lambda_7)
,
\nonumber\\
\bm{6}_4 &:&
(\lambda_1,\lambda_1,\lambda_3,\lambda_3,\lambda_5,\lambda_7)
,
\nonumber\\
\bm{6}_6 &:&
(\lambda_1,\lambda_3) \label{eq:c1lambda}
.
\end{eqnarray}
In the SU(6) limit, $\lambda_1=\lambda_2$, $\lambda_3=\lambda_4$,
$\lambda_5=\lambda_6$, and $ \lambda_7=\lambda_8$. If only SU(3) symmetry is
assumed, we expect the following number of states: up to four states in
$\bm{3}^*_2$, three in $\bm{3}^*_4$, one in $\bm{3}^*_6$, four in $\bm{6}_2$,
four in $\bm{6}_4$, and one in $\bm{6}_6$.

\subsubsection{$C=2$}

For the hidden charm sector with $C=2$ there are two baryon-meson quark
structures, $(\ell cc) (c\bc)$ and $(c c c) (\ell\bc)$. The space has
  dimension 120, with the following reduction and eigenvalues:
\begin{eqnarray}
\mathcal{H}_{C=2} &=& 
\bm{6}_{2,2}\oplus\bm{6}_{4,2}\oplus \bm{6}_{4,2}
\qquad
(\SU(6)\times\HQSS)
\nonumber \\
\lambda_{\bm{6}_{2,2}} &=& 
\lambda_{\bm{6}_{4,2}} = -2
,\quad
\lambda^\prime_{\bm{6}_{4,2}} = 6
.
\end{eqnarray}

The $\SU(3)\times\HQSS$ multiplets are as follows:
\begin{eqnarray}
\bm{6}_{2,2}
&=&
(\bm{3}_2 )_{2,2}
=
(\bm{3}_2 \oplus \bm{3}_2 \oplus \bm{3}_4 )
,
\nonumber \\
\bm{6}_{4,2}
&=&
(\bm{3}_2)_{4,2}
=
(\bm{3}_2 \oplus \bm{3}_4 \oplus \bm{3}_4 \oplus \bm{3}_6)
.
\end{eqnarray}
In this case the $\SU(6)\times\HQSS$ multiplets are not reduced further under
$\SU(3)\times\HQSS$. Calling $\lambda_1$ the eigenvalue of the repulsive
$\bm{6}_{4,2}$, $\lambda_2$ that of the attractive $\bm{6}_{4,2}$, and
$\lambda_3$ the one of $\bm{6}_{2,2}$, yields the following spectra for the
three $(C,{\bm r},J)$ sectors:
\begin{eqnarray}
\bm{3}_2 &:&
(\lambda_1,\lambda_2,\lambda_3,\lambda_3)
,
\nonumber\\
\bm{3}_4 &:&
(\lambda_1,\lambda_1,\lambda_2,\lambda_2,\lambda_3)
,
\nonumber\\
\bm{3}_6 &:&
(\lambda_1,\lambda_2)
.\label{eq:c2lambda}
\end{eqnarray}
This produces the following expected maximum number of states: three in
$\bm{3}_2$, three in $\bm{3}_4$ and one in $\bm{3}_6$

It can be noted in the present case, $C=2$ with hidden charm, the assumption
of SU(6) invariance does not add anything (does not reduce the number of
parameters) on top of that of SU(3). The reason is that here HQSS
automatically implies SF: there is just one light quark and rotations of it
can be produced by combining global rotations with heavy quark rotations.

\subsubsection{$C=3$}

For $C=3$ there is just one quark structure, $(ccc)c\bc$. The dimension is
$16$.  The $\SU(6)\times\HQSS$ reduction is
\begin{eqnarray}
\mathcal{H}_{C=3}
&=&
\bm{1}_{3,3} \oplus \bm{1}_{5,3}
\qquad
(\SU(6)\times\HQSS)
,
\end{eqnarray}
with eigenvalues
\begin{equation}
\lambda_{\bm{1}_{3,3}} = -2
,\qquad
\lambda_{\bm{1}_{5,3}} = 6
.
\end{equation}

The $\SU(3)\times\SU(2)_J$ reduction is:
\begin{eqnarray}
\bm{1}_{3,3}
&=&
(\bm{1}_1 )_{3,3}
=
(\bm{1}_2 \oplus \bm{1}_4 )
,
\nonumber \\
\bm{1}_{5,3}
&=&
(\bm{1}_1)_{5,3}
=
(\bm{1}_4 \oplus \bm{1}_6)
.
\end{eqnarray}
Once again, the $\SU(6)\times\HQSS$ multiplets are not reduced further under
$\SU(3)\times\HQSS$.

The following spectra is obtained for the various $(C,{\bm r},J)$ sectors
\begin{eqnarray}
\bm{1}_2 &:&
(\lambda_2)
,
\nonumber\\
\bm{1}_4 &:&
(\lambda_1,\lambda_2)
,
\nonumber\\
\bm{1}_6 &:&
(\lambda_1)\label{eq:c3lambda}
,
\end{eqnarray}
where $\lambda_1$ denotes the eigenvalue corresponding to $\bm{1}_{5,3}$, and
$\lambda_2$ that of $\bm{1}_{3,3}$. So no states will be produced in
$\bm{1}_6$, and up to one state is expected in $\bm{1}_2$ and $\bm{1}_4$.

Some of the results of this subsection are summarized in Table
\ref{tab:sectors}.

%%%%%%%%%%%%%%%%%%%%%%%% TABLE
\begin{table*}[ht]
\begin{center}
\begin{tabular}{| c | c | ccc | cc | c | c | c |}
\hline 
    & $C$ & \multicolumn{3}{c|}{$0$} & \multicolumn{2}{c|}{$1$} & $2$ & $3$ \\
\hline
$J$ & $\SU(3)$ & $\bm{1}$ & $\bm{8}$ & $\bm{10}$ & $\bm{3}^*$ & $\bm{6}$ 
& $\bm{3}$  & $\bm{1}$ \\ 
\hline
\multirow{2}{*}{$1/2$} & &  2  &  7  &  4  &  5  &  6  &  4  &  1 \\
       & & (2) & (5) & (3) & (4) & (4) & (3) & (1) \\
\hline
\multirow{2}{*}{$3/2$} & &  1  &  5  &  5  &  4  &  6  &  5  &  2 \\
      & & (1) & (4) & (3) & (3) & (4) & (3) & (1) \\
\hline
\multirow{2}{*}{$5/2$} & &  0  &  1  &  2  &  1  &  2  &  2  &  1 \\
      & & (0) & (1) & (1) & (1) & (1) & (1) & (0) \\
\hline
\end{tabular}
\end{center}
\caption{Total number of channels for each $J$ and each $\SU(3)$ irrep, for
  the various hidden charm sectors. Here each channel represents a full
  $\SU(3)$ and spin multiplet $(C,\bm{r},J)$ (rather than a isospin-spin
  multiplet, $(C,S,I,J)$.) The expected number of resonances in each sector is
  shown between parenthesis.The actual number of resonances will depend on the
  values of the physical masses and meson decay constants.}
\label{tab:sectors}
\end{table*}

\subsection{Lagrangian form of the interaction}
\label{sec:2.F}

In the strict heavy quark limit, the total spin $J$, and the separate spins of
the light, heavy quarks and heavy antiquarks subsystems are conserved, when
only $S$-wave interactions are considered. Moreover, the matrix elements of
QCD Hamiltonian depend only on the spin-flavor quantum numbers of the light
degrees of freedom.  Each $n$-fold degenerated $\SU(6)$ or $\SU(3)$ multiplet
implies the specification of $n(n+1)/2$ parameters which are coupling
constants of the corresponding operators present in the interaction. From the
previous analysis it follows that the number of independent operators in the
hidden charm sectors with $C=0,1,2,3$ is $4,5,4,2$, respectively, for a
generic interaction if $\SU(6)\times\HQSS$ invariance is assumed and
$12,10,4,2$ if the symmetry is reduced to $\SU(3)\times\HQSS$.

In what follows we will focus on the sector with hidden charm and $C=0$. For
this sector we will write down the most general (modulo kinematical factors)
$S$-wave Lagrangian consistent with $\SU(3)\times\HQSS$ for the baryon-meson
coupled-channels space. This Lagrangian contains 12 operators and our model
gives well defined values for the corresponding coupling constants.  We
should note that we actually compute the matrix elements of our interaction
using directly the previous expressions, either in terms of projectors using
Clebsch-Gordan coefficients, or of hadron creation and annihilation operators
in spin-flavor space, by taking Wick contractions.  Nevertheless, writing the
interaction in field-theoretical Lagrangian form is highly interesting in
order to make contact with alternative approaches in the literature.

For this purpose it is convenient to organize the hadrons forming multiplets
of HQSS into building blocks with well defined HQSS transformation properties
\cite{Falk:1990yz,Grinstein:1992qt,Jenkins:1992nb,Manohar:2000dt}. Specifically,
consider a HQSS doublet composed of pseudoscalar meson and vector meson with
one heavy quark (e.g. $D$ and $D^*$). Let $M^{(c)}$ and $M^{(c)}_\mu$ be the
corresponding fields, then
\begin{eqnarray}
\bm{M}^{(c)} &=& Q_+(M^{(c)}_\mu{}^{(+)} \gamma^\mu + M^{(c)}{}^{(+)}  \gamma_5),
\nonumber \\
\overline{\bm{M}}{}^{(c)} &=& \gamma_0 \bm{M}^{(c)}{}^\dagger \gamma_0 
= (M^{(c)}{}^\dagger_\mu{}^{(-)} \gamma^\mu - M^{(c)}{}^\dagger{}^{(-)} \gamma_5)Q_+
.
\end{eqnarray}
As usual $(\pm)$ represent the positive and negative frequency part of the
fields, corresponding to purely annihilation for $(+)$ and purely creation for
$(-)$. Therefore, $\bm{M}^{(c)}$ [$\overline{\bm{M}}{}^{(c)}$] annihilates
[creates] the meson with one heavy quark but it does not create [annihilate]
the corresponding antimeson with a heavy antiquark. A similar proviso is
applied in all the other fields for hadrons carrying heavy quarks and/or
antiquarks. Besides,
\begin{equation}
Q_\pm = 
\frac{1\pm \thrur{v}}{2}
\end{equation}
where $v^\mu$ is the heavy hadron velocity ($v^2=1$). We use Bjorken and Drell
\cite{Bjorken:1965zz} conventions for the Dirac gammas. For the hadron fields
we use the conventions of \cite{GarciaRecio:2010vf} and this fixes the
relative sign between pseudoscalar and vector (see Appendix
\ref{app:spin-flavor}).

Likewise, for a HQSS meson  doublet with one heavy antiquark
\begin{eqnarray}
\bm{M}^{(\bc)} &=& (M^{(\bc)}_\mu{}^{(+)} \gamma^\mu + M^{(\bc)}{}^{(+)}  \gamma_5)Q_-,
\nonumber \\
\overline{\bm{M}}{}^{(\bc)} &=& \gamma_0 \bm{M}^{(\bc)}{}^\dagger \gamma_0 
= Q_- (M^{(\bc)}{}^\dagger_\mu{}^{(-)} \gamma^\mu - M^{(\bc)}{}^\dagger{}^{(-)} \gamma_5)
.
\end{eqnarray}
For a HQSS meson  doublet with one heavy quark and one heavy antiquark
(e.g., $\eta_c$ and $J/\psi$):
\begin{eqnarray}
\bm{M}^{(c\bc)} &=& Q_+(M^{(c\bc)}_\mu{}^{(+)} \gamma^\mu + M^{(c\bc)}{}^{(+)}  \gamma_5)Q_-,
\nonumber \\
\overline{\bm{M}}{}^{(c\bc)} &=& \gamma_0 \bm{M}^{(c\bc)}{}^\dagger \gamma_0 
= Q_- (M^{(c\bc)}{}^\dagger_\mu{}^{(-)} \gamma^\mu - M^{(c\bc)}{}^\dagger{}^{(-)} \gamma_5) Q_+
.
\end{eqnarray}

For a HQSS baryon doublet with exactly one heavy quark (e.g., $\Sigma_c$ and
$\Sigma_c^*$),
\begin{eqnarray}
\bm{B}^{(c)}{}^\mu &=& B^{(c)}{}^\mu{}^{(+)} +
\frac{1}{\sqrt{3}}(\gamma^\mu + v^\mu)\gamma_5 B^{(c)}{}^{(+)},
\nonumber \\
\overline{\bm{B}}{}^{(c)}{}^\mu  &=& \bm{B}^{(c)}{}^\dagger{}^\mu \gamma_0 
= \overline{B}{}^{(c)}{}^\mu{}^{(-)}
+ \frac{1}{\sqrt{3}} \overline{B}{}^{(c)}{}^{(-)} (\gamma^\mu - v^\mu)\gamma_5 
.
\end{eqnarray}
Here $B^{(c)}$ is the Dirac spinor of the $1/2^+$ baryon in the doublet while
$\overline{B}{}^{(c)}{}^\mu$ is the Rarita-Schwinger field for the $3/2^+$
baryon: $v_\mu B^{(c)}{}^\mu{}=\gamma_\mu B^{(c)}{}^\mu=0$.

Finally, for a HQSS singlet baryon with exactly one heavy quark (e.g., $\Lambda_c$)
\begin{eqnarray}
\bm{B}^{(c)} = B^{(c)}{}^{(+)}
,\qquad
\overline{\bm{B}}{}^{(c)} = \bm{B}^{(c)}{}^\dagger \gamma_0 = \overline{B}{}^{(c)}{}^{(-)}
.
\end{eqnarray}
In addition, the polarization of the baryons carrying heavy quarks is such
that $Q_- B^{(c)}{}^{(+)}=Q_- B^{(c)}{}^\mu{}^{(+)}=0$, hence
\begin{equation}
Q_-\bm{B}^{(c)}=Q_-\bm{B}^{(c)}{}^\mu = 0
.
\end{equation}

Under HQSS rotations these fields transform as follows
\begin{eqnarray}
&&
S_c\bm{M}^{(c)}, \quad
 \bm{M}^{(\bc)}S_\bc^{-1}, \quad
S_c \bm{M}^{(c\bc)}S_\bc^{-1}, \quad
S_c\bm{B}^{(c)}, \quad
S_c\bm{B}^{(c)}{}^\mu
\nonumber \\
&&
\overline{\bm{M}}{}^{(c)}S_c^{-1}, \quad
S_\bc \overline{\bm{M}}{}^{(\bc)}, \quad
S_\bc \overline{\bm{M}}{}^{(c\bc)}S_c^{-1}, \quad
\overline{\bm{B}}{}^{(c)}S_c^{-1}, \quad
\overline{\bm{B}}{}^{(c)}{}^\mu S_c^{-1}
.
\end{eqnarray}
Here $S_c$ and $S_\bc$ are the matrices in Dirac space representing the $c$ or
$\bc$ spin rotation and satisfy $S_{c,\bc}^\dagger = \gamma_0
S_{c,\bc}^{-1}\gamma_0$.

All other hadrons without heavy quarks nor antiquarks are HQSS singlets. They
have complete fields (positive and negative frequency parts) and are denoted
without boldface type.

The hadrons are also organized into SU(3) multiplets. We use fields with
labels $a,b,c,\ldots= 1,2,3$ (or up, down, strange) in the fundamental or
anti-fundamental representations of SU(3), in such a way that
\begin{equation}
T^{a \cdots}_{b \cdots} 
\to
U^\dagger{}^a{}_{a^\prime} U^{b^\prime}{}_b \cdots
T^{a^\prime \cdots}_{b^\prime \cdots}
,\qquad U \in\SU(3) 
\end{equation}

For $C=0$ with hidden charm, the following SU(3) multiplets are needed
\begin{equation}
\bm{\bar{D}}^a = 
\left( \begin{matrix} 
\bm{\bar{D}}^0 & \bm{\bar{D}}^- & \bm{\bar{D}_s} 
\end{matrix} \right)
\end{equation}
Here $\bm{\bar{D}}^0$ represents the HQSS doublet formed by $\bar{D}^0$
and $\bar{D}^*{}^0$, etc.
\begin{equation}
\bm{\Sigma_c}^\mu{}^{ab} =
\left( \begin{matrix} 
\sqrt{2}\bm{\Sigma_c}^\mu{}^{++} & \bm{\Sigma_c}^\mu{}^+ & \bm{\Xi_c}^\mu{}^+ \cr 
\bm{\Sigma_c}^\mu{}^+ & \sqrt{2}\bm{\Sigma_c}^\mu{}^0 & \bm{\Xi_c}^\mu{}^0 \cr
 \bm{\Xi_c}^\mu{}^+ & \bm{\Xi_c}^\mu{}^0 & \sqrt{2}\bm{\Omega_c}^\mu
\end{matrix} \right)
\end{equation}
This is a symmetric tensor (irrep $\bm{6}$ of $\SU(3)$).
$\bm{\Sigma_c}^\mu{}$, $\bm{\Xi_c}^\mu$, and $\bm{\Omega_c}^\mu$ are the HQSS
doublets with $(\Sigma_c,\Sigma_c^*)$, $(\Xi_c^\prime,\Xi_c^*)$, and
$(\Omega_c,\Omega_c^*)$, respectively.
\begin{equation}
\bm{\Xi_c}{}_a = 
\left( \begin{matrix} 
 -\bm{\Xi_c}^0 & \bm{\Xi_c}^+ & -\bm{\Lambda_c}
\end{matrix} \right)
\end{equation}
In this case the members of the SU(3) multiplet are HQSS singlets. Further we
define $\bm{\psi}$ as the SU(3) singlet and HQSS doublet containing
$(\eta_c,J/\psi)$.

In addition, the following light baryon multiplets appear: with $J^P=1/2^+$
\begin{equation}
\Sigma^a{}_b =
\left( \begin{matrix} 
\frac{1}{\sqrt{6}} \Lambda - \frac{1}{\sqrt{2}} \Sigma^0  & \Sigma^+ & -p \cr 
-\Sigma^- & \frac{1}{\sqrt{6}} \Lambda + \frac{1}{\sqrt{2}} \Sigma^0 & -n \cr 
-\Xi^- & \Xi^0 & -\sqrt{\frac{2}{3}} \Lambda 
\end{matrix} \right)
,
\end{equation}
and with $J^P=3/2^+$, $\Delta_\mu^{abc}$, a Rarita-Schwinger field and a
symmetric tensor normalized as
\begin{eqnarray}
&&
\Delta_\mu^{111} = \sqrt{6}\Delta_\mu^{++}, \quad
\Delta_\mu^{112} = \sqrt{2}\Delta_\mu^+, \quad
\Delta_\mu^{122} = \sqrt{2}\Delta_\mu^0, \quad
\Delta_\mu^{222} = \sqrt{6}\Delta_\mu^-, \quad
\nonumber \\
&&
\Delta_\mu^{113} = \sqrt{2}\Sigma^*_\mu{}^+, \quad
\Delta_\mu^{123} = \Sigma^*_\mu{}^0, \quad
\Delta_\mu^{223} = \sqrt{2}\Sigma^*_\mu{}^-,
\nonumber \\
&&
\Delta_\mu^{133} = \sqrt{2}\Xi^*_\mu{}^0, \quad
\Delta_\mu^{233} = \sqrt{2} \Xi^*_\mu{}^-, 
\nonumber \\
&&
\Delta_\mu^{333} = \sqrt{6}\Omega_\mu
.
\end{eqnarray}

The relative phases of all fields here are standard with respect to the
conventions adopted in \cite{GarciaRecio:2010vf} for the rotation, flavor and
spin-flavor groups. So for instance, $(\Xi_c^+,\Xi_c^0)$ is a standard isospin
doublet and $(\Sigma^+,\Sigma^0,\Sigma^-)$ is a standard isovector. For SU(3)
(and SU(4)) the convention in \cite{Baird:1964zm} is used instead of that in
\cite{de Swart:1963gc}.\footnote{The matrix elements between standard states
  of the step operators $u\leftrightarrow d$, $d\leftrightarrow s$, and
  $s\leftrightarrow c$ are required to be non negative \cite{Baird:1964zm},
  rather than those of $u\leftrightarrow d$ and $u\leftrightarrow s$ \cite{de
    Swart:1963gc}.}

Regarding parity, we note that the hadrons with spin-parity $1^-$ or $1/2^+$
have normal parity, whereas those with $0^-$ or $3/2^+$ have abnormal
parity. So $\Sigma^b{}_a$, $\bm{\psi}$, $\bm{\Xi_c}{}_a$, and $\bm{\bar{D}}^a$
are true tensors while $\Delta^{abc}_\mu$, and $\bm{\Sigma_c}^\mu{}_{ab}$ are
pseudotensors. Also, $\Sigma^b{}_a$, $\bm{\Xi_c}{}_a$, $\Delta^{abc}_i$, and
$\bm{\Sigma_c}^i{}_{ab}$ have large upper components while $\bm{\psi}$ and
$\bm{\bar{D}}^a$ have large off-diagonal blocks in Dirac space.  The
$\gamma_5$ matrices introduced to preserve parity fit in this scheme.

Next, we write down the 12 most general operators allowed by
$\SU(3)\times\HQSS$ in the baryon-meson coupled-channels space, in $S$-wave
and preserving parity. The operator $\overset{\leftrightarrow}\partial\!{}_v =
v^\mu(\overset{\rightarrow}\partial\!{}_\mu
-\overset{\leftarrow}\partial\!{}_\mu)$ acts on the mesons only and it is
introduced in order to produce the correct kinematical dependence in the
amplitudes.

\begin{eqnarray}
\mathcal{L}_1(x) &=&
g_1\, \overline{\Sigma}{}^a{}_b
\Sigma^b{}_a
\,\tr(\overline{\bm{\psi}}\,\pv\bm{\psi}) 
, \label{eq:g1}
\\
\mathcal{L}_2(x) &=&
g_2\, \frac{1}{3!}\overline{\Delta}{}_{abc}^\mu \Delta^{abc}_\mu 
\,\tr( \overline{\bm{\psi}} \,\pv \bm{\psi} )
,
\\
\mathcal{L}_3(x) &=&
g_3\, \overline{\bm{\Xi}}{}_{\bm{c}}^a \bm{\psi}(-\pv)
\overline{\bm{\bar{D}}}_b \Sigma^b{}_a 
+\mbox{h.c.}
,
\\
\mathcal{L}_4(x) &=&
g_4\, \epsilon^{bcd}
\overline{\bm{\Sigma}}{}^\mu_{\bm{c}}{}_{ab}\bm{\psi}(-\pv)
\overline{\bm{\bar{D}}}_c \gamma_\mu\gamma_5 \Sigma^a{}_d
+\mbox{h.c.}
,
\\
\mathcal{L}_5(x) &=&
g_5\, \frac{1}{2}
\overline{\bm{\Sigma}}{}^\mu_{\bm{c}}{}_{ab}
\bm{\psi} (-\pv)
\overline{\bm{\bar{D}}}_c
\,\Delta^{abc}_\mu 
+\mbox{h.c.}
,
\\
\mathcal{L}_6(x) &=&
g_6\, \overline{\bm{\Xi}}{}_{\bm{c}}^a \bm{\Xi}_{\bm{c}}{}_a 
\,\tr(\overline{\bm{\bar{D}}}_b \,\pv \bm{\bar{D}}^b) 
,
\\
\mathcal{L}_7(x) &=&
g_7 \,\overline{\bm{\Xi}}{}_{\bm{c}}^a \bm{\Xi}_{\bm{c}}{}_b 
\,\tr(\overline{\bm{\bar{D}}}_a \,\pv \bm{\bar{D}}^b) \label{eq:eqg7}
,
\\
\mathcal{L}_8(x) &=&
g_8 \,
 \epsilon^{bcd}
\overline{\bm{\Sigma}}{}^\mu_{\bm{c}}{}_{ab}
\bm{\Xi}_{\bm{c}}{}_d
\,\tr(
\overline{\bm{\bar{D}}}_c
\gamma_\mu\gamma_5
\,\pv
\bm{\bar{D}}^a
)
+\mbox{h.c.}
,
\\
\mathcal{L}_9(x)+\mathcal{L}_{10}(x) &=&
\frac{1}{2}
\overline{\bm{\Sigma}}{}^\mu_{\bm{c}}{}_{ab}
\bm{\Sigma}_{\bm{c}}^\nu{}^{ab}
\,\tr(
\overline{\bm{\bar{D}}}_c
 (g_9 \,g_{\mu\nu} + g_{10} \,i\sigma_{\mu\nu})
\,\pv \bm{\bar{D}}^c
)
,
\\
\mathcal{L}_{11}(x)+\mathcal{L}_{12}(x) &=&
\overline{\bm{\Sigma}}{}^\mu_{\bm{c}}{}_{ac}
\bm{\Sigma}_{\bm{c}}^\nu{}^{bc}
\,\tr(
\overline{\bm{\bar{D}}}_b
 (g_{11} \,g_{\mu\nu} + g_{12} \,i\sigma_{\mu\nu})
\,\pv \bm{\bar{D}}^a
)
. \label{eq:g11}
\end{eqnarray}
The traces refer to Dirac space.

The reduction of these Lagrangians when no strangeness is involved is as
follows:
\begin{eqnarray}
\mathcal{L}_1(x) &=&
g_1\, \overline{N}N \,\tr( \overline{\bm{\psi}} \,\pv \bm{\psi} ) 
,
\\
\mathcal{L}_2(x) &=&
g_2\, \overline{\Delta}{}^\mu \Delta_\mu \,\tr( \overline{\bm{\psi}} \,\pv \bm{\psi} )
,
\\
\mathcal{L}_3(x) &=&
g_3\, \overline{\bm{\Lambda}}_{\bm{c}} \bm{\psi} (-\pv) \overline{\bm{\bar{D}}} N
+\mbox{h.c.}
,
\\
\mathcal{L}_4(x) &=&
g_4\, \overline{\bm{\Sigma}}{}^\mu_{\bm{c}}{}_j \bm{\psi} (-\pv) 
\overline{\bm{\bar{D}}} \,\tau_j \gamma_\mu \gamma_5 N
+\mbox{h.c.}
,
\\
\mathcal{L}_5(x) &=&
\sqrt{3} \,g_5\, \overline{\bm{\Sigma}}{}^\mu_{\bm{c}}{}_j \overline{\bm{\psi}}
(-\pv) \bm{\psi} \,T_j \Delta_\mu
+\mbox{h.c.}
,
\\
\mathcal{L}_6(x) &=&
g_6\, \overline{\bm{\Lambda}}_{\bm{c}} \bm{\Lambda_c}
\,\tr( \overline{\bm{\bar{D}}} \,\pv \bm{\bar{D}} )
,
\\
\mathcal{L}_7(x) &=& 0 
,
\\
\mathcal{L}_8(x) &=&
g_8\,\overline{\bm{\Sigma}}{}^\mu_{\bm{c}}{}_j  \bm{\Lambda_c}
\,\tr( \overline{\bm{\bar{D}}} \,\tau_j \gamma_\mu \gamma_5 \,\pv \bm{\bar{D}} )
+\mbox{h.c.}
,
\\
\mathcal{L}_9(x)+\mathcal{L}_{10}(x) 
+ \mathcal{L}_{11}(x)+\mathcal{L}_{12}(x)
&=&
\overline{\bm{\Sigma}}{}^\mu_{\bm{c}}{}_j \bm{\Sigma_c}^\nu{}_j
\,\tr\big( \overline{\bm{\bar{D}}} 
( G_9 \,g_{\mu\nu} + G_{10} \,i\sigma_{\mu\nu} )
\,\pv \bm{\bar{D}} \big)
\nonumber \\ && +
\overline{\bm{\Sigma}}{}^\mu_{\bm{c}}{}_j \bm{\Sigma_c}^\nu{}_k
\,\tr\big( \overline{\bm{\bar{D}}} \,\tau_j \tau_k
( G_{11} \,g_{\mu\nu} + G_{12} \,i\sigma_{\mu\nu} )
\,\pv \bm{\bar{D}} \big)
.
\end{eqnarray}
Here $j,k$ are isovector indices, $\vec{\tau}$ are the Pauli matrices and
$\langle 3/2,M|\vec{\epsilon}_\lambda\vec{T}^\dagger|1/2,m\rangle= 
C(1/2,1,3/2;m,\lambda,M)$. Further,
\begin{equation}
G_9= g_9+2g_{11},\quad
G_{10}= g_{10}+2g_{12},\quad
G_{11}=-g_{11},\quad
G_{12}=-g_{12} .
\end{equation}

Our WT model with SF and HQSS gives the following values for the parameters:
\begin{eqnarray}
&&
\hat{g}_1 = 0 ,\quad
\hat{g}_2 = 0 ,\quad
\hat{g}_3 = \sqrt{\frac{3}{2}} ,\quad
\hat{g}_4 =  \sqrt{\frac{1}{6}} ,\quad
\hat{g}_5 =  -1 ,\quad
\hat{g}_6 =  \frac{1}{2} ,
\nonumber \\
&&
\hat{g}_7 = -\frac{1}{2} ,\quad
\hat{g}_8 = \frac{1}{2} ,\quad
\hat{g}_9 = 0 ,\quad
\hat{g}_{10} =  0 ,\quad
\hat{g}_{11} =  -\frac{1}{2} ,\quad
\hat{g}_{12} = -\frac{1}{2} \label{eq:gs}
.
\end{eqnarray}
where we have defined $\hat{g}_i = 4f^2 g_i$.

The vanishing of $g_1$ and $g_2$ follows from the OZI rule, which is fulfilled
by the model.

\section{Coupled-channels unitarization and symmetry breaking}
\label{sec:uni}

\subsection{Unitarization and renormalization scheme}

As previously discussed, the baryon-meson interaction is mediated by the
extended WT interaction of \Eq{2.21} that fulfills $\SU(6)\times \HQSS$ and it
is consistent with chiral symmetry in the light sector. The final expression
for the potential to be used throughout this work is
\begin{equation}
V_{ij}^{CSIJ} =
D_{ij}^{CSIJ}\,\frac{1}{4f_if_j} (k_i^0+k_j^\prime{}^0) \ ,
\label{eq:pot}
\end{equation}
 where $k_i^0$ and $k_j^\prime{}^0$ are the CM energies of the incoming and
 outgoing mesons, respectively, and $f_i$ and $f_j$ are the decay constants of
 the meson in the $i$-channel and $j$-channel.\footnote{As compared to our
   previous work of Ref.~\cite{Romanets:2012hm}, we have
\begin{itemize}
\item approximated $(2\sqrt{s}-M_i-M_j)$, with $M_i$ and $M_j$ the incoming
  and outgoing baryon masses, by the sum of the CM energies of the incoming
  and outgoing mesons.  In the present case, the non relativistic
  approximation tends to increase the binding by up to few tens of MeV.  This
  non-relativistic approximation for the baryons is consistent with the
  treatment for the baryons adopted in the previous section to implement the
  HQSS constraints, and it makes easier to connect with the effective HQSS
  Lagrangians introduced in Eqs.~(\ref{eq:g1})--(\ref{eq:g11}).
\item and, also for this latter reason, moved the $\sqrt{(E+M)/(2M)}$ factors
  included in the potential used in \cite{Romanets:2012hm} to the definition
  of the loop function in \Eq{loop}.
\end{itemize}
}We use the hadron masses and meson decay constants compiled in Table II of
 Ref.~\cite{Romanets:2012hm}. In particular, $f_{J/\Psi}$ is taken from the
 width of the $J/\Psi \rightarrow e^- e^+$ decay, that is, $f_{J/\Psi}=$ 290
 \MeV $~$ and we set $f_{\eta_c}=f_{J/\Psi}$, as predicted by HQSS and
 corroborated in the lattice evaluation of Ref.~\cite{Dudek:2006ej}.  The
 $D_{ij}$ are the matrix elements of $H^\prime_\WT$, \Eq{2.21}, for the
 various hidden charm $CSIJ$ sectors previously discussed. Those for the
 strangeless hidden charm $C=0$ case, for which $H^\prime_\WT= H_{\rm ex}$,
 and that we will discuss in what follows are given in Appendix
 \ref{app:tables}.
\footnote{For the sake of completeness, and to make possible the determination
  (see \Eq{gs}) of the coupling $g_7$ of the HQSS effective Lagrangian of
  \Eq{eqg7}, we also give in the Appendix \ref{app:tables} the coefficients
  for the rest of hidden charm sectors with explicit strangeness.}

In order to calculate the scattering amplitudes, $T_{ij}$, we solve the
on-shell Bethe-Salpeter equation (BSE), using the matrix $V^{CSIJ}$ as kernel:
\begin{equation}
T^{CSIJ}=(1-V^{CSIJ}G^{CSIJ})^{-1}V^{CSIJ}\label{eq:bse} \ ,
\end{equation}
where $G^{CSIJ}$ is a diagonal matrix containing the baryon-meson propagator
for each channel.  Explicitly
\begin{equation}
G^{CSIJ}_{ii} (s) = %\sqrt{E_i+M_i} \,
\frac{(\sqrt{s}+M_i)^2-m_i^2}{2\sqrt{s}}
\left(\bar{J}_0(\sqrt{s};M_i,m_i) - \bar{J}_0(\mu^{SI};M_i,m_i)\right), \label{eq:loop}
\end{equation}
$M_i$ ($m_i$) is the mass of the baryon (meson) in the channel $i$. The loop
function $\bar{J}_0$ can be found in the appendix of \cite{Nieves:2001wt}
(Eq.~A9) for the different possible Riemann sheets.  The baryon-meson
propagator is logarithmically ultraviolet divergent, thus, the loop needed to
be renormalized. This has been done by a subtraction point regularization such
that
\begin{equation}
G_{ii}^{CSIJ} (s)=0 \quad\text{at~~} \sqrt{s}=\mu^{CSI},
\label{eq:musi}
\end{equation}
with $\mu^{CSI} =\sqrt{m_{\rm{th}}^2+M_{\rm{th}}^2}$, where $m_{\rm{th}}$ and
$M_{\rm{th}}$, are, respectively, the masses of the meson and baryon producing
the lowest threshold (minimal value of $m_{\rm{th}}+M_{\rm{th}}$) for each
$CSI$ sector, independent of the angular momentum $J$.  This renormalization
scheme was first proposed in Refs.~\cite{Hofmann:2005sw,Hofmann:2006qx} and it
was successfully used in
Refs.~\cite{GarciaRecio:2003ks,GarciaRecio:2008dp,Romanets:2012hm}. A recent
discussion on the regularization method can be found in
Ref.~\cite{Hyodo:2008xr}.

The dynamically-generated baryon resonances appear as poles of the scattering
amplitudes on the complex energy $\sqrt{s}$ plane. One has to check both first
and second Riemann sheets. The poles of the scattering amplitude on the first
Riemann sheet that appear on the real axis below threshold are interpreted as
bound states. The poles that are found on the second Riemann sheet below the
real axis and above threshold are identified with resonances\footnote{Often we
  refer to all poles generically as resonances, regardless of their concrete
  nature, since usually they can decay through other channels not included in
  the model space.}.  The mass and the width of the state can be found from
the position of the pole on the complex energy plane. Close to the pole, the
scattering amplitude behaves as
\begin{equation} \label{Tfit} 
T^{CSIJ}_{ij} (s) \approx \frac{g_i
e^{i\phi_i}\,g_je^{i\phi_j}}{\sqrt{s}-\sqrt{s_R}} \,.  
\end{equation} %
The mass $M_R$ and width $\Gamma_R$ of the resonance result from
$\sqrt{s_R}=M_R - \rm{i}\, \Gamma_R/2$, while $g_j e^{i\phi_j}$ (modulus and
phase) is the coupling of the resonance to the $j$-channel.

\subsection{Symmetry breaking}

As it was already pointed out in the subsection~\ref{subsec:SymStr}, we
classify states under the symmetry group $\SU(6)\times\HQSS$, and consider the
breaking of the light SF SU(6) to $\SU(3) \times \SU_{J_l}(2)$.  Subsequently
we break the SU(3) light flavor group to SU(2) isospin symmetry group,
preserving the HQSS, and finally we break the HQSS.  Thus, we assume exact
isospin, total spin and flavor conservation. The symmetry breaking is
performed by adiabatic change of hadron masses and meson weak decay constants,
as it was previously done in Ref.~\cite{Romanets:2012hm}. At each symmetric
point, the hadron masses and meson decay constants are averaged over the
corresponding group multiplets. Further, we introduce three parameters, $x,
x'$ and $x''$ that are changed from 0 to 1, to gradually break the symmetry
from $\SU(6)\times\HQSS$ down to $\SU(3)\times\HQSS$, then to
$\SU(2)\times\HQSS$, and finally down to SU(2) isospin, respectively:
\begin{eqnarray}
\label{XSymBr}
m(x)=(1-x)~m_{{\rm SU(6)} \times {\rm HQSS}} + x~m_{{\rm SU(3)} \times {\rm HQSS} },
\nonumber\\
f(x)=(1-x)~f_{{\rm SU(6)} \times {\rm HQSS}} + x~f_{{\rm SU(3)} \times {\rm HQSS} },
\nonumber\\
m(x')=(1-x')~m_{{\rm SU(3)} \times {\rm HQSS}} + x'~m_{{\rm SU(2)} \times {\rm HQSS} }, 
\nonumber\\
f(x')=(1-x')~f_{{\rm SU(3)} \times {\rm HQSS}} + x'~f_{{\rm SU(2)} \times {\rm HQSS} },
\nonumber\\
m(x'')=(1-x'')~m_{{\rm SU(2)} \times {\rm HQSS}} + x''~m_{\rm SU(2)}, 
\nonumber\\
f(x'')=(1-x'')~f_{{\rm SU(2)} \times {\rm HQSS}} + x''~f_{\rm SU(2)}.
\end{eqnarray}
In this way we can assign SU(3) and SU(6) representation labels to each found
resonance, and also identify the HQSS multiplets.  We will show below a
diagram (Fig.~\ref{fig_evolution}) with the evolution of the hidden charm $N$
and $\Delta$ pole positions as the various symmetries are gradually broken.

\section{Charmless and Strangeless Hidden Charm Sector: 
The $N$ and $\Delta$ states}
\label{sec:results}

In this work we will only discuss results on hidden charm baryon resonances
with total charm $C=0$ and strangeness $S=0$. Other sectors with charm
different from zero will be studied elsewhere.

In this sector, we find several $I=1/2$ and $I=3/2$ states, which correspond
to $N$-like and $\Delta$-like states, respectively (here we use the same
notation as in Refs.~\cite{Wu:2010jy,Wu:2010vk}).  All these states have odd
parity and different values ($J=1/2,~3/2$ and $5/2$) of total angular
momentum. The list of coupled-channels and the corresponding coefficients
$D^{IJ}_{ij}$ can be found in the first six tables of Appendix
\ref{app:tables}.

In this hidden charm sector and in the $\SU(6)\times\HQSS$ limit, we saw
(Eqs.~(\ref{eq:2.26bis}) and (\ref{eq:2.26})) that the group structure of the
HQSS-constrained extension of the WT interaction developed in this work
consists of two $\bf{56_{2,0}}$ and one $\bf{70_{2,0}}$ representations. One
of the $\bf{56_{2,0}}$ multiplets and the $\bf{70_{2,0}}$ one are
attractive. Thus, from the decomposition in \Eq{2.29} (see also
Table~\ref{tab:sectors}), we could expect up to a total of ten $N$-like and
seven $\Delta$-like resonances.\footnote{Those lie in the $\SU(3)$ octets and
  decuplets irreps, respectively, contained in the attractive $\bf{56_{2,0}}$
  and $\bf{70_{2,0}}$ multiplets.} Because of the breaking of the
$\SU(6)\times\HQSS$ symmetry, due to the use of physical hadron masses and
meson decay constants, we only find seven heavy $N$ and five heavy $\Delta$
states in the physical Riemann sheets. They have masses around $4\GeV$ and
most of them turn out to be bound. The remaining missing states show up in
unphysical Riemann sheets.  The evolution of all states as we gradually break
the symmetry from $\SU(6)\times\HQSS$ down to $\SU(3)\times\HQSS$, then to
$\SU(2)\times\HQSS$, and finally down to SU(2) isospin, is depicted in
Fig.~\ref{fig_evolution}. Thanks to this latter study, we could assign
$\SU(6)\times\HQSS$ and $\SU(3)\times\HQSS$ labels to each of the predicted
resonances, which are all of them collected in Tables~\ref{tab0011} and
~\ref{tab002}, and could also identify two HQSS multiplets in each isospin
sector.

%%%%%%%%%%%%%%%%%%%%%% C=0  S=0  I=1/2 J=1/2

\subsection{${\bm N}$  states 
($\bm{  C = 0}$, $\bm{ S = 0} $, $\bm{ I=1/2}$ ) }
\label{subsect:N}

As mentioned above, the model predicts the existence of seven heavy nucleon
resonances: three states with the spin-parity $J^P= \frac{1}{2}^-$, also three
states with $\frac{3}{2}^-$ sectors, and one state with $J^P=
\frac{5}{2}^-$. Their masses, widths and couplings to the different channels
are compiled in Table~\ref{tab0011}.

%%%%%%%%%%%%%%%%%%%%%%  COUPLED CHANNELS:

%
\begin{itemize}
\item $ J=1/2$: In this sector, there are seven coupled
channels, with the following threshold energies (in MeV):

\begin{center}
\noindent
\begin{tabular}{ccccccc}

$N \eta_c$   &$N J/\psi$   &$\Lambda_c \bar{D}$   &$\Lambda_c \bar{D}^*$ 
&$\Sigma_c \bar{D}$   &$\Sigma_c \bar{D}^*$   &$\Sigma_c^* \bar{D}^*$  \\

3918.6   & 4035.8   & 4153.7   & 4294.8   & 4320.8   & 4461.9   & 4526.3
\end{tabular}
\smallskip
\end{center}

\item $ J=3/2$: In this sector, there are five coupled
channels, with the following threshold energies:
\begin{center}
\smallskip
\noindent
\begin{tabular}{ccccc}

$N J/\psi$    &$\Lambda_c \bar{D}^*$   &$\Sigma_c^* \bar{D}$
&$\Sigma_c \bar{D}^*$   &$\Sigma_c^* \bar{D}^*$ \\

4035.8  &4294.8  &4385.2  &4461.9  &4526.3
\end{tabular}
\smallskip
\end{center}

\item $ J=5/2$: In this sector there is only one channel, $\Sigma_c^* \bar
  D^*$, with threshold equal to $4526.3\MeV$.

\end{itemize}

\begin{figure*}[h]
\begin{center}
\includegraphics{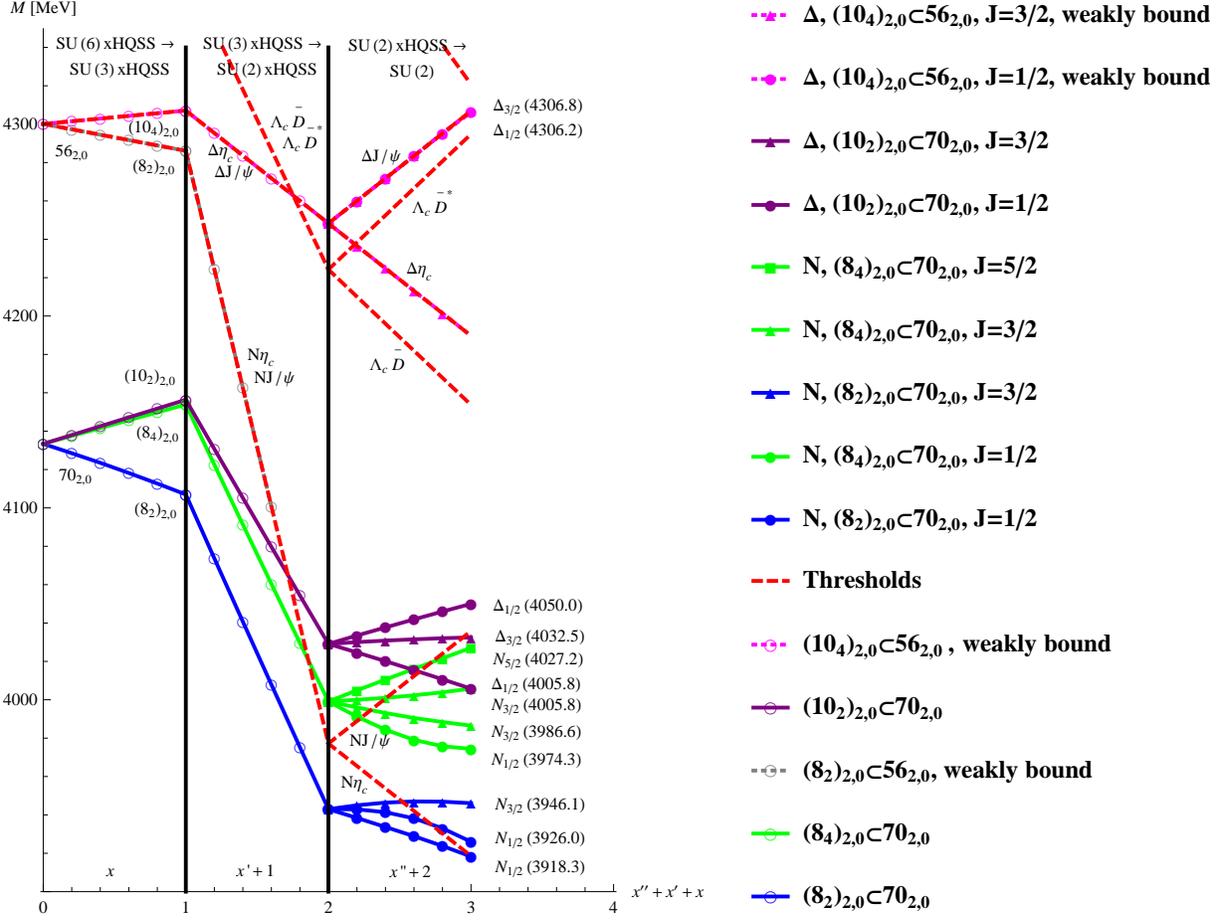}
\medskip
\vskip0.3cm 
\end{center}
\caption{Evolution of the poles as symmetries, starting from $\SU(6)\times
  \HQSS$, are sequentially broken to reach the isospin symmetric final
  crypto-exotic $N$ and $\Delta$ odd-parity resonances.  The meanings of $x$,
  $x'$ and $x''$ can be found in~(\ref{XSymBr}). The lower index of the final
  states stands for the spin $J$ of the corresponding resonance. The
  thresholds (red dashed lines) are marked together with the respective
  baryon-meson channel.  The $\SU(6)\times\HQSS$ labels ${\bf 70_{2,0}}$ and
  ${\bf 56_{2,0}}$, and the $\SU(3 )\times\HQSS$ labels ${\bf (8_2)_{2,0} }$,
  ${\bf (8_4)_{2,0} }$, ${\bf (10_2)_{2,0} }$, ${\bf (10_4)_{2,0} }$ are also
  shown at the corresponding symmetric points.}
\label{fig_evolution}
\end{figure*}

From the group decomposition of the $\SU(6)\times\HQSS$ representations, we
could expect up to a maximum of five states with spin $J=1/2$ (see
Table~\ref{tab:sectors}): one state from each of the two $J=1/2$ octets
encoded in the attractive $\bf{56_{2,0}}$ representation, and three states
corresponding to the each of the $\bf{8_2}$ octets that appear in the
reduction of the $\bf{70_{2,0}}$ representation [\Eq{2.29}].  However, the two
poles related to the $\bf{56_{2,0}}$ representation appear in an unphysical
Riemann sheet, at the physical point (i.e., at the point of the evolution when
the hadron masses and meson decay constants attain their physical values).  As
it can be seen from Fig.~\ref{fig_evolution}, these poles dissappear from the
physical sheet when we pass from the $\SU(3)\times\HQSS$ limit to the
$\SU(2)\times\HQSS$ one. Indeed, we could observe how the $\bf{ (8_2)_{2,0}
  \subset (56_2)_{2,0}}$ pole almost coincides with the threshold value of the
degenerated $N \eta_c$ and $N J/\psi$ channels in the first steps of this
evolution until it finally disappears.  On the other hand, the
$\bf{(8_2)_{2,0} \subset (56_2)_{2,0}}$ pole also gives rise to an octet of
$J=3/2$ states (see \Eq{2.29}), which is also lost at the physical
point. Thus, for $J=3/2$ we are also left only with the three baryon
resonances stemming from the ${ \bf 70_{2,0} }$ representation, one from ${\bf
  (8_2)_{2,0} }$, and two from ${\bf (8_4)_{2,0} }$. The $J=5/2$ state is
originated also from this latter multiplet.

From the above discussion, it is clear that the $N$-like resonances found in
this work, and collected in Table~\ref{tab0011}, form two HQSS multiplets. In
the first one the light degrees of freedom have quantum numbers
$\bf{(8_2)_{2,0} \subset (70_2)_{2,0}}$. This multiplet is formed by the three
first resonances of the table (two with spin 1/2 and third one with spin 3/2)
that correspond to the blue states in Fig.~\ref{fig_evolution}. They only
differ in how the light sector spin is coupled to the spin of the $c\bar c$
pair. The second HQSS multiplet corresponds to $\bf{(8_4)_{2,0} \subset
  (70_2)_{2,0}}$ quantum numbers for the light sector, and it consists of the
four remaining states in Table~\ref{tab0011} (displayed in green in
Fig.~\ref{fig_evolution}) one with spin 1/2, two with spin 3/2, and another one
with spin 5/2.

The members of each HQSS multiplet are nearly degenerate, but not totally
because we also break the HQSS by the use of physical hadron masses.

A word of caution is needed here. The mass of the $J=5/2$ resonance is around
$4027.2\MeV$. In this sector there is only one channel ($\Sigma_c^* \bar D^*$,
with threshold equal to $4526.3\MeV$), thus this state is around five hundred
MeV bound. We expect our model to work well close to threshold, and therefore,
in this case, interaction mechanisms neglected here and involving higher
partial waves could be relevant for determining the actual properties of this
resonance.

%N
\begin{table*}[ht]
\begin{center}
\begin{tabular}{| c | c | c | c | c | c | c |}
\hline 
 $\SU(6)\times\HQSS$  & $\SU(3)\times\HQSS$  &  	           & 		        & Couplings  	         &  \\
 irrep             & irrep            & $M_{R} [\MeV]$   & $\Gamma_{R} [\MeV]$ &  to main channels  & $J$ \\ 
\hline
%% $\bf{g_{}=}$ - possible decaying channel

 $\bf{70_{2,0}}$  & $\bf{(8_2)_{2,0}}$  &  3918.3  &  0.0 & $g_{N \eta_c}=0.5$, $g_{N J/\psi}=0.6$,
 $g_{\Lambda_c \bar{D}}=3.1$,  $g_{\Lambda_c \bar{D}^*}=0.5$,
& $1/2$ \\
 & & & &   $g_{\Sigma_c \bar{D}}=0.2$ ,
$g_{\Sigma_c~\bar{D}^*}=2.6$, 
$g_{\Sigma_c^*~\bar{D}^*}=2.6$
  &  \\
\hline

 $\bf{70_{2,0}}$ &   $\bf{(8_2)_{2,0}}$  & 3926.0 & 0.1 &  $\bf{g_{N \eta_c}=0.2}$, $g_{N J/\psi}=0.04$,
$g_{\Lambda_c \bar{D}}=0.4$,  $g_{\Lambda_c \bar{D}^*}=3.0$, 
 & $1/2$ \\
 &  &     &      &    $g_{\Sigma_c \bar{D}}=4.2$,  $g_{\Sigma_c \bar{D}^*}=0.2$,
 $g_{\Sigma_c^* \bar{D}^*}=0.7$
&  \\
\hline

 $\bf{70_{2,0}}$ & $\bf{(8_2)_{2,0}}$ &  3946.1  & 0.  &  $g_{N J/\psi}=0.2$,  $g_{\Lambda_c \bar{D}^*}=3.4$,  
  $g_{\Sigma_c^* \bar{D}}=3.6$,  $g_{\Sigma_c \bar{D}^*}=1.1$,
& $3/2$ \\
 &  & &  &  $g_{\Sigma_c^* \bar{D}^*}=1.5$
&  \\
\hline

    $\bf{70_{2,0}}$  &  $\bf{(8_4)_{2,0}}$  & 3974.3 & 2.8 &  $\bf {g_{N \eta_c}=0.5}$,  $g_{N J/\psi} \sim 0.05$,
  $g_{\Lambda_c \bar{D}}=0.4$,  $g_{\Lambda_c \bar{D}^*}=2.2$,
  & $1/2$ \\
 &  &             &        &  $g_{\Sigma_c \bar{D}}=2.1$,  $g_{\Sigma_c \bar{D}^*}=3.4$,
$g_{\Sigma_c^* \bar{D}^*}=3.1$	&  \\
\hline

    $\bf{70_{2,0}}$  &  $\bf{(8_4)_{2,0}}$ &   3986.5  & 0.  & $g_{N J/\psi}=0.2$,   $g_{\Lambda_c \bar{D}^*}=1.0$,
  $g_{\Sigma_c^* \bar{D}}=2.7$,   $g_{\Sigma_c \bar{D}^*}=4.3$,
& $3/2$ \\
 & & & &  $g_{\Sigma_c^* \bar{D}^*}=1.8$
&  \\
\hline

    $\bf{70_{2,0}}$  &  $\bf{(8_4)_{2,0}}$ &  4005.8 &  0. &  $g_{N J/\psi}=0.3$,   $g_{\Lambda_c \bar{D}^*}=1.0$,
  $g_{\Sigma_c^* \bar{D}}=1.6$,  $g_{\Sigma_c \bar{D}^*}=3.2$,
& $3/2$ \\
 & & & &  $g_{\Sigma_c^* \bar{D}^*}=4.2$
&  \\
\hline

    $\bf{70_{2,0}}$  &  $\bf{(8_4)_{2,0}}$ &   4027.1  & 0. &  $g_{\Sigma_c^*~\bar{D}^*}=5.6$
& $5/2$ \\
\hline

\end{tabular}
\end{center}
\caption{Odd parity hidden charm $N$ ($J=1/2$, $J=3/2$ and $J=5/2$) resonances
  found in this work.  The first two columns contain the $\SU(6)\times\HQSS$
  and $\SU(3)\times\HQSS$ quantum numbers of each state, while $M_R$ and
  $\Gamma_{R}$ stand for its mass and width (in MeV). The largest couplings of
  each pole, ordered by their threshold energies, are collected in the next
  column. In boldface, we highlight the channels which are open for
  decay. Finally, the spin of the state is given in the last column.
  Resonances with equal $\SU(6)\times\HQSS$ and $\SU(3)\times\HQSS$ labels
  form HQSS multiplets, and they are collected in consecutive rows.  }
\label{tab0011}
\end{table*}
%%%%%%%%%%%%%%%

%%%%%%%%%%%%%%%%%%%%%%%% graphics:

%------------------------- Comparing with Lutz,...-------------------
%\section{Comparing with other crypto-exotic states}
%%%%%%%%%%%%%%%%%%%%%%%% TABLE
\begin{table}[ht]
\begin{center}
\begin{tabular}{|l l  | l r c c c l | l r c c l |l c|}
\hline & &\multicolumn{6}{|c|}{$N(1/2^-)$}
         &\multicolumn{5}{|c|}{$N(3/2^-)$}
         &\multicolumn{2}{|c|}{$N(5/2^-)$}\\
             &  & 
 $M_R$       &\multicolumn{5}{c|}{ $g$ to main-channels} &
 $M_R$       & \multicolumn{4}{c|}{ $g$ to
  main-channels~~}&
 $M_R$       & $g$
\\ 
Ref. & Model  &  
 [MeV]       & $\Lambda_c\bar{D}$&
 $\Lambda_c\bar{D}^*$&${\Sigma_c\bar{D}}$&${\Sigma_c\bar{D}^*}$
&${\Sigma_c^*\bar{D}^*}$ &
 [MeV]       &
 $\Lambda_c\bar{D}^*$&${\Sigma_c^*\bar{D}}$&$\Sigma_c\bar{D}^*$ 
& ${\Sigma_c^*\bar{D}^*}$ &
 [MeV]       & ${\Sigma_c^*\bar{D}^*}$
\\ \hline
This work &   $\bf{(8_2)_{2,0} \subset (70_2)_{2,0}}$ %$\SU(6)\times HQSS$ 
            & 3918  & 3.1 & 0.5 & 0.2 & 2.6 & 2.6
            &  &  &  &  & 
            &  & \\
         &  & 3926 & 0.4 &3.0 & 4.2& 0.2 & 0.7
            & 3946 & 3.4 & 3.6 & 1.1 & 1.5
            &  & 
\\ \  
      &  &\multicolumn{6}{|l|}{\hrulefill }
&\multicolumn{5}{|l|}{\hrulefill
  }&\multicolumn{2}{|l|}{\hrulefill } \\
         &  $\bf{(8_4)_{2,0} \subset (70_2)_{2,0}}$ 
            & 3974 &0.4 & 2.2 & 2.1 & 3.4 & 3.1
            & 3987 &1.0 & 2.7 & 4.3 & 1.8 
            & & \\
         &  &  &   &  &  &  & 
            & 4006 & 1.0 & 1.6 & 3.2 & 4.2
            & 4027 & 5.6 \\
\hline
\cite{Hofmann:2005sw}\cite{Hofmann:2006qx} & zero-range &
 3520 &  &  & 5.3 &  &  &
 3430 &  &  & 5.6 &  
            & &
\\
 & vector exchange &
  &  &  &  &  &  &
  &  &  &  &  
            & &
\\
\hline
\cite{Wu:2010vk} & hidden-gauge &
4265 & 0.1 &  & 3.0 &   &   &
     &     &  &   &  
           &  &
\\
 &  &
4415 &   & 0.1 &   & 2.8 &   &
4415 & 0.1 &  & 2.8 &  
            &  & 
\\
\hline
\cite{Wu:2012md} & hidden-gauge &
4315 & {\sf X} &  & {\sf X} &   &   &
     &     &  &   &  
           &  &
\\
 &  &
4454 &   & {\sf X} &   & {\sf X} &   &
4454 & {\sf X} &  & {\sf X} &  
            &  & 
\\
\hline
 \cite{Yuan:2012wz}& quark model &
FS$-$CM  &  & & & &
         & FS$-$CM &  &  &  &  & &
\\
&  $uudc\bar{c}$     &
3933$-$4267 &   &  & & & &
            &  &  &  &  &  &
\\
 & &
4013$-$4363 &    &  & & & &
4013$-$4389 &    &  &    
         &     &  &

\\
 & & 
4119$-$4377 &   &   &  &   & & 
4119$-$4445 &   &   &    
         &    &   &

\\

 & &
4136$-$4471 &   &  & & & &
4136$-$4476 &   &  &    
         &     &  &

\\
 & &
4156$-$4541 &   &  & & & &
4236$-$4526 &   &  & 
            & & 4236$-$4616    &

\\\hline
\end{tabular}
\end{center}
\caption{Comparison of the whole spectrum of hidden-charm nucleons (or
  crypto-exotic nucleons) with odd-parity and angular momentum $L=0$ predicted
  by our model with some results from previous models. In all cases, masses
  and couplings ($g$) to the dominant channels (when available) are shown in
  sequential rows. In the hidden gauge model of Ref.~\cite{Wu:2012md}, the
  numerical values of the couplings are not given. In this case, we indicate
  with a symbol {\sf X} the elements of the coupled channel space used to
  generate each resonance.  On the other hand, in the case of the predictions
  of this work, in the column ``Model'' we give the HQSS multiplet. Besides,
  we have also omitted the small couplings to the $N\eta_c$ and $N J/\psi$
  channels that can be seen in Table~\ref{tab0011}. }
\label{tab:comp}
\end{table}

There exist previous works on hidden-charm odd-parity nucleon states, also
named crypto-exotic hadronic states.  These studies can be divided in two
types. Namely, those based on a constituent quark description of the
resonances, and those where they are described as baryon-meson bound molecules
or resonating states.  Some of the predictions of these other models are
compiled in Table~\ref{tab:comp}.

The baryon-meson coupled-channel calculations by Hofmann and Lutz for
$J^P=1/2^-$ in Ref.~\cite{Hofmann:2005sw} and for $J^P=3/2^-$ in
Ref.~\cite{Hofmann:2006qx} rely on a model of zero-range $t$-channel exchange
of light vector mesons, based on chiral and large $N_C$ considerations, and
supplemented with SU(4) input in some vertices.  This model is used as the
driving interaction of pseudo-scalar mesons with the $J^P=1/2^+, 3/2^+$ baryon
ground states. After solving the BSE using a renormalization scheme similar to
that proposed here, some $1/2^-,3/2^-$--resonances were dynamically generated.
Vector mesons in the coupled-channel space were omitted in those early
studies, thus channels like $\Sigma_c \bar D^*$ or $\Lambda_c \bar D^*$ were
not considered.

More recently baryon-meson calculations using a hidden-gauge model have been
carried out in Refs.~\cite{Wu:2010jy,Wu:2010vk,Wu:2012md}. These works
consider $1/2^+$ baryons interacting with pseudoscalar mesons and dynamically
generate $J^P=1/2^-$ hidden-charm nucleon resonances as poles in the
$T$-matrix.  Yet, the interaction of vector mesons with $1/2^+$ baryons
($VB\to VB$) is also taken into account in
\cite{Wu:2010jy,Wu:2010vk,Wu:2012md}, which leads to additional and degenerate
$J^P=1/2^-$ and $3/2^-$ hidden-charm nucleons. However, the $J=3/2^+$ baryons
are not included at all, and thus some channels like $\Sigma_c^* \bar D^*$ are
excluded.

The main difference among our scheme and the hidden-gauge models is the
definition of the coupled-channels space. We consider simultaneously
pseudoscalar meson--baryon ($PB$) and vector meson--baryon ($VB$) channels,
with $J^P=1/2^+$ and $3/2^+$ baryons. However in the approaches of
Refs.~\cite{Wu:2010jy,Wu:2010vk,Wu:2012md} all interaction terms of the type
$PB\to VB$ are neglected. Furthermore, channels with $J^P=3/2^+$ baryons are
not considered either. The potential used in
~\cite{Wu:2010jy,Wu:2010vk,Wu:2012md} for the $PB\to PB$ transitions, with
$J^P=1/2^+$ baryons, is similar to that derived here. However, there exist
important differences in all transitions involving vector mesons.  When
restricting our model to the $PB\to PB$ sector, we still do not obtain the
same results as in Refs.~\cite{Wu:2010jy,Wu:2010vk,Wu:2012md}. This is mainly
due to i) the use of a different renormalization scheme and, ii) the presence
in these latter works of a suppression factor in those transitions that
involve a $t$-channel exchange of a heavy charm vector meson.

However, when we use our full space, the inclusion of a similar suppression
factor in our HQSS kernel is not quantitatively relevant for the dynamical
generation of the resonances.  Note that HQSS does not require the presence
of such suppression factor. In summary, when we compare our approach with the
other molecular-type ones, we observe in our model a rich structure of
resonances due to the many channels cooperating to create them. Our states are
much more lighter that those predicted in
Refs.~\cite{Wu:2010jy,Wu:2010vk,Wu:2012md}, though significantly less bound
that the crypto-exotic baryons reported in
Refs.~\cite{Hofmann:2005sw,Hofmann:2006qx}.

Finally, we will pay attention to the recent work of
Ref.~\cite{Yuan:2012wz}. There a constituent quark model is used to describe
isospin $I=1/2$ baryons with $uudc\bar{c}$ quark content. The mass spectra is
evaluated with three types of hyperfine interactions: color-magnetic
interaction (CM) based on one-gluon exchange, chiral interaction (FS) based on
meson exchange, and instanton-induced interaction (INST) based on the
non-perturbative QCD vacuum structure.  The FS (CM) model predicts the lowest
(highest) mass for each state. Results for the FS and CM models are displayed
in Table~\ref{tab:comp}.  In all cases, the mass predicted by the INST model
(not displayed in the table) lies between the values predicted by the other
two models. Our results are closer to those predicted by the FS model,
specially for the lowest lying states.

% ~~~~~~~~~~~~~~~~~~~~~~~~~~~~~~~~~~~~~~~~~~~~C=0  S=0  I=3/2  J=1/2~~~~~~~~~~
\subsection{$\Delta$  states 
($\bm{  C = 0}$, $\bm{ S = 0} $, $\bm{ I=3/2}$ ) }

The model predicts in this sector the existence of five heavy resonances
(bound states; all of them appear below threshold): three with spin-parity
$J^P= \frac{1}{2}^-$ and another two with $J^P=\frac{3}{2}^-$. Their masses,
widths and couplings to the different channels are compiled in
Table~\ref{tab002}.

\begin{itemize}
\item $ J=1/2$: In this sector, there are four coupled
channels, with the following threshold energies (in MeV):

\begin{center}
\noindent
\begin{tabular}{cccc}
$\Delta J/\psi$    & $\Sigmac \bar D$   &   $\Sigma_c \bar D^*$   &  $\Sigma_c^* \bar D^*$ \\
4306.9    &   4320.8   &  4461.9   &   4526.3
\end{tabular}
\smallskip
\end{center}

\item $ J=3/2$: In this sector, there are five coupled channels, with the
  following threshold energies:

\begin{center}
\noindent
\begin{tabular}{ccccc}
$\Delta \eta_c$   & $\Delta J/\psi$ 
   & $\SigmacS \bar D$    & $\Sigma_c \bar D^*$   & $\Sigma_c^* \bar D^*$ \\
4189.7   & 4306.9   & 4385.2   & 4461.9   & 4526.3
\end{tabular}
\end{center}

\item $ J=5/2$: In this sector there are only two channels, with the following
  threshold energies:

\begin{center}
\noindent
\begin{tabular}{cc}
$\Delta J/\psi$     &  $\Sigma_c^* \bar D^*$ \\
4306.9   & 4526.3
\end{tabular}
\end{center}

\end{itemize}

We obtain three $\Delta$($J=1/2$) states as expected from the group
decomposition of the $\SU(6)\times\HQSS$ representations (see
Table~\ref{tab:sectors}): one state from each of the two $J=1/2$ decuplets
encoded in the attractive $\bf{70_{2,0}}$ representation, and a further state
corresponding to the $J=1/2$ decuplet that appears in the reduction of the
$\bf{56_{2,0}}$ representation [\Eq{2.29}].  The evolution of the
corresponding poles is shown in Fig.~\ref{fig_evolution}.

The pole that corresponds to ${\bf (10_4)_{2,0} \subset 56_{2,0} }$ (light
magenta circles in Fig.~\ref{fig_evolution}) has a mass quite close to the
$\Delta \eta_c$ and $\Delta J/\psi$ degenerated thresholds, between the
$\SU(6)\times\HQSS$ and the $\SU(2)\times\HQSS$ symmetric points. Later, while
moving to the SU(2) isospin symmetric point, the spin 1/2 $\Delta$ resonance
keeps having a mass close to the $\Delta J/\psi$ threshold, and ends up with a
final mass of $4306.2\MeV$ (the $\Delta J/\psi$ threshold is at $4306.9\MeV$).
However, the spin 5/2 and the two spin 3/2 states, that are also originated
from this ${\bf (10_4)_{2,0} \subset 56_{2,0} }$ pole, essentially
disappear. One of the $J=3/2$ states still shows up as a cusp very close to
the $\Delta J/\psi$ threshold, and it has been included in the table. The
second state with spin 3/2 (light magenta triangles in
Fig.~\ref{fig_evolution}) and the spin 5/2 one appear as small unnoticeable
peaks right at the $\Delta \eta_c$ and $\Delta J/\Psi$ thresholds,
respectively.

From the discussion above, the ${\bf (10_4)_{2,0} \subset 56_{2,0} }$ HQSS
multiplet could be incomplete.
 
However the three $\Delta$ states (dark magenta circles for the two $J^P=
1/2^-$ states and dark magenta triangles for the $J^P= 3/2^-$ resonance in
Fig.~\ref{fig_evolution}) that stem from the ${\bf (10_2)_{2,0} \subset
  70_{2,0} }$ configuration of the light degrees of freedom turn out to be
quite bound. Indeed, we find binding energies of at least 250 (150) MeV in the
spin 1/2 (3/2) sector. These three states, nearly degenerate, form a clear
HQSS multiplet.

% \Delta
\begin{table*}[ht]
\begin{center}
\begin{tabular}{| c | c | c | c | c | c | c |}
\hline 
 $\SU(6)\times\HQSS$  & $\SU(3)\times\HQSS$  &  	           & 		        & Couplings  	         &  \\
 irrep             & irrep            & $M_{R}$   & $\Gamma_{R}$ &  to main channels  & $J$ \\ 
\hline
%% $\bf{g_{}=}$ - possible decaying channel

 $\bf{70_{2,0}}$ &  $\bf{(10_2)_{2,0}}$  &4005.8  & 0. & $g_{\Delta J/\psi}=0.3$,  $g_{\Sigma_c \bar{D}}=2.7$,
 $g_{\Sigma_c \bar{D}^*}=4.4$, $g_{\Sigma_c^* \bar{D}^*}=1.2$
  &   1/2 \\

\hline

 $\bf{70_{2,0}}$ &  $\bf{(10_2)_{2,0}}$   &  4032.5  & 0. &  $g_{\Delta \eta_c}=0.2$,  $g_{\Delta J/\psi}=0.1$,
$g_{\Sigma_c^* \bar{D}}=2.9$,  $g_{\Sigma_c \bar{D}^*}=1.8$
   &  3/2  \\
 &  &  &  &  $g_{\Sigma_c^* \bar{D}^*}=4.1$  
  & \\

\hline

 $\bf{70_{2,0}}$ & $\bf{(10_2)_{2,0}}$  & 4050.0  & 0. & $g_{\Delta J/\psi}=0.2$,  $g_{\Sigma_c \bar{D}}=0.8$,
 $g_{\Sigma_c \bar{D}^*}=1.9$,  $g_{\Sigma_c^* \bar{D}^*}=5.1$
 & $1/2$ \\

\hline

 $\bf{56_{2,0}}$ & $\bf{(10_4)_{2,0}}$  & 4306.2  & 0. & $g_{\Delta J/\psi}=1.3$,  $g_{\Sigma_c \bar{D}}=0.3$,
$g_{\Sigma_c \bar{D}^*}=0.3$,  $g_{\Sigma_c^* \bar{D}^*}=0.3$
 & $1/2$ \\
 &  &  &  (cusp)  & 
  & \\

\hline

 $\bf{56_{2,0}}$ & $\bf{(10_4)_{2,0}}$ &  4306.8  & 0. & $g_{\Delta \eta_c} \sim 0.1$,  $g_{\Delta J/\psi}=0.8$,
$g_{\Sigma_c^* \bar{D}}=0.2$,  $g_{\Sigma_c \bar{D}^*}=0.2$,
     &  $3/2$  \\
 &  &  &  (cusp) & $g_{\Sigma_c^* \bar{D}^*}=0.1$
  & \\

\hline

\end{tabular}
\end{center}
\caption{ As in Table \ref{tab0011}, for the $\Delta$ ($J=1/2$,$J=3/2$)
  resonances with hidden charm content.  }
\label{tab002}
\end{table*}
%%%%%%%%%%%%%%%

The models based on vector meson exchange, naturally predict a suppression
factor in the baryon-meson amplitudes involving exchange of charm, from the
propagator of the exchanged heavy vector meson. In the heavy quark limit, the
suppression factor is of the order of $1/m_H$.\footnote{The boson propagator
  is approximately $1/(2 M_H(E_H-M_H))$, with $M_H$ the mass of the heavy
  vector meson and $E_H$ its energy, and $E_H-M_H$ is $O(1)$ in the heavy
  quark limit.} Therefore, in that limit, one expects a quenching of order
$M_V/M_{D^*}$ for the charm exchanging amplitudes of those models. (Of course,
the true factor for large but finite physical heavy hadron masses needs not
exactly coincide with this heavy quark limit estimate.) Our model is not
directly based on exchange of vector mesons. Nevertheless, as commented above,
we have verified that adding such suppression by hand in the charm exchanging
amplitudes does not have an impact on our results. Even a factor
$(M_V/M_{D^*})^2$, proposed in the literature \cite{Mizutani:2006vq} has a
very small effect in the position of the resonances we find.  Presumably, this
is due to the fact that the relevant channels have a small coupling. An
exception comes from the two very weakly bound $\Delta$ resonances from the
$\bm{56}_{2,0}$ irrep, which disappear due to the suppression of their
dominant channel $\Delta J/\psi$.

\section{Conclusions}
\label{sec:conc}

In the present work we develop a model for the interaction of lowest-lying
$1/2^+$ and $3/2^+$ baryons with $0^-$ and $1^-$ mesons, including light and
heavy flavors. The interaction is of zero range and it is modeled as a
suitable extension of the Weinberg-Tomozawa term to include, besides chiral
symmetry, spin-flavor symmetry in the light sector and heavy quark spin
symmetry. These symmetries are only broken in our model to the extent that
they are broken at level of the physical masses and meson decay constants. The
OZI rule is also automatically implemented. Our extended WT model, \Eq{2.21},
contains no adjustable parameters, although some ambiguity is present through
the choice of renormalization prescription, as in all other hadronic molecular
approaches.  The model has been applied previously to the light sector and to
charm or bottom sectors with a single heavy quark. Here we show that it admits
a natural realization in sectors with hidden charm in such a way that HQSS is
preserved. In particular, the spin of $c$ quarks and the spin of $\bc$
antiquarks are separately conserved.

We have carried out a detailed analysis of the hidden charm sectors (i.e.,
with $c\bc$ pairs) with $C=0,1,2,3$ and their breaking as the symmetry is
lifted from $\SU(6)\times\HQSS$ to $\SU(3)\times\HQSS$ (and then to
$\SU(2)\times\HQSS$ and $\SU(2)$ of isospin). This allows to count the
expected number of bound states or resonances, and to classify them into
multiplets corresponding to the various symmetries. Taking the eigenvalues
$\lambda's$ as undetermined (free) parameters, the results of
Eqs.~(\ref{eq:c0lambda}), (\ref{eq:c1lambda}), (\ref{eq:c2lambda}) and
(\ref{eq:c3lambda}), for the $C=0$, $C=1$, $C=2$ and $C=3$ hidden charm
sectors respectively, are general. Indeed, these equations fixes the most
general structure of eigenvalues that can be deduced from
$\SU(3)\times\HQSS$. The rest of undetermined parameters, not fixed by this
latter symmetry, accounts for non-diagonal transitions between multiplets with
the same $\SU(3)\times \SU_{J_\ell}(2)$ SF quantum numbers for the light
degrees of freedom. Further, we have translated this general discussion of the
group structure allowed by $\SU(3)\times\HQSS$ into Lagrangian form, for the
charmless hidden-charm sector. This makes the HQSS of the model explicit and
it allows to compare with other models in the literature. Finally, we have
found the couplings of the HQSS effective Lagrangians of
Eqs.~(\ref{eq:g1})--(\ref{eq:g11}) for the particular case of our extended WT
model. This constitutes an additional check of its compatibility with HQSS.

We have analyzed the charmless and strangeless sector, where we have
dynamically generated several $N$ and $\Delta$ states.  Actually, we predict
the existence of seven $N$-like and five $\Delta$-like states with masses
around $4\GeV$, most of them as bound states.  These states form heavy-quark
spin multiplets, which are almost degenerate in mass.  The $N$ states form two
HQSS multiplets. The lowest one has the light quark flavor-spin content
coupled to $\bf{8_2} $. Since the $\bar{c}c$ pair can couple to spin $S_{c\bar
  c}=0,1$, this HQSS multiplet consists of three nucleon states with
$J=1/2,~1/2,$ and $3/2$, and masses around $3930\MeV$.  On the other hand, the
highest HQSS nucleon-like multiplet contains four resonances with
$J=1/2,~3/2,~3/2$ and $5/2$, and masses around $4000\MeV$. In this case, these
states are originated from the $\bf{8_4}$ SF light configuration.  These two
$\SU(3)\times\HQSS$ multiplets arise from the {\bf 70}-plet of
$\SU(6)\times\HQSS$.  There are no $N$ physical states coming from the {\bf
  56}-plet.  With regards to $\Delta$ states, we find two multiples with very
different average masses, because in this case they are originated from
different $\SU(6)\times\HQSS$ representations. The $\Delta$ multiplet coming
from the ${\bf (10_2)_{2,0} \subset 70_{2,0} }$ irrep is formed by 3 states
($J=1/2,~1/2,~3/2$) with an average mass of $4035\MeV$. Besides, we find only
two ($J=1/2,~3/2$) $\Delta$ resonances at the physical point out of the four
states originated from the ${\bf (10_4)_{2,0} \subset 56_{2,0} }$ in the
$\SU(6)$ limit. These two states are nearly degenerate, with a mass of
$4306\MeV$.

When we compare our approach with the other molecular-type ones, we observe in
our model a rich structure of resonances due to the many channels cooperating
to create them. Our states are much more lighter that those predicted in the
hidden-gauge scheme~\cite{Wu:2010jy,Wu:2010vk,Wu:2012md}, though significantly
less bound that the crypto-exotic baryons reported in the zero range vector
meson exchange model of Refs.~\cite{Hofmann:2005sw,Hofmann:2006qx}.  Moreover,
we have presented the first prediction for exotic hidden-charm $\Delta$-like
resonances within a molecular baryon-meson scheme.

In comparison with the quark model of Ref.~\cite{Yuan:2012wz}, we find that
our results are closer to those predicted by the FS hyperfine interaction
discussed in Ref.~\cite{Yuan:2012wz}, specially for the lowest lying states.

The predicted new resonances definitely cannot be accommodated by quark models
with three constituent quarks and they might be looked for in the forthcoming
$\bar {\mathrm P}$ANDA experiment at the future FAIR facility.

\section*{Acknowledgments}

This research was supported by Spanish DGI and FEDER funds, under contracts
FIS2011-28853-C02-02, FIS2011-24149, FPA2010-16963 and the Spanish
Consolider-Ingenio 2010 Programme CPAN (CSD2007-00042), by Junta de
Andaluc\'{\i}a grant FQM-225, by Generalitat Valenciana under contract
PROMETEO/2009/0090 and by the EU HadronPhysics2 project, grant agreement
n. 227431. O.R. wishes to acknowledge support from the Rosalind Franklin
Fellowship. L.T. acknowledges support from Ramon y Cajal Research Programme,
and from FP7-PEOPLE-2011-CIG under contract PCIG09-GA-2011-291679.

\appendix
\section{Spin-flavor states}
\label{app:spin-flavor}

In this appendix we give details regarding the construction of the tensors
$M^A{}_B$ and $B^{ABC}$, the SU(3) multiplets $\Sigma^a{}_b$, $\bar{D}^a$,
etc, and the computation of the matrix elements of the interaction.

The wave-functions in spin-flavor space of the basic mesons and baryons are
constructed in terms of bosonic quark and antiquark operators with spin and
flavor labels, namely, $Q^\dagger_{f\uparrow}$,$Q^\dagger_{f\downarrow}$,
$Q^\dagger_{\bar{f} \uparrow}$ $Q^\dagger_{\bar{f} \downarrow}$, $f={\rm
  u,d,s,c}$.  The concrete wave-functions are those given in the Appendix A of
\cite{GarciaRecio:2008dp} with the following modification: a minus sign is to
be applied to all $1/2^+$ baryons, to all $0^-$ mesons except $\eta$,
$\eta^\prime$ and $\eta_c$, and to $\phi$, $\omega$ and $J/\psi$ (denoted
$\psi$ in \cite{GarciaRecio:2008dp}).  No change of sign is to be applied to
$3/2^+$ baryons, nor to $\eta$, $\eta^\prime$ and $\eta_c$, nor to $1^-$
mesons (except $\phi$, $\omega$ and $J/\psi$).

The states just defined are standard with respect to the flavor and
spin-flavor groups conventions of \cite{GarciaRecio:2010vf}. In particular
they are $\SU(2)_J$, $\SU(2)_I$ standard and follow the convention of
\cite{Baird:1964zm} for flavor SU(3) and SU(4). The only exceptions come from
the neutral mesons for which we use ideal mixing. In terms of these, the
standard states of \cite{GarciaRecio:2010vf} are given by:
\begin{eqnarray}
|\eta^\prime\rangle_{\rm stan} &=& |\eta^\prime\rangle
\qquad (\SU(3))
,\\
|\eta^\prime\rangle_{\rm stan} &=&
\sqrt{\frac{3}{4}}|\eta^\prime\rangle
+\frac{1}{2} |\eta_c\rangle 
\qquad (\SU(4))
,\\
|\eta_c\rangle_{\rm stan} &=& 
-\frac{1}{2}|\eta^\prime\rangle
+\sqrt{\frac{3}{4}} |\eta_c\rangle
\qquad (\SU(4))
,\\
|\omega_8\rangle &=&
\sqrt{\frac{1}{3}} |\omega \rangle 
+\sqrt{\frac{2}{3}} |\phi \rangle 
\qquad (\SU(3) \mbox{~and~} \SU(4))
,\\
|\omega_1\rangle &=&
\sqrt{\frac{2}{3}} |\omega \rangle 
-\sqrt{\frac{1}{3}} |\phi \rangle 
\qquad (\SU(3))
,\\
|\omega_1\rangle &=&
\sqrt{\frac{1}{2}} |\omega \rangle 
-\frac{1}{2} |\phi \rangle 
+\frac{1}{2} |J/\psi\rangle
\qquad (\SU(4))
,\\
|\psi\rangle &=&
-\sqrt{\frac{1}{6}} |\omega \rangle 
+\sqrt{\frac{1}{12}} |\phi \rangle 
+\sqrt{\frac{3}{4}} |J/\psi\rangle
\qquad (\SU(4))
\label{eq:A7}
.
\end{eqnarray}
In these formulas, the right-hand sides contain the physical (or rather ideal
mixing) neutral mesons that we use in this work. Their wave-functions are
constructed as indicated above (i.e., from those in
\cite{GarciaRecio:2008dp}). The left-hand sides contain the standard or
mathematical states used in \cite{GarciaRecio:2010vf}. They have good quantum
numbers with respect to SU(6) or SU(8) (and their corresponding chain of
subgroups).\footnote{Note that $|\psi\rangle$ of \cite{GarciaRecio:2008dp}
  corresponds to $-|J/\psi\rangle$ here, not to $|\psi\rangle$ of
  \cite{GarciaRecio:2010vf} and of \Eq{A7}.}

In order to construct the tensors $M^A{}_B$ and $B^{ABC}$, $M^\dagger{}^A{}_B$
and $B^\dagger_{ABC}$, with good transformations properties, the following
procedure is used: For all flavors $f={\rm u,d,s,c}$, and for the various
creation operators $Q^\dagger_{f\uparrow}$, $Q^\dagger_{f\downarrow}$,
$Q^\dagger_{\bar{f} \uparrow}$, and $Q^\dagger_{\bar{f}\downarrow}$, appearing
in the wave-functions of the hadrons, the following replacements are to be
applied:
\begin{equation}
Q^\dagger_{f\uparrow} \to +Q^\dagger_{f1}, 
\quad
Q^\dagger_{f\downarrow} \to +Q^\dagger_{f2},
\quad Q^\dagger_{\bar{f} \uparrow} \to -\bQ^\dagger{}^{f2},
\quad
Q^\dagger_{\bar{f}\downarrow} \to +Q^\dagger_{f1}
\end{equation}
Note i) the minus sign in $Q^\dagger_{\bar{f} \uparrow}$, and ii) for quarks,
the labels 1 and 2 correspond to spin up and down, respectively, but for
antiquarks they correspond to spin down and up, respectively.

After the replacement, there are only operators $Q^\dagger_A$, and
$\bQ^\dagger{}^A$ for creation (and $Q^A$, and $\bQ_A$ for annihilation),
carrying any of the eight labels $A={\rm u1,d1,s1,c1,u2,d2,s2,c2}$. These
operators transform under SU(8) in the way indicated in \Eq{2.17}.

The meson matrix is then obtained by replacing $\bQ^\dagger{}^A Q^\dagger_B$
with $M^\dagger{}^A{}_B$ and expressing it in terms of meson operators by
inverting the wave-function equations. Similarly, for the baryons,
$Q^\dagger_A Q^\dagger_B Q^\dagger_C$ is replaced with $B^\dagger_{ABC}$ and
then expressed in terms of baryon operators. The fields $\Phi^A{}_B(x)$ and
$\mathcal{B}^{ABC}(x)$ of Section \ref{sec:2.B} are constructed in the usual
way from these annihilation and creation operators.

The SU(3) multiplets introduced in Section \ref{sec:2.F} are obtained as
follows.
\begin{equation}
\Delta^{abc} = B^{(as_1,bs_2,cs_3)_{3/2}}
\end{equation}
Here $A=as_1$, $B=bs_2$, $C=cs_3$, with $a,b,c\in\{{\rm u,d,s}\}$,
$s_1,s_2,s_3\in\{1,2\}$, and the notation $(ABC)_{3/2}$ indicates that the
spin part is coupled to $J=3/2$ with $1=\uparrow$ and $2=\downarrow$. Besides
we refer here to the (annihilation) operator; the field $\Delta^{abc}_\mu(x)$
is constructed out of it.
\begin{eqnarray}
\Sigma^a{}_b &=& \sqrt{\frac{1}{6}} \epsilon_{bcd} B^{(as_1(cs_2,ds_3)_0)_{1/2}}
,
\\
\Sigma_c^{ab} &=& B^{((as_1,bs_2)_1 {\rm c}s_3)_{1/2}}, 
,
\\
\Sigma_c^*{}^{ab} &=& B^{((as_1,bs_2)_1 {\rm c}s_3)_{3/2}}, 
,
\\
\Xi_c {}_a &=& \frac{1}{2}\epsilon_{abc}B^{((bs_1,cs_2)_0 {\rm c}s_3)_{1/2}}, 
.
\end{eqnarray}

For the mesons
\begin{eqnarray}
\bar{D}^a &=& M^{(as_1}{}_{{\rm c}s_2)_0}
,
\\
\bar{D}^*{}^a &=&  - M^{(as_1}{}_{{\rm c} s_2)_1}
,
\\
\eta_c &=& M^{({\rm c}s_1}{}_{{\rm c}s_2)_0}
,
\\
J/\psi &=& - M^{({\rm c}s_1}{}_{{\rm c} s_2)_1}
.
\end{eqnarray}
Recalling that $\bar{Q}_{f1}= Q_{\bar{f}\downarrow}$ and $\bar{Q}_{f2}=
-Q_{\bar{f}\uparrow}$, it follows that $M^{(as_1}{}_{b s_2)_0} = (
M^{a1}{}_{b1} + M^{a2}{}_{b2})/\sqrt{2}$, while $M^{(as_1}{}_{b s_2)_1}$
equals $-M^{a1}{}_{b2}$, $( M^{a1}{}_{b1} - M^{a2}{}_{b2})/\sqrt{2}$ and
$M^{a2}{}_{b1}$, for $J_3=+1,0,-1$, respectively.

Finally, we remark that we systematically take the coupling of baryon and
meson in the order ${\rm |baryon\rangle\otimes|meson\rangle}$, rather than
${\rm |meson\rangle\otimes|baryon\rangle}$.

\section{Baryon-meson matrix elements}
\label{app:tables}

The coefficients $D_{ij}$, appearing in \Eq{pot}, for the charmless ($C=0$)
and strangeless ($S=0$) sector are compiled in this Appendix
(Tables~\ref{tab:0m022}, \ref{tab:0m024}, \ref{tab:0m026}, \ref{tab:0m042},
\ref{tab:0m044} and \ref{tab:0m046}). In addition, we also provide here the
corresponding coefficients for the $C=0$ and $S\ne 0$ sectors (Tables
~\ref{tab:0m112}, ~\ref{tab:0m114}, ~\ref{tab:0m116}, ~\ref{tab:0m132},
~\ref{tab:0m134}, ~\ref{tab:0m136} ~\ref{tab:0m222}, ~\ref{tab:0m224},
~\ref{tab:0m226}, ~\ref{tab:0m312}, ~\ref{tab:0m314},~\ref{tab:0m316}).

%----------------------------------------------------------------------

\bibliographystyle{h-physrev4}

\begin{table}
\caption{
  $ C=0$, $ S=0$, $ I=1/2$, $ J= 1/2$.
}
\label{tab:0m022}
\begin{tabular}{r|rrrrrrrrrrrrrrrrrrrrrrrrrrrrrrrrrrrrrrrrrrrrrrrr}
 &  $ \Nucleon \etac $ 
 &  $ \Nucleon \Jpsi $ &  $ \Lambdac 
  \Db $ &  $ \Lambdac \DSb $ 
 &  $ \Sigmac \Db $ &  $ \Sigmac \DSb 
  $ &  $ \SigmacS \DSb $\\
\hline
$ \Nucleon \etac $& $ 0 $ 
 & $ 0 $ 
 & $ \sqrt{\frac{ 3 }{ 2 }} 
   $ & $ -\sqrt{\frac{ 9 }{ 2 
     }} $ & $ \sqrt{\frac{ 3 
    }{ 2 }} $ 
 & $ \sqrt{\frac{ 1 }{ 2 }} 
   $ & $ 2 $ \\
$ \Nucleon \Jpsi $& $ 0 $ 
 & $ 0 $ 
 & $ -\sqrt{\frac{ 9 }{ 2 }} 
   $ & $ -\sqrt{\frac{ 3 }{ 2 
     }} $ & $ \sqrt{\frac{ 1 
    }{ 2 }} $ 
 & $ \sqrt{\frac{ 25 }{ 6 }} 
   $ & $ -\sqrt{\frac{ 4 }{ 3 
     }} $ \\
$ \Lambdac \Db $
 & $ \sqrt{\frac{ 3 }{ 2 }} 
   $ & $ -\sqrt{\frac{ 9 }{ 2 
     }} $ & $ 1 $ 
 & $ 0 $ & $ 0 $ 
 & $ -\sqrt{ 3 } $ 
 & $ \sqrt{ 6 } $ \\
$ \Lambdac \DSb $
 & $ -\sqrt{\frac{ 9 }{ 2 }} 
   $ & $ -\sqrt{\frac{ 3 }{ 2 
     }} $ & $ 0 $ 
 & $ 1 $ 
 & $ -\sqrt{ 3 } $ 
 & $ -2 $ 
 & $ -\sqrt{ 2 } $ \\
$ \Sigmac \Db $
 & $ \sqrt{\frac{ 3 }{ 2 }} 
   $ & $ \sqrt{\frac{ 1 }{ 2 
    }} $ & $ 0 $ 
 & $ -\sqrt{ 3 } $ 
 & $ -1 $ 
 & $ \sqrt{\frac{ 4 }{ 3 }} 
   $ & $ \sqrt{\frac{ 2 }{ 3 
    }} $ \\
$ \Sigmac \DSb $
 & $ \sqrt{\frac{ 1 }{ 2 }} 
   $ & $ \sqrt{\frac{ 25 }{ 6 
    }} $ & $ -\sqrt{ 3 } 
   $ & $ -2 $ 
 & $ \sqrt{\frac{ 4 }{ 3 }} 
   $ & $ \frac{ 1 }{ 3  } 
   $ & $ -\sqrt{\frac{ 2 }{ 9 
     }} $ \\
$ \SigmacS \DSb $& $ 2 $ 
 & $ -\sqrt{\frac{ 4 }{ 3 }} 
   $ & $ \sqrt{ 6 } $ 
 & $ -\sqrt{ 2 } $ 
 & $ \sqrt{\frac{ 2 }{ 3 }} 
   $ & $ -\sqrt{\frac{ 2 }{ 9 
     }} $ & $ \frac{ 2 }{ 3 
     } $ \\
\end{tabular}
\label{tabD11}
\end{table}

\begin{table}
\caption{
  $ C=0$, $ S=0$, $ I=1/2$, $ J= 3/2$.
}
\label{tab:0m024}
\begin{tabular}{r|rrrrrrrrrrrrrrrrrrrrrrrrrrrrrrrrrrrrrrrrrrrrrrrr}
 &  $ \Nucleon \Jpsi $ 
 &  $ \Lambdac \DSb $ &  $ \SigmacS \Db 
  $ &  $ \Sigmac \DSb $ 
 &  $ \SigmacS \DSb $\\
\hline
$ \Nucleon \Jpsi $& $ 0 $ 
 & $ \sqrt{ 6 } $ 
 & $ -\sqrt{ 2 } $ 
 & $ \sqrt{\frac{ 2 }{ 3 }} 
   $ & $ -\sqrt{\frac{ 10 }{ 3 
     }} $ \\
$ \Lambdac \DSb $
 & $ \sqrt{ 6 } $ 
 & $ 1 $ 
 & $ -\sqrt{ 3 } $ 
 & $ 1 $ 
 & $ -\sqrt{ 5 } $ \\
$ \SigmacS \Db $
 & $ -\sqrt{ 2 } $ 
 & $ -\sqrt{ 3 } $ 
 & $ -1 $ 
 & $ -\sqrt{\frac{ 1 }{ 3 }} 
   $ & $ \sqrt{\frac{ 5 }{ 3 
    }} $ \\
$ \Sigmac \DSb $
 & $ \sqrt{\frac{ 2 }{ 3 }} 
   $ & $ 1 $ 
 & $ -\sqrt{\frac{ 1 }{ 3 }} 
   $ & $ -\frac{ 5 }{ 3  } 
   $ & $ -\sqrt{\frac{ 5 }{ 9 
     }} $ \\
$ \SigmacS \DSb $
 & $ -\sqrt{\frac{ 10 }{ 3 }} 
   $ & $ -\sqrt{ 5 } $ 
 & $ \sqrt{\frac{ 5 }{ 3 }} 
   $ & $ -\sqrt{\frac{ 5 }{ 9 
     }} $ 
 & $ -\frac{ 1 }{ 3  } $ \\
\end{tabular}
\label{tabD12}
\end{table}

\begin{table}
\caption{
  $ C=0$, $ S=0$, $ I=1/2$, $ J= 5/2$.
}
\label{tab:0m026}
\begin{tabular}{r|rrrrrrrrrrrrrrrrrrrrrrrrrrrrrrrrrrrrrrrrrrrrrrrr}
 &  $ \SigmacS \DSb $\\
\hline
$ \SigmacS \DSb $& $ -2 $ \\
\end{tabular}
\label{tabD13}
\end{table}

\begin{table}
\caption{
  $ C=0$, $ S=0$, $ I=3/2$, $ J= 1/2$.
}
\label{tab:0m042}
\begin{tabular}{r|rrrrrrrrrrrrrrrrrrrrrrrrrrrrrrrrrrrrrrrrrrrrrrrr}
 &  $ \Delta \Jpsi $ &  $ \Sigmac 
  \Db $ &  $ \Sigmac \DSb $ 
 &  $ \SigmacS \DSb $\\
\hline
$ \Delta \Jpsi $& $ 0 $ 
 & $ \sqrt{ 8 } $ 
 & $ -\sqrt{\frac{ 8 }{ 3 }} 
   $ & $ -\sqrt{\frac{ 4 }{ 3 
     }} $ \\
$ \Sigmac \Db $
 & $ \sqrt{ 8 } $ 
 & $ 2 $ 
 & $ -\sqrt{\frac{ 16 }{ 3 }} 
   $ & $ -\sqrt{\frac{ 8 }{ 3 
     }} $ \\
$ \Sigmac \DSb $
 & $ -\sqrt{\frac{ 8 }{ 3 }} 
   $ & $ -\sqrt{\frac{ 16 }{ 3 
     }} $ 
 & $ -\frac{ 2 }{ 3  } $ 
 & $ \sqrt{\frac{ 8 }{ 9 }} 
   $ \\
$ \SigmacS \DSb $
 & $ -\sqrt{\frac{ 4 }{ 3 }} 
   $ & $ -\sqrt{\frac{ 8 }{ 3 
     }} $ & $ \sqrt{\frac{ 8 
    }{ 9 }} $ 
 & $ -\frac{ 4 }{ 3  } $ \\
\end{tabular}
\label{tabD21}
\end{table}

\begin{table}
\caption{
  $ C=0$, $ S=0$, $ I=3/2$, $ J= 3/2$.
}
\label{tab:0m044}
\begin{tabular}{r|rrrrrrrrrrrrrrrrrrrrrrrrrrrrrrrrrrrrrrrrrrrrrrrr}
 &  $ \Delta \etac $ &  $ \Delta 
  \Jpsi $ &  $ \SigmacS \Db $ 
 &  $ \Sigmac \DSb $ &  $ \SigmacS \DSb 
  $\\
\hline
$ \Delta \etac $& $ 0 $ 
 & $ 0 $ & $ \sqrt{ 3 } 
   $ & $ -2 $ 
 & $ -\sqrt{ 5 } $ \\
$ \Delta \Jpsi $& $ 0 $ 
 & $ 0 $ 
 & $ -\sqrt{ 5 } $ 
 & $ -\sqrt{\frac{ 20 }{ 3 }} 
   $ & $ \sqrt{\frac{ 1 }{ 3 
    }} $ \\
$ \SigmacS \Db $
 & $ \sqrt{ 3 } $ 
 & $ -\sqrt{ 5 } $ 
 & $ 2 $ 
 & $ \sqrt{\frac{ 4 }{ 3 }} 
   $ & $ -\sqrt{\frac{ 20 }{ 3 
     }} $ \\
$ \Sigmac \DSb $& $ -2 $ 
 & $ -\sqrt{\frac{ 20 }{ 3 }} 
   $ & $ \sqrt{\frac{ 4 }{ 3 
    }} $ & $ \frac{ 10 }{ 3 
     } $ & $ \sqrt{\frac{ 20 
    }{ 9 }} $ \\
$ \SigmacS \DSb $
 & $ -\sqrt{ 5 } $ 
 & $ \sqrt{\frac{ 1 }{ 3 }} 
   $ & $ -\sqrt{\frac{ 20 }{ 3 
     }} $ & $ \sqrt{\frac{ 20 
    }{ 9 }} $ 
 & $ \frac{ 2 }{ 3  } $ \\
\end{tabular}
\label{tabD22}
\end{table}

\begin{table}
\caption{
  $ C=0$, $ S=0$, $ I=3/2$, $ J= 5/2$.
}
\label{tab:0m046}
\begin{tabular}{r|rrrrrrrrrrrrrrrrrrrrrrrrrrrrrrrrrrrrrrrrrrrrrrrr}
 &  $ \Delta \Jpsi $ &  $ \SigmacS 
  \DSb $\\
\hline
$ \Delta \Jpsi $& $ 0 $ 
 & $ \sqrt{ 12 } $ \\
$ \SigmacS \DSb $
 & $ \sqrt{ 12 } $ 
 & $ 4 $ \\
\end{tabular}
\label{tabD23}
\end{table}

\begin{table}
\caption{
  $ C=0$, $ S=-1$, $ I=0$, $ J= 1/2$.
}
\label{tab:0m112}
\begin{tabular}{r|rrrrrrrrrrrrrrrrrrrrrrrrrrrrrrrrrrrrrrrrrrrrrrrr}
 &  $ \Lambda \etac $ 
 &  $ \Lambda \Jpsi $ &  $ \Lambdac 
  \Dsb $ &  $ \Cascadac \Db $ 
 &  $ \Lambdac \DsSb $ &  $ \Cascadacp 
  \Db $ &  $ \Cascadac \DSb $ 
 &  $ \Cascadacp \DSb $ &  $ \CascadacS 
  \DSb $\\
\hline
$ \Lambda \etac $& $ 0 $ 
 & $ 0 $ & $ 1 $ 
 & $ \sqrt{\frac{ 1 }{ 2 }} 
   $ & $ -\sqrt{ 3 } $ 
 & $ \sqrt{\frac{ 3 }{ 2 }} 
   $ & $ -\sqrt{\frac{ 3 }{ 2 
     }} $ & $ \sqrt{\frac{ 1 
    }{ 2 }} $ & $ 2 $ \\
$ \Lambda \Jpsi $& $ 0 $ 
 & $ 0 $ 
 & $ -\sqrt{ 3 } $ 
 & $ -\sqrt{\frac{ 3 }{ 2 }} 
   $ & $ -1 $ 
 & $ \sqrt{\frac{ 1 }{ 2 }} 
   $ & $ -\sqrt{\frac{ 1 }{ 2 
     }} $ & $ \sqrt{\frac{ 25 
    }{ 6 }} $ 
 & $ -\sqrt{\frac{ 4 }{ 3 }} 
   $ \\
$ \Lambdac \Dsb $& $ 1 $ 
 & $ -\sqrt{ 3 } $ 
 & $ 0 $ & $ \sqrt{ 2 } 
   $ & $ 0 $ & $ 0 $ 
 & $ 0 $ 
 & $ -\sqrt{ 2 } $ 
 & $ 2 $ \\
$ \Cascadac \Db $
 & $ \sqrt{\frac{ 1 }{ 2 }} 
   $ & $ -\sqrt{\frac{ 3 }{ 2 
     }} $ & $ \sqrt{ 2 } 
   $ & $ -1 $ & $ 0 $ 
 & $ 0 $ & $ 0 $ 
 & $ -1 $ 
 & $ \sqrt{ 2 } $ \\
$ \Lambdac \DsSb $
 & $ -\sqrt{ 3 } $ 
 & $ -1 $ & $ 0 $ 
 & $ 0 $ & $ 0 $ 
 & $ -\sqrt{ 2 } $ 
 & $ \sqrt{ 2 } $ 
 & $ -\sqrt{\frac{ 8 }{ 3 }} 
   $ & $ -\sqrt{\frac{ 4 }{ 3 
     }} $ \\
$ \Cascadacp \Db $
 & $ \sqrt{\frac{ 3 }{ 2 }} 
   $ & $ \sqrt{\frac{ 1 }{ 2 
    }} $ & $ 0 $ & $ 0 $ 
 & $ -\sqrt{ 2 } $ 
 & $ -1 $ & $ -1 $ 
 & $ \sqrt{\frac{ 4 }{ 3 }} 
   $ & $ \sqrt{\frac{ 2 }{ 3 
    }} $ \\
$ \Cascadac \DSb $
 & $ -\sqrt{\frac{ 3 }{ 2 }} 
   $ & $ -\sqrt{\frac{ 1 }{ 2 
     }} $ & $ 0 $ 
 & $ 0 $ & $ \sqrt{ 2 } 
   $ & $ -1 $ & $ -1 $ 
 & $ -\sqrt{\frac{ 4 }{ 3 }} 
   $ & $ -\sqrt{\frac{ 2 }{ 3 
     }} $ \\
$ \Cascadacp \DSb $
 & $ \sqrt{\frac{ 1 }{ 2 }} 
   $ & $ \sqrt{\frac{ 25 }{ 6 
    }} $ & $ -\sqrt{ 2 } 
   $ & $ -1 $ 
 & $ -\sqrt{\frac{ 8 }{ 3 }} 
   $ & $ \sqrt{\frac{ 4 }{ 3 
    }} $ 
 & $ -\sqrt{\frac{ 4 }{ 3 }} 
   $ & $ \frac{ 1 }{ 3  } 
   $ & $ -\sqrt{\frac{ 2 }{ 9 
     }} $ \\
$ \CascadacS \DSb $& $ 2 $ 
 & $ -\sqrt{\frac{ 4 }{ 3 }} 
   $ & $ 2 $ 
 & $ \sqrt{ 2 } $ 
 & $ -\sqrt{\frac{ 4 }{ 3 }} 
   $ & $ \sqrt{\frac{ 2 }{ 3 
    }} $ 
 & $ -\sqrt{\frac{ 2 }{ 3 }} 
   $ & $ -\sqrt{\frac{ 2 }{ 9 
     }} $ & $ \frac{ 2 }{ 3 
     } $ \\
\end{tabular}
\end{table}

\begin{table}
\caption{
  $ C=0$, $ S=-1$, $ I=0$, $ J= 3/2$.
}
\label{tab:0m114}
\begin{tabular}{r|rrrrrrrrrrrrrrrrrrrrrrrrrrrrrrrrrrrrrrrrrrrrrrrr}
 &  $ \Lambda \Jpsi $ 
 &  $ \Lambdac \DsSb $ &  $ \Cascadac 
  \DSb $ &  $ \CascadacS \Db $ 
 &  $ \Cascadacp \DSb $ &  $ \CascadacS 
  \DSb $\\
\hline
$ \Lambda \Jpsi $& $ 0 $ 
 & $ 2 $ & $ \sqrt{ 2 } 
   $ & $ -\sqrt{ 2 } $ 
 & $ \sqrt{\frac{ 2 }{ 3 }} 
   $ & $ -\sqrt{\frac{ 10 }{ 3 
     }} $ \\
$ \Lambdac \DsSb $& $ 2 $ 
 & $ 0 $ & $ \sqrt{ 2 } 
   $ & $ -\sqrt{ 2 } $ 
 & $ \sqrt{\frac{ 2 }{ 3 }} 
   $ & $ -\sqrt{\frac{ 10 }{ 3 
     }} $ \\
$ \Cascadac \DSb $
 & $ \sqrt{ 2 } $ 
 & $ \sqrt{ 2 } $ 
 & $ -1 $ & $ -1 $ 
 & $ \sqrt{\frac{ 1 }{ 3 }} 
   $ & $ -\sqrt{\frac{ 5 }{ 3 
     }} $ \\
$ \CascadacS \Db $
 & $ -\sqrt{ 2 } $ 
 & $ -\sqrt{ 2 } $ 
 & $ -1 $ & $ -1 $ 
 & $ -\sqrt{\frac{ 1 }{ 3 }} 
   $ & $ \sqrt{\frac{ 5 }{ 3 
    }} $ \\
$ \Cascadacp \DSb $
 & $ \sqrt{\frac{ 2 }{ 3 }} 
   $ & $ \sqrt{\frac{ 2 }{ 3 
    }} $ & $ \sqrt{\frac{ 1 
    }{ 3 }} $ 
 & $ -\sqrt{\frac{ 1 }{ 3 }} 
   $ & $ -\frac{ 5 }{ 3  } 
   $ & $ -\sqrt{\frac{ 5 }{ 9 
     }} $ \\
$ \CascadacS \DSb $
 & $ -\sqrt{\frac{ 10 }{ 3 }} 
   $ & $ -\sqrt{\frac{ 10 }{ 3 
     }} $ 
 & $ -\sqrt{\frac{ 5 }{ 3 }} 
   $ & $ \sqrt{\frac{ 5 }{ 3 
    }} $ 
 & $ -\sqrt{\frac{ 5 }{ 9 }} 
   $ & $ -\frac{ 1 }{ 3  } 
   $ \\
\end{tabular}
\end{table}

\begin{table}
\caption{
  $ C=0$, $ S=-1$, $ I=0$, $ J= 5/2$.
}
\label{tab:0m116}
\begin{tabular}{r|rrrrrrrrrrrrrrrrrrrrrrrrrrrrrrrrrrrrrrrrrrrrrrrr}
 &  $ \CascadacS \DSb $\\
\hline
$ \CascadacS \DSb $& $ -2 $ \\
\end{tabular}
\end{table}

\begin{table}
\caption{
  $ C=0$, $ S=-1$, $ I=1$, $ J= 1/2$.
}
\label{tab:0m132}
\begin{tabular}{r|rrrrrrrrrrrrrrrrrrrrrrrrrrrrrrrrrrrrrrrrrrrrrrrr}
 &  $ \Sigma \etac $ &  $ \Sigma 
  \Jpsi $ &  $ \Cascadac \Db $ 
 &  $ \Sigmac \Dsb $ &  $ \Cascadacp 
  \Db $ &  $ \Cascadac \DSb $ 
 &  $ \SigmaS \Jpsi $ &  $ \Sigmac 
  \DsSb $ &  $ \Cascadacp \DSb $ 
 &  $ \SigmacS \DsSb $ &  $ \CascadacS 
  \DSb $\\
\hline
$ \Sigma \etac $& $ 0 $ 
 & $ 0 $ 
 & $ \sqrt{\frac{ 3 }{ 2 }} 
   $ & $ 1 $ 
 & $ -\sqrt{\frac{ 1 }{ 2 }} 
   $ & $ -\sqrt{\frac{ 9 }{ 2 
     }} $ & $ 0 $ 
 & $ \sqrt{\frac{ 1 }{ 3 }} 
   $ & $ -\sqrt{\frac{ 1 }{ 6 
     }} $ & $ \sqrt{\frac{ 8 
    }{ 3 }} $ 
 & $ -\sqrt{\frac{ 4 }{ 3 }} 
   $ \\
$ \Sigma \Jpsi $& $ 0 $ 
 & $ 0 $ 
 & $ -\sqrt{\frac{ 9 }{ 2 }} 
   $ & $ \sqrt{\frac{ 1 }{ 3 
    }} $ 
 & $ -\sqrt{\frac{ 1 }{ 6 }} 
   $ & $ -\sqrt{\frac{ 3 }{ 2 
     }} $ & $ 0 $ 
 & $ \frac{ 5 }{ 3  } $ 
 & $ -\sqrt{\frac{ 25 }{ 18 }} 
   $ & $ -\sqrt{\frac{ 8 }{ 9 
     }} $ & $ \frac{ 2 }{ 3 
     } $ \\
$ \Cascadac \Db $
 & $ \sqrt{\frac{ 3 }{ 2 }} 
   $ & $ -\sqrt{\frac{ 9 }{ 2 
     }} $ & $ 1 $ 
 & $ 0 $ & $ 0 $ 
 & $ 0 $ & $ 0 $ 
 & $ -\sqrt{ 2 } $ 
 & $ 1 $ & $ 2 $ 
 & $ -\sqrt{ 2 } $ \\
$ \Sigmac \Dsb $& $ 1 $ 
 & $ \sqrt{\frac{ 1 }{ 3 }} 
   $ & $ 0 $ & $ 0 $ 
 & $ \sqrt{ 2 } $ 
 & $ -\sqrt{ 2 } $ 
 & $ \sqrt{\frac{ 8 }{ 3 }} 
   $ & $ 0 $ 
 & $ -\sqrt{\frac{ 8 }{ 3 }} 
   $ & $ 0 $ 
 & $ -\sqrt{\frac{ 4 }{ 3 }} 
   $ \\
$ \Cascadacp \Db $
 & $ -\sqrt{\frac{ 1 }{ 2 }} 
   $ & $ -\sqrt{\frac{ 1 }{ 6 
     }} $ & $ 0 $ 
 & $ \sqrt{ 2 } $ 
 & $ 1 $ & $ 1 $ 
 & $ \sqrt{\frac{ 16 }{ 3 }} 
   $ & $ -\sqrt{\frac{ 8 }{ 3 
     }} $ 
 & $ -\sqrt{\frac{ 4 }{ 3 }} 
   $ & $ -\sqrt{\frac{ 4 }{ 3 
     }} $ 
 & $ -\sqrt{\frac{ 2 }{ 3 }} 
   $ \\
$ \Cascadac \DSb $
 & $ -\sqrt{\frac{ 9 }{ 2 }} 
   $ & $ -\sqrt{\frac{ 3 }{ 2 
     }} $ & $ 0 $ 
 & $ -\sqrt{ 2 } $ 
 & $ 1 $ & $ 1 $ 
 & $ 0 $ 
 & $ -\sqrt{\frac{ 8 }{ 3 }} 
   $ & $ \sqrt{\frac{ 4 }{ 3 
    }} $ 
 & $ -\sqrt{\frac{ 4 }{ 3 }} 
   $ & $ \sqrt{\frac{ 2 }{ 3 
    }} $ \\
$ \SigmaS \Jpsi $& $ 0 $ 
 & $ 0 $ & $ 0 $ 
 & $ \sqrt{\frac{ 8 }{ 3 }} 
   $ & $ \sqrt{\frac{ 16 }{ 3 
    }} $ & $ 0 $ & $ 0 $ 
 & $ -\sqrt{\frac{ 8 }{ 9 }} 
   $ & $ -\frac{ 4 }{ 3  } 
   $ & $ -\frac{ 2 }{ 3  } 
   $ & $ -\sqrt{\frac{ 8 }{ 9 
     }} $ \\
$ \Sigmac \DsSb $
 & $ \sqrt{\frac{ 1 }{ 3 }} 
   $ & $ \frac{ 5 }{ 3  } 
   $ & $ -\sqrt{ 2 } $ 
 & $ 0 $ 
 & $ -\sqrt{\frac{ 8 }{ 3 }} 
   $ & $ -\sqrt{\frac{ 8 }{ 3 
     }} $ 
 & $ -\sqrt{\frac{ 8 }{ 9 }} 
   $ & $ 0 $ 
 & $ -\sqrt{\frac{ 2 }{ 9 }} 
   $ & $ 0 $ 
 & $ \frac{ 2 }{ 3  } $ \\
$ \Cascadacp \DSb $
 & $ -\sqrt{\frac{ 1 }{ 6 }} 
   $ & $ -\sqrt{\frac{ 25 }{ 18 
     }} $ & $ 1 $ 
 & $ -\sqrt{\frac{ 8 }{ 3 }} 
   $ & $ -\sqrt{\frac{ 4 }{ 3 
     }} $ & $ \sqrt{\frac{ 4 
    }{ 3 }} $ 
 & $ -\frac{ 4 }{ 3  } $ 
 & $ -\sqrt{\frac{ 2 }{ 9 }} 
   $ & $ -\frac{ 1 }{ 3  } 
   $ & $ \frac{ 2 }{ 3  } 
   $ & $ \sqrt{\frac{ 2 }{ 9 
    }} $ \\
$ \SigmacS \DsSb $
 & $ \sqrt{\frac{ 8 }{ 3 }} 
   $ & $ -\sqrt{\frac{ 8 }{ 9 
     }} $ & $ 2 $ 
 & $ 0 $ 
 & $ -\sqrt{\frac{ 4 }{ 3 }} 
   $ & $ -\sqrt{\frac{ 4 }{ 3 
     }} $ 
 & $ -\frac{ 2 }{ 3  } $ 
 & $ 0 $ 
 & $ \frac{ 2 }{ 3  } $ 
 & $ 0 $ 
 & $ -\sqrt{\frac{ 8 }{ 9 }} 
   $ \\
$ \CascadacS \DSb $
 & $ -\sqrt{\frac{ 4 }{ 3 }} 
   $ & $ \frac{ 2 }{ 3  } 
   $ & $ -\sqrt{ 2 } $ 
 & $ -\sqrt{\frac{ 4 }{ 3 }} 
   $ & $ -\sqrt{\frac{ 2 }{ 3 
     }} $ & $ \sqrt{\frac{ 2 
    }{ 3 }} $ 
 & $ -\sqrt{\frac{ 8 }{ 9 }} 
   $ & $ \frac{ 2 }{ 3  } 
   $ & $ \sqrt{\frac{ 2 }{ 9 
    }} $ 
 & $ -\sqrt{\frac{ 8 }{ 9 }} 
   $ & $ -\frac{ 2 }{ 3  } 
   $ \\
\end{tabular}
\end{table}

\begin{table}
\caption{
  $ C=0$, $ S=-1$, $ I=1$, $ J= 3/2$.
}
\label{tab:0m134}
\begin{tabular}{r|rrrrrrrrrrrrrrrrrrrrrrrrrrrrrrrrrrrrrrrrrrrrrrrr}
 &  $ \Sigma \Jpsi $ &  $ \SigmaS 
  \etac $ &  $ \Cascadac \DSb $ 
 &  $ \SigmaS \Jpsi $ &  $ \SigmacS 
  \Dsb $ &  $ \CascadacS \Db $ 
 &  $ \Sigmac \DsSb $ &  $ \Cascadacp 
  \DSb $ &  $ \SigmacS \DsSb $ 
 &  $ \CascadacS \DSb $\\
\hline
$ \Sigma \Jpsi $& $ 0 $ 
 & $ 0 $ & $ \sqrt{ 6 } 
   $ & $ 0 $ 
 & $ -\sqrt{\frac{ 4 }{ 3 }} 
   $ & $ \sqrt{\frac{ 2 }{ 3 
    }} $ & $ \frac{ 2 }{ 3 
     } $ 
 & $ -\sqrt{\frac{ 2 }{ 9 }} 
   $ & $ -\sqrt{\frac{ 20 }{ 9 
     }} $ & $ \sqrt{\frac{ 10 
    }{ 9 }} $ \\
$ \SigmaS \etac $& $ 0 $ 
 & $ 0 $ & $ 0 $ 
 & $ 0 $ & $ 1 $ 
 & $ \sqrt{ 2 } $ 
 & $ -\sqrt{\frac{ 4 }{ 3 }} 
   $ & $ -\sqrt{\frac{ 8 }{ 3 
     }} $ 
 & $ -\sqrt{\frac{ 5 }{ 3 }} 
   $ & $ -\sqrt{\frac{ 10 }{ 3 
     }} $ \\
$ \Cascadac \DSb $
 & $ \sqrt{ 6 } $ 
 & $ 0 $ & $ 1 $ 
 & $ 0 $ 
 & $ -\sqrt{ 2 } $ 
 & $ 1 $ 
 & $ \sqrt{\frac{ 2 }{ 3 }} 
   $ & $ -\sqrt{\frac{ 1 }{ 3 
     }} $ 
 & $ -\sqrt{\frac{ 10 }{ 3 }} 
   $ & $ \sqrt{\frac{ 5 }{ 3 
    }} $ \\
$ \SigmaS \Jpsi $& $ 0 $ 
 & $ 0 $ & $ 0 $ 
 & $ 0 $ 
 & $ -\sqrt{\frac{ 5 }{ 3 }} 
   $ & $ -\sqrt{\frac{ 10 }{ 3 
     }} $ 
 & $ -\sqrt{\frac{ 20 }{ 9 }} 
   $ & $ -\sqrt{\frac{ 40 }{ 9 
     }} $ & $ \frac{ 1 }{ 3 
     } $ & $ \sqrt{\frac{ 2 
    }{ 9 }} $ \\
$ \SigmacS \Dsb $
 & $ -\sqrt{\frac{ 4 }{ 3 }} 
   $ & $ 1 $ 
 & $ -\sqrt{ 2 } $ 
 & $ -\sqrt{\frac{ 5 }{ 3 }} 
   $ & $ 0 $ 
 & $ \sqrt{ 2 } $ 
 & $ 0 $ 
 & $ \sqrt{\frac{ 2 }{ 3 }} 
   $ & $ 0 $ 
 & $ -\sqrt{\frac{ 10 }{ 3 }} 
   $ \\
$ \CascadacS \Db $
 & $ \sqrt{\frac{ 2 }{ 3 }} 
   $ & $ \sqrt{ 2 } $ 
 & $ 1 $ 
 & $ -\sqrt{\frac{ 10 }{ 3 }} 
   $ & $ \sqrt{ 2 } $ 
 & $ 1 $ 
 & $ \sqrt{\frac{ 2 }{ 3 }} 
   $ & $ \sqrt{\frac{ 1 }{ 3 
    }} $ 
 & $ -\sqrt{\frac{ 10 }{ 3 }} 
   $ & $ -\sqrt{\frac{ 5 }{ 3 
     }} $ \\
$ \Sigmac \DsSb $
 & $ \frac{ 2 }{ 3  } $ 
 & $ -\sqrt{\frac{ 4 }{ 3 }} 
   $ & $ \sqrt{\frac{ 2 }{ 3 
    }} $ 
 & $ -\sqrt{\frac{ 20 }{ 9 }} 
   $ & $ 0 $ 
 & $ \sqrt{\frac{ 2 }{ 3 }} 
   $ & $ 0 $ 
 & $ \sqrt{\frac{ 50 }{ 9 }} 
   $ & $ 0 $ 
 & $ \sqrt{\frac{ 10 }{ 9 }} 
   $ \\
$ \Cascadacp \DSb $
 & $ -\sqrt{\frac{ 2 }{ 9 }} 
   $ & $ -\sqrt{\frac{ 8 }{ 3 
     }} $ 
 & $ -\sqrt{\frac{ 1 }{ 3 }} 
   $ & $ -\sqrt{\frac{ 40 }{ 9 
     }} $ & $ \sqrt{\frac{ 2 
    }{ 3 }} $ 
 & $ \sqrt{\frac{ 1 }{ 3 }} 
   $ & $ \sqrt{\frac{ 50 }{ 9 
    }} $ & $ \frac{ 5 }{ 3 
     } $ & $ \sqrt{\frac{ 10 
    }{ 9 }} $ 
 & $ \sqrt{\frac{ 5 }{ 9 }} 
   $ \\
$ \SigmacS \DsSb $
 & $ -\sqrt{\frac{ 20 }{ 9 }} 
   $ & $ -\sqrt{\frac{ 5 }{ 3 
     }} $ 
 & $ -\sqrt{\frac{ 10 }{ 3 }} 
   $ & $ \frac{ 1 }{ 3  } 
   $ & $ 0 $ 
 & $ -\sqrt{\frac{ 10 }{ 3 }} 
   $ & $ 0 $ 
 & $ \sqrt{\frac{ 10 }{ 9 }} 
   $ & $ 0 $ 
 & $ \sqrt{\frac{ 2 }{ 9 }} 
   $ \\
$ \CascadacS \DSb $
 & $ \sqrt{\frac{ 10 }{ 9 }} 
   $ & $ -\sqrt{\frac{ 10 }{ 3 
     }} $ & $ \sqrt{\frac{ 5 
    }{ 3 }} $ 
 & $ \sqrt{\frac{ 2 }{ 9 }} 
   $ & $ -\sqrt{\frac{ 10 }{ 3 
     }} $ 
 & $ -\sqrt{\frac{ 5 }{ 3 }} 
   $ & $ \sqrt{\frac{ 10 }{ 9 
    }} $ & $ \sqrt{\frac{ 5 
    }{ 9 }} $ 
 & $ \sqrt{\frac{ 2 }{ 9 }} 
   $ & $ \frac{ 1 }{ 3  } 
   $ \\
\end{tabular}
\end{table}

\begin{table}
\caption{
  $ C=0$, $ S=-1$, $ I=1$, $ J= 5/2$.
}
\label{tab:0m136}
\begin{tabular}{r|rrrrrrrrrrrrrrrrrrrrrrrrrrrrrrrrrrrrrrrrrrrrrrrr}
 &  $ \SigmaS \Jpsi $ 
 &  $ \SigmacS \DsSb $ &  $ \CascadacS 
  \DSb $\\
\hline
$ \SigmaS \Jpsi $& $ 0 $ 
 & $ 2 $ & $ \sqrt{ 8 } 
   $ \\
$ \SigmacS \DsSb $& $ 2 $ 
 & $ 0 $ & $ \sqrt{ 8 } 
   $ \\
$ \CascadacS \DSb $
 & $ \sqrt{ 8 } $ 
 & $ \sqrt{ 8 } $ 
 & $ 2 $ \\
\end{tabular}
\end{table}

\begin{table}
\caption{
  $ C=0$, $ S=-2$, $ I=1/2$, $ J= 1/2$.
}
\label{tab:0m222}
\begin{tabular}{r|rrrrrrrrrrrrrrrrrrrrrrrrrrrrrrrrrrrrrrrrrrrrrrrr}
 &  $ \Cascada \etac $ 
 &  $ \Cascada \Jpsi $ &  $ \Cascadac 
  \Dsb $ &  $ \Cascadacp \Dsb $ 
 &  $ \Omegac \Db $ &  $ \Cascadac 
  \DsSb $ &  $ \CascadaS \Jpsi $ 
 &  $ \Cascadacp \DsSb $ 
 &  $ \Omegac \DSb $ &  $ \CascadacS 
  \DsSb $ &  $ \OmegacS \DSb $\\
\hline
$ \Cascada \etac $& $ 0 $ 
 & $ 0 $ 
 & $ \sqrt{\frac{ 3 }{ 2 }} 
   $ & $ \sqrt{\frac{ 1 }{ 2 
    }} $ & $ -1 $ 
 & $ -\sqrt{\frac{ 9 }{ 2 }} 
   $ & $ 0 $ 
 & $ \sqrt{\frac{ 1 }{ 6 }} 
   $ & $ -\sqrt{\frac{ 1 }{ 3 
     }} $ & $ \sqrt{\frac{ 4 
    }{ 3 }} $ 
 & $ -\sqrt{\frac{ 8 }{ 3 }} 
   $ \\
$ \Cascada \Jpsi $& $ 0 $ 
 & $ 0 $ 
 & $ -\sqrt{\frac{ 9 }{ 2 }} 
   $ & $ \sqrt{\frac{ 1 }{ 6 
    }} $ 
 & $ -\sqrt{\frac{ 1 }{ 3 }} 
   $ & $ -\sqrt{\frac{ 3 }{ 2 
     }} $ & $ 0 $ 
 & $ \sqrt{\frac{ 25 }{ 18 }} 
   $ & $ -\frac{ 5 }{ 3  } 
   $ & $ -\frac{ 2 }{ 3  } 
   $ & $ \sqrt{\frac{ 8 }{ 9 
    }} $ \\
$ \Cascadac \Dsb $
 & $ \sqrt{\frac{ 3 }{ 2 }} 
   $ & $ -\sqrt{\frac{ 9 }{ 2 
     }} $ & $ 1 $ 
 & $ 0 $ & $ 0 $ 
 & $ 0 $ & $ 0 $ 
 & $ -1 $ 
 & $ \sqrt{ 2 } $ 
 & $ \sqrt{ 2 } $ 
 & $ -2 $ \\
$ \Cascadacp \Dsb $
 & $ \sqrt{\frac{ 1 }{ 2 }} 
   $ & $ \sqrt{\frac{ 1 }{ 6 
    }} $ & $ 0 $ & $ 1 $ 
 & $ \sqrt{ 2 } $ 
 & $ -1 $ 
 & $ \sqrt{\frac{ 16 }{ 3 }} 
   $ & $ -\sqrt{\frac{ 4 }{ 3 
     }} $ 
 & $ -\sqrt{\frac{ 8 }{ 3 }} 
   $ & $ -\sqrt{\frac{ 2 }{ 3 
     }} $ 
 & $ -\sqrt{\frac{ 4 }{ 3 }} 
   $ \\
$ \Omegac \Db $& $ -1 $ 
 & $ -\sqrt{\frac{ 1 }{ 3 }} 
   $ & $ 0 $ 
 & $ \sqrt{ 2 } $ 
 & $ 0 $ & $ \sqrt{ 2 } 
   $ & $ \sqrt{\frac{ 8 }{ 3 
    }} $ 
 & $ -\sqrt{\frac{ 8 }{ 3 }} 
   $ & $ 0 $ 
 & $ -\sqrt{\frac{ 4 }{ 3 }} 
   $ & $ 0 $ \\
$ \Cascadac \DsSb $
 & $ -\sqrt{\frac{ 9 }{ 2 }} 
   $ & $ -\sqrt{\frac{ 3 }{ 2 
     }} $ & $ 0 $ 
 & $ -1 $ 
 & $ \sqrt{ 2 } $ 
 & $ 1 $ & $ 0 $ 
 & $ -\sqrt{\frac{ 4 }{ 3 }} 
   $ & $ \sqrt{\frac{ 8 }{ 3 
    }} $ 
 & $ -\sqrt{\frac{ 2 }{ 3 }} 
   $ & $ \sqrt{\frac{ 4 }{ 3 
    }} $ \\
$ \CascadaS \Jpsi $& $ 0 $ 
 & $ 0 $ & $ 0 $ 
 & $ \sqrt{\frac{ 16 }{ 3 }} 
   $ & $ \sqrt{\frac{ 8 }{ 3 
    }} $ & $ 0 $ & $ 0 $ 
 & $ -\frac{ 4 }{ 3  } $ 
 & $ -\sqrt{\frac{ 8 }{ 9 }} 
   $ & $ -\sqrt{\frac{ 8 }{ 9 
     }} $ 
 & $ -\frac{ 2 }{ 3  } $ \\
$ \Cascadacp \DsSb $
 & $ \sqrt{\frac{ 1 }{ 6 }} 
   $ & $ \sqrt{\frac{ 25 }{ 18 
    }} $ & $ -1 $ 
 & $ -\sqrt{\frac{ 4 }{ 3 }} 
   $ & $ -\sqrt{\frac{ 8 }{ 3 
     }} $ 
 & $ -\sqrt{\frac{ 4 }{ 3 }} 
   $ & $ -\frac{ 4 }{ 3  } 
   $ & $ -\frac{ 1 }{ 3  } 
   $ & $ -\sqrt{\frac{ 2 }{ 9 
     }} $ & $ \sqrt{\frac{ 2 
    }{ 9 }} $ 
 & $ \frac{ 2 }{ 3  } $ \\
$ \Omegac \DSb $
 & $ -\sqrt{\frac{ 1 }{ 3 }} 
   $ & $ -\frac{ 5 }{ 3  } 
   $ & $ \sqrt{ 2 } $ 
 & $ -\sqrt{\frac{ 8 }{ 3 }} 
   $ & $ 0 $ 
 & $ \sqrt{\frac{ 8 }{ 3 }} 
   $ & $ -\sqrt{\frac{ 8 }{ 9 
     }} $ 
 & $ -\sqrt{\frac{ 2 }{ 9 }} 
   $ & $ 0 $ 
 & $ \frac{ 2 }{ 3  } $ 
 & $ 0 $ \\
$ \CascadacS \DsSb $
 & $ \sqrt{\frac{ 4 }{ 3 }} 
   $ & $ -\frac{ 2 }{ 3  } 
   $ & $ \sqrt{ 2 } $ 
 & $ -\sqrt{\frac{ 2 }{ 3 }} 
   $ & $ -\sqrt{\frac{ 4 }{ 3 
     }} $ 
 & $ -\sqrt{\frac{ 2 }{ 3 }} 
   $ & $ -\sqrt{\frac{ 8 }{ 9 
     }} $ & $ \sqrt{\frac{ 2 
    }{ 9 }} $ 
 & $ \frac{ 2 }{ 3  } $ 
 & $ -\frac{ 2 }{ 3  } $ 
 & $ -\sqrt{\frac{ 8 }{ 9 }} 
   $ \\
$ \OmegacS \DSb $
 & $ -\sqrt{\frac{ 8 }{ 3 }} 
   $ & $ \sqrt{\frac{ 8 }{ 9 
    }} $ & $ -2 $ 
 & $ -\sqrt{\frac{ 4 }{ 3 }} 
   $ & $ 0 $ 
 & $ \sqrt{\frac{ 4 }{ 3 }} 
   $ & $ -\frac{ 2 }{ 3  } 
   $ & $ \frac{ 2 }{ 3  } 
   $ & $ 0 $ 
 & $ -\sqrt{\frac{ 8 }{ 9 }} 
   $ & $ 0 $ \\
\end{tabular}
\end{table}

\begin{table}
\caption{
  $ C=0$, $ S=-2$, $ I=1/2$, $ J= 3/2$.
}
\label{tab:0m224}
\begin{tabular}{r|rrrrrrrrrrrrrrrrrrrrrrrrrrrrrrrrrrrrrrrrrrrrrrrr}
 &  $ \Cascada \Jpsi $ 
 &  $ \CascadaS \etac $ &  $ \Cascadac 
  \DsSb $ &  $ \CascadacS \Dsb $ 
 &  $ \CascadaS \Jpsi $ &  $ \OmegacS 
  \Db $ &  $ \Cascadacp \DsSb $ 
 &  $ \Omegac \DSb $ &  $ \CascadacS 
  \DsSb $ &  $ \OmegacS \DSb $\\
\hline
$ \Cascada \Jpsi $& $ 0 $ 
 & $ 0 $ & $ \sqrt{ 6 } 
   $ & $ -\sqrt{\frac{ 2 }{ 3 
     }} $ & $ 0 $ 
 & $ \sqrt{\frac{ 4 }{ 3 }} 
   $ & $ \sqrt{\frac{ 2 }{ 9 
    }} $ 
 & $ -\frac{ 2 }{ 3  } $ 
 & $ -\sqrt{\frac{ 10 }{ 9 }} 
   $ & $ \sqrt{\frac{ 20 }{ 9 
    }} $ \\
$ \CascadaS \etac $& $ 0 $ 
 & $ 0 $ & $ 0 $ 
 & $ \sqrt{ 2 } $ 
 & $ 0 $ & $ 1 $ 
 & $ -\sqrt{\frac{ 8 }{ 3 }} 
   $ & $ -\sqrt{\frac{ 4 }{ 3 
     }} $ 
 & $ -\sqrt{\frac{ 10 }{ 3 }} 
   $ & $ -\sqrt{\frac{ 5 }{ 3 
     }} $ \\
$ \Cascadac \DsSb $
 & $ \sqrt{ 6 } $ 
 & $ 0 $ & $ 1 $ 
 & $ -1 $ & $ 0 $ 
 & $ \sqrt{ 2 } $ 
 & $ \sqrt{\frac{ 1 }{ 3 }} 
   $ & $ -\sqrt{\frac{ 2 }{ 3 
     }} $ 
 & $ -\sqrt{\frac{ 5 }{ 3 }} 
   $ & $ \sqrt{\frac{ 10 }{ 3 
    }} $ \\
$ \CascadacS \Dsb $
 & $ -\sqrt{\frac{ 2 }{ 3 }} 
   $ & $ \sqrt{ 2 } $ 
 & $ -1 $ & $ 1 $ 
 & $ -\sqrt{\frac{ 10 }{ 3 }} 
   $ & $ \sqrt{ 2 } $ 
 & $ \sqrt{\frac{ 1 }{ 3 }} 
   $ & $ \sqrt{\frac{ 2 }{ 3 
    }} $ 
 & $ -\sqrt{\frac{ 5 }{ 3 }} 
   $ & $ -\sqrt{\frac{ 10 }{ 3 
     }} $ \\
$ \CascadaS \Jpsi $& $ 0 $ 
 & $ 0 $ & $ 0 $ 
 & $ -\sqrt{\frac{ 10 }{ 3 }} 
   $ & $ 0 $ 
 & $ -\sqrt{\frac{ 5 }{ 3 }} 
   $ & $ -\sqrt{\frac{ 40 }{ 9 
     }} $ 
 & $ -\sqrt{\frac{ 20 }{ 9 }} 
   $ & $ \sqrt{\frac{ 2 }{ 9 
    }} $ & $ \frac{ 1 }{ 3 
     } $ \\
$ \OmegacS \Db $
 & $ \sqrt{\frac{ 4 }{ 3 }} 
   $ & $ 1 $ 
 & $ \sqrt{ 2 } $ 
 & $ \sqrt{ 2 } $ 
 & $ -\sqrt{\frac{ 5 }{ 3 }} 
   $ & $ 0 $ 
 & $ \sqrt{\frac{ 2 }{ 3 }} 
   $ & $ 0 $ 
 & $ -\sqrt{\frac{ 10 }{ 3 }} 
   $ & $ 0 $ \\
$ \Cascadacp \DsSb $
 & $ \sqrt{\frac{ 2 }{ 9 }} 
   $ & $ -\sqrt{\frac{ 8 }{ 3 
     }} $ & $ \sqrt{\frac{ 1 
    }{ 3 }} $ 
 & $ \sqrt{\frac{ 1 }{ 3 }} 
   $ & $ -\sqrt{\frac{ 40 }{ 9 
     }} $ & $ \sqrt{\frac{ 2 
    }{ 3 }} $ 
 & $ \frac{ 5 }{ 3  } $ 
 & $ \sqrt{\frac{ 50 }{ 9 }} 
   $ & $ \sqrt{\frac{ 5 }{ 9 
    }} $ & $ \sqrt{\frac{ 10 
    }{ 9 }} $ \\
$ \Omegac \DSb $
 & $ -\frac{ 2 }{ 3  } $ 
 & $ -\sqrt{\frac{ 4 }{ 3 }} 
   $ & $ -\sqrt{\frac{ 2 }{ 3 
     }} $ & $ \sqrt{\frac{ 2 
    }{ 3 }} $ 
 & $ -\sqrt{\frac{ 20 }{ 9 }} 
   $ & $ 0 $ 
 & $ \sqrt{\frac{ 50 }{ 9 }} 
   $ & $ 0 $ 
 & $ \sqrt{\frac{ 10 }{ 9 }} 
   $ & $ 0 $ \\
$ \CascadacS \DsSb $
 & $ -\sqrt{\frac{ 10 }{ 9 }} 
   $ & $ -\sqrt{\frac{ 10 }{ 3 
     }} $ 
 & $ -\sqrt{\frac{ 5 }{ 3 }} 
   $ & $ -\sqrt{\frac{ 5 }{ 3 
     }} $ & $ \sqrt{\frac{ 2 
    }{ 9 }} $ 
 & $ -\sqrt{\frac{ 10 }{ 3 }} 
   $ & $ \sqrt{\frac{ 5 }{ 9 
    }} $ & $ \sqrt{\frac{ 10 
    }{ 9 }} $ 
 & $ \frac{ 1 }{ 3  } $ 
 & $ \sqrt{\frac{ 2 }{ 9 }} 
   $ \\
$ \OmegacS \DSb $
 & $ \sqrt{\frac{ 20 }{ 9 }} 
   $ & $ -\sqrt{\frac{ 5 }{ 3 
     }} $ & $ \sqrt{\frac{ 10 
    }{ 3 }} $ 
 & $ -\sqrt{\frac{ 10 }{ 3 }} 
   $ & $ \frac{ 1 }{ 3  } 
   $ & $ 0 $ 
 & $ \sqrt{\frac{ 10 }{ 9 }} 
   $ & $ 0 $ 
 & $ \sqrt{\frac{ 2 }{ 9 }} 
   $ & $ 0 $ \\
\end{tabular}
\end{table}

\begin{table}
\caption{
  $ C=0$, $ S=-2$, $ I=1/2$, $ J= 5/2$.
}
\label{tab:0m226}
\begin{tabular}{r|rrrrrrrrrrrrrrrrrrrrrrrrrrrrrrrrrrrrrrrrrrrrrrrr}
 &  $ \CascadaS \Jpsi $ 
 &  $ \CascadacS \DsSb $ 
 &  $ \OmegacS \DSb $\\
\hline
$ \CascadaS \Jpsi $& $ 0 $ 
 & $ \sqrt{ 8 } $ 
 & $ 2 $ \\
$ \CascadacS \DsSb $
 & $ \sqrt{ 8 } $ 
 & $ 2 $ & $ \sqrt{ 8 } 
   $ \\
$ \OmegacS \DSb $& $ 2 $ 
 & $ \sqrt{ 8 } $ 
 & $ 0 $ \\
\end{tabular}
\end{table}

\begin{table}
\caption{
  $ C=0$, $ S=-3$, $ I=0$, $ J= 1/2$.
}
\label{tab:0m312}
\begin{tabular}{r|rrrrrrrrrrrrrrrrrrrrrrrrrrrrrrrrrrrrrrrrrrrrrrrr}
 &  $ \Omegac \Dsb $ &  $ \Omega 
  \Jpsi $ &  $ \Omegac \DsSb $ 
 &  $ \OmegacS \DsSb $\\
\hline
$ \Omegac \Dsb $& $ 2 $ 
 & $ \sqrt{ 8 } $ 
 & $ -\sqrt{\frac{ 16 }{ 3 }} 
   $ & $ -\sqrt{\frac{ 8 }{ 3 
     }} $ \\
$ \Omega \Jpsi $
 & $ \sqrt{ 8 } $ 
 & $ 0 $ 
 & $ -\sqrt{\frac{ 8 }{ 3 }} 
   $ & $ -\sqrt{\frac{ 4 }{ 3 
     }} $ \\
$ \Omegac \DsSb $
 & $ -\sqrt{\frac{ 16 }{ 3 }} 
   $ & $ -\sqrt{\frac{ 8 }{ 3 
     }} $ 
 & $ -\frac{ 2 }{ 3  } $ 
 & $ \sqrt{\frac{ 8 }{ 9 }} 
   $ \\
$ \OmegacS \DsSb $
 & $ -\sqrt{\frac{ 8 }{ 3 }} 
   $ & $ -\sqrt{\frac{ 4 }{ 3 
     }} $ & $ \sqrt{\frac{ 8 
    }{ 9 }} $ 
 & $ -\frac{ 4 }{ 3  } $ \\
\end{tabular}
\end{table}

\begin{table}
\caption{
  $ C=0$, $ S=-3$, $ I=0$, $ J= 3/2$.
}
\label{tab:0m314}
\begin{tabular}{r|rrrrrrrrrrrrrrrrrrrrrrrrrrrrrrrrrrrrrrrrrrrrrrrr}
 &  $ \Omega \etac $ &  $ \OmegacS 
  \Dsb $ &  $ \Omega \Jpsi $ 
 &  $ \Omegac \DsSb $ &  $ \OmegacS 
  \DsSb $\\
\hline
$ \Omega \etac $& $ 0 $ 
 & $ \sqrt{ 3 } $ 
 & $ 0 $ & $ -2 $ 
 & $ -\sqrt{ 5 } $ \\
$ \OmegacS \Dsb $
 & $ \sqrt{ 3 } $ 
 & $ 2 $ 
 & $ -\sqrt{ 5 } $ 
 & $ \sqrt{\frac{ 4 }{ 3 }} 
   $ & $ -\sqrt{\frac{ 20 }{ 3 
     }} $ \\
$ \Omega \Jpsi $& $ 0 $ 
 & $ -\sqrt{ 5 } $ 
 & $ 0 $ 
 & $ -\sqrt{\frac{ 20 }{ 3 }} 
   $ & $ \sqrt{\frac{ 1 }{ 3 
    }} $ \\
$ \Omegac \DsSb $& $ -2 $ 
 & $ \sqrt{\frac{ 4 }{ 3 }} 
   $ & $ -\sqrt{\frac{ 20 }{ 3 
     }} $ & $ \frac{ 10 }{ 3 
     } $ & $ \sqrt{\frac{ 20 
    }{ 9 }} $ \\
$ \OmegacS \DsSb $
 & $ -\sqrt{ 5 } $ 
 & $ -\sqrt{\frac{ 20 }{ 3 }} 
   $ & $ \sqrt{\frac{ 1 }{ 3 
    }} $ & $ \sqrt{\frac{ 20 
    }{ 9 }} $ 
 & $ \frac{ 2 }{ 3  } $ \\
\end{tabular}
\end{table}

\begin{table}
\caption{
  $ C=0$, $ S=-3$, $ I=0$, $ J= 5/2$.
}
\label{tab:0m316}
\begin{tabular}{r|rrrrrrrrrrrrrrrrrrrrrrrrrrrrrrrrrrrrrrrrrrrrrrrr}
 &  $ \Omega \Jpsi $ &  $ \OmegacS 
  \DsSb $\\
\hline
$ \Omega \Jpsi $& $ 0 $ 
 & $ \sqrt{ 12 } $ \\
$ \OmegacS \DsSb $
 & $ \sqrt{ 12 } $ 
 & $ 4 $ \\
\end{tabular}
\end{table}

\end{document}